\documentclass[review]{elsarticle}
\pdfoutput=1
\usepackage{lineno}
\modulolinenumbers[5]
\usepackage{graphicx}
\usepackage{lipsum}
\makeatletter
\def\ps@pprintTitle{%
 \let\@oddhead\@empty
 \let\@evenhead\@empty
 \def\@oddfoot{}%
 \let\@evenfoot\@oddfoot}
\makeatother
\usepackage{float}
\usepackage{geometry}
\usepackage[caption=false]{subfig}
\makeatletter
\renewcommand{\fnum@figure}{Fig. \thefigure}
\makeatother
\usepackage{subfig}
\usepackage{caption}
\usepackage{comment}
\usepackage{physics}
\usepackage{amsmath}
\usepackage{changepage}
\usepackage{array}
\usepackage{siunitx}
\newcolumntype{C}[1]{>{\centering}m{#1}}

\begin{document}

\begin{frontmatter}

\title{Modeling of pressure drop and heat transfer for flow boiling in a mini/micro-channel of rectangular cross-section}

\author{Shashwat Jain, Prasanna Jayaramu and Sateesh Gedupudi\fnref{myfootnote}}
\address{Heat Transfer and Thermal Power Laboratory, Department of Mechanical Engineering, IIT Madras, Chennai, India}
\fntext[myfootnote]{Corresponding author, Email: sateeshg@iitm.ac.in}




\begin{abstract}
In the present study, a 1-dimensional model is proposed to estimate the pressure drop and heat transfer coefficient for flow boiling in a rectangular microchannel. The present work takes into account the pressure fluctuations caused due to the confined bubble growth and the effect of pressure fluctuations on the heat transfer characteristics. The heat transfer model considers five zones, namely, liquid slug, partially confined bubble, fully confined (elongated) bubble, partial dryout and full dry-out. The model incorporates the thinning of liquid film due to shear stress at liquid-vapour interface in addition to evaporation. The transient fluctuations in pressure and heat transfer coefficient, along with the time-averaged ones, are verified with the experimental data available in the literature. Heat transfer characteristics with flow reversal caused by inlet compressibility are also presented.
\end{abstract}


\end{frontmatter}

\setlength{\parindent}{0in}

\section{Introduction}

 Flow boiling in microchannel has been extensively researched in recent times due to its higher heat removal capacity, caused by its high heat transfer surface area to fluid flow volume ratio coupled with the utilization of latent heat of vaporization.  \citet{Harirchian2010} observed single phase, bubbly region, slug - plug, churn and annular flow patterns based on the operating conditions. The heat transfer capabilities are highly affected by the flow patterns. In general, the heat transfer coefficient associated with the slug-plug or the elongated bubble  is higher due to the presence of thin liquid film surrounding the bubble. \citet{Mirmanto2016} experimentally investigated flow boiling in a single mini/micro-channel and observed a decrease in heat transfer coefficient with the increase in the local thermodynamic quality. \citet{Yin2016} observed an increasing and then a decreasing trend in heat transfer coefficient with local thermodynamic quality. \citet{sobierska2007heat} observed single phase, bubble and slug flow during boiling in a narrow channel. The maximum heat transfer coefficient was found in the slug region.
 
\subsection{Transient variation}
  
 \citet{Jagirdar2016} investigated the transient local heat transfer coefficient at different locations in a single rectangular channel with dimensions as \(0.42 \times 2.54 \times 25.4 \, mm\) during subcooled flow boiling. The synchronized flow visualization showed that the variation in the heat transfer coefficient is due to the passing of liquid, vapour bubble (surrounded by liquid) and dry-out, cyclically over the given location. The increase in the local heat transfer coefficient was observed due to the passing of bubble with evaporating thin film. Similar observations were also reported by \citet{Bigham2015} for flow boiling experiment conducted in a microchannel with nanostructures. A sudden drop was observed in wall temperature during the passing of the bubble. \citet{Jafari2015} performed Computational Fluid Dynamics (CFD) simulations using the phase-field method to study the hydrodynamics and heat transfer characteristics of an elongated bubble (present in superheated liquid) inside the microchannel. The results showed high heat transfer coefficient on the bottom and side walls in the confined bubble region. \citet{sun2018transient} demonstrated and formulated the effect of shear stress on the evaporating thin film inside a microtube. The results showed variation in the heat transfer coefficient due to the cyclic process of elongated bubble growth, dry out and liquid slug. The residence time of each region was also measured and it was found that with the increase in the heat flux, liquid slug time decreases and the elongated bubble or dry region time increases. Transient pressure fluctuations and heat transfer coefficient were observed by \citet{Barber2011} for a single microchannel using n-Pentane as the working fluid. The confined bubble was observed due to the small aspect ratio and the bubble expanded both in the upstream and downstream directions. Pressure fluctuations were observed due to confinement of the bubble inside the channel. \citet{Rao2015a} experimentally studied transient heat transfer coefficient in multiple flow regimes such as bubbly, slug and annular in a rectangular microchannel. The periodic variation in the heat transfer coefficient was observed with the peak due to explosive bubble growth and presence of evaporating thin film which is followed by a sharp decrease due to the emergence of local dry out.
 
 Rapid confined bubble growth causes pressure fluctuations during flow boiling in mini/micro-channels. These fluctuations in the presence of upstream compressibility such as trapped non-condensables, pump characteristics, subcooled boiling in the upstream region/preheater \citet{Gedupudi2011} and interaction between the neighbouring channels (in the case of parallel microchannels), as summarized by \citet{Prajapati2017}, leads to flow reversal. \citet{Liu2013} extensively studied the effect of upstream compressibility on flow boiling. Heat transfer coefficient was found to increase with increase in the compressible volume, for certain range of operating conditions. On the other hand, the inlet pressure was found to be oscillating with lower frequency. On further increasing the compressible volume, there was a decrease in pressure fluctuation frequency.  \citet{Karayiannis2017} pointed out the lack of data on the effect of flow reversal on heat transfer coefficient. \citet{kandlikar2001high} observed pressure fluctuation and flow reversal due to the interaction between the neighbouring channels in a parallel microchannel heat sink. \citet{Kenning2001} and \citet{Yan1999} observed pressure fluctuations under both negligible and high upstream compressibility conditions. \citet{Barber2010} studied bubble dynamics inside a rectangular microchannel with FC-72 as the working fluid and reported a sharp rise in pressure along with flow reversal. The experimental investigations carried by \citet{Gedupudi2011} indicates that in the presence of upstream compressibility, the upstream end moved towards the  inlet of the channel. On the other hand, no flow reversal was observed when the compressibility was removed. However, pressure fluctuations were observed in both the cases. \citet{Yin2017} experimentally observed pressure fluctuations in the slug and bubbly flow patterns. \citet{Wang2014} and \citet{Singh2009a} experimentally observed transient two-phase pressure drop, with the amplitude and frequency dependent on heat flux and mass flux.

\subsection{Average heat transfer coefficient}
 
\citet{Wang2012} experimentally studied the effect of heat flux, vapor quality and hydraulic diameter on flow boiling heat transfer in a rectangular microchannel using FC-72 as the working fluid. The local heat transfer coefficient initially increases due to thin film evaporation and then decreases due to the emergence of dry patches. \citet{Diaz2007} studied the variation of local heat transfer coefficient for different heat fluxes and working fluids. The heat transfer coefficient was found to increase with heat flux, indicating the dominance of the nucleation boiling regime. Temperature fluctuation of high amplitude was also observed for water in the low-quality region. \citet{Mirmanto2016} experimentally investigated the local pressure drop and heat transfer coefficient for flow boiling of water in a micro-channel. Non-linear variation in pressure drop was observed along the channel length. Local heat transfer coefficient was found to  increase and then decrease along the channel. Effect of local pressure on the estimation of local heat transfer coefficient was also studied. 
 
\subsection{Prediction of heat transfer coefficient} 
 \citet{Thome2004} proposed a three-zone model consisting of the liquid zone, bubble with the thin liquid film and vapor zone for small tubes supplied with constant heat flux on the boundary and evaluated time-averaged heat transfer coefficient. The model utilized three adjustable parameters for initial film thickness, nucleation frequency and minimum film thickness, which were obtained from the experimental data set as explained by \citet{Dupont2004}. This model does not consider the effect of local transient pressure during the bubble growth and the effect of shear stress on the thin film depletion.  \citet{Wang2010} extended the three-zone model (\citet{Thome2004}) to a rectangular channel. The four zones they considered are liquid zone, elongated bubble zone, partial dry-out with liquid present in the corners and full dry-out region. The model was verified with the experiments conducted in a single rectangular channel with constant heat flux supplied through the channel walls. This model eliminated minimum thin film thickness as a parameter. This model also neglected the effect of local pressure conditions and effect of shear stress on the depletion of the thin film. The three-zone and four-zone models mentioned above were used to predict heat transfer coefficient over the entire range of thermodynamic quality. \citet{Harirchian2012} estimated heat transfer coefficient based on the flow patterns observed during flow boiling. For the bubbly region, the correlation is used. For confined annular flow, the model has been developed based on the conservation of mass, momentum (along with the interfacial stresses) and energy equation. Similarly, three-zone model developed by \citet{Thome2004} was utilized for slug-plug region but with a different initial film thickness relationship and the associated parameters. A similar flow pattern based approach was also utilized for the prediction of pressure drop across the channel. \citet{antonsen2015pragmatic} developed numerical models for the evaporation and condensation of refrigerants in a microtube with the combination of correlations, numerical methods and analytical model. Bubble growth model based on the heat flux is calculated for the cases with and without axial conduction. The model utilized the three-zone model for the estimation of heat transfer coefficient. Effects of different parameters such as wall material, nucleation frequency and mass flux have been studied. \citet{Jafari2016} simulated flow boiling process in a microchannel considering an artificial cavity at the channel wall using Cahn - Hilliard method. The bubble is nucleated from the artificial cavity and it grows up till it reaches the channel height and then departs from the nucleation site. The velocity, temperature, and pressure distributions were observed inside the channel. The heat transfer performance increased by 8 times the single phase value due to the presence of the thin liquid film layer, which, later, upon the dry-out, led to a reduction in heat transfer capability. Periodic fluctuations in heat transfer coefficient were observed based on the bubble location. The heat removal capacity was observed to increase with the increase in the mass flux. \citet{cerza2007analytical} performed an analytical investigation of the bubble growth and estimation of heat transfer. The model assumed a moving liquid layer along with the bubble with non-zero velocity and presence of the developing thermal profile. The model utilized the film thickness for the estimation of the heat transfer coefficient under various bubble velocity conditions. The model estimated the temperature profile of the liquid film below the bubble which caused the bubble growth. \citet{Magnini2017} improvised the three-zone model initially developed by \citet{Thome2004} for the two-phase system with refrigerant as a fluid. The improved model considered the time dependent variation of the local film thickness along with the inclusion of thermal inertia of the liquid film. The recirculation of the liquid in between the trapped bubbles is also considered for the liquid slug heat transfer coefficient calculation. \citet{li2017effect} proposed a model for the bubble dynamics in a micro-tube under flow reversal condition and its effect on the heat transfer coefficient for R134a. 

\subsection{Objective of the current work}
Most of the heat transfer and pressure drop models in the literature focus on microtubes (circular in cross section). But rectangular microchannel heat sinks, machined in copper or aluminum or etched in silicon, are important from the application point of view, considering the dissipation of high heat fluxes and the ease of manufacture. The four-zone model proposed by \citet{Wang2010} for a rectangular channel does not consider partially confined bubble region and  the influence of local pressure fluctuations, shear stress and flow reversals on heat transfer characteristics.    The current work is aimed at predicting the transient heat transfer coefficient under the influence of local pressure fluctuation caused due to the confined growth of the bubble inside a rectangular micro-channel. Transient and time-averaged pressure drop and heat transfer coefficient are compared with the experimental data available in the literature for a single rectangular microchannel. Flow reversal caused due to explosive bubble growth and presence of upstream compressible volume and its effect on the heat transfer coefficient is studied. Effect of shear stress on transient heat transfer characteristics is also modeled.

\section{Model description}

\begin{figure}[H]
	\centering
	\includegraphics[width=\linewidth]{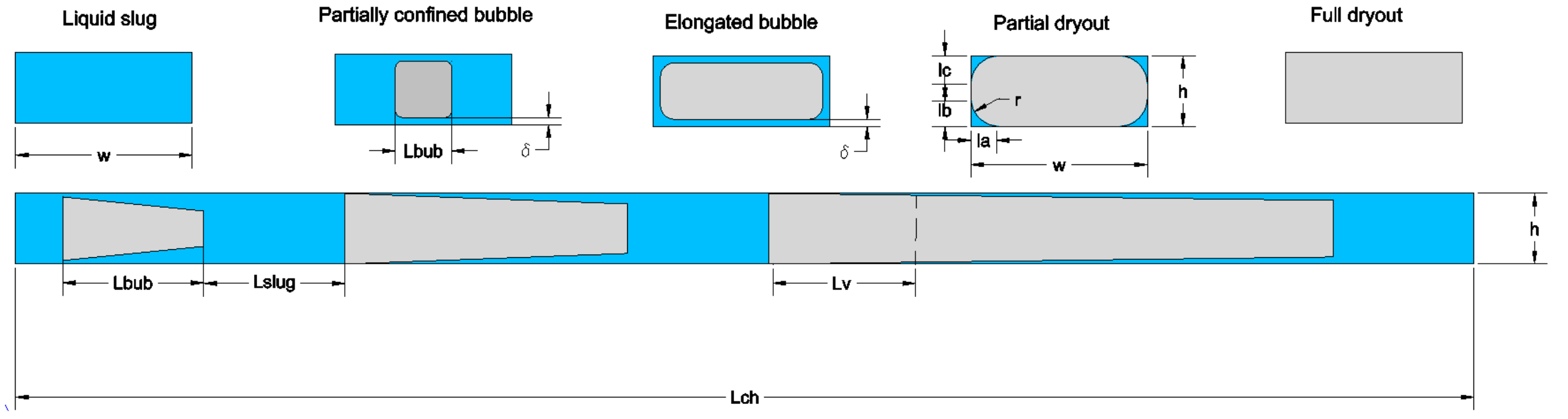}
	\caption{Various zones considered in the present study.}
	\label{fig:1a}
\end{figure}

The current model consists of two modules i.e., the bubble growth module and the heat transfer module. The first module estimates bubble length, bubble velocity, and pressure drop and the second module calculates the heat transfer coefficient based on the four zones, namely, (1) the liquid zone, (2) elongated bubble zone, (3) partial dry-out zone and (4) full dry-out zone as proposed by \citet{Wang2010} and additionally considers partially confined bubble zone, as shown in Fig. \ref{fig:1a}. The liquid zone consists of the liquid trapped between the two bubbles. The second zone (partially confined bubble zone) consists of liquid bulk on either side the bubble as shown in Fig. \ref{fig:1a}. The third zone consists of thin liquid film surrounding bubble along with the bulk liquid present in the corner of the rectangular channel between the bubble and the wall. As the thin film depletes the third zone appears with the dry patch area along with the bulk liquid in the corner. Finally, complete evaporation of the corner liquid leads to full dry-out zone/vapor zone.

\subsection{Assumptions}
In order to simplify the model few assumptions are considered. These assumptions are in line with the assumptions made by \citet{Thome2004}, \citet{Wang2010} and \citet{Gedupudi2011}.
\begin{enumerate}
\item	Current model is 1-dimensional i.e. variations are only considered in the axial direction, i.e., along the flow direction. 
\item	The heat sink consist of a single channel with constant heat supplied through the three sides (two sides and the base) of the channel.
\item	The bubble in the elongated region (liquid as thin film and liquid in the corner) is considered as a rectangle with rounded corners for simplification.
\item	Fluid is considered to be saturated from the point of nucleation; this is possible for the cases with the flow boiling in the near saturated condition.
\item	The bubbles are considered to be in the saturated condition, hence properties corresponding to local pressure are considered.
\item	Initial bubble is assumed to be a cube with the length, breadth and height equal to the minimum dimension of the cross-section as assumed by \citet{Gedupudi2011}.
\item	Single nucleation site is considered in the present study. In the current case, the bubble grows in axial direction exponentially leading to increased velocity which reduces the thermal boundary layer thickness suppressing the bubble growth. Moreover passing of a long bubble in a shorter channel also decreases the possibility of multiple nucleation sites. The models developed by \citet{Thome2004} and \citet{Wang2010} also consider single nucleation site.
\item	The length of liquid slug is equivalent to the liquid passing over the nucleation site during the interval between the successive nucleation, i.e., the liquid slug length is equal to the liquid velocity multiplied by the nucleation time period.
\item	Minimum film thickness before the occurrence of the dry-out is of the order of surface roughness. This value varies with surface, so, if available, the specified value would be used. Else, it would be considered as a constant similar to the value proposed by \citet{Thome2004}.
\item	Change in property of fluid due to viscous dissipation as explained by \citet{Han2007} does not affect the properties of liquid significantly.
\item	The simulation ends if there is no bubble in the channel, or the upstream end of the bubble reaches the channel entrance, possible in the case of inlet compressibility.
\item	The contribution of surface tension to the pressure drop is negligible. The surface tension of water at \ang{100}C is 0.058 N/m and the minor dimension of the channel considered in the study is 0.00024 m. Therefore the approximate pressure drop due to surface tension is \(\sigma/R = 0.058/0.00012 = 0.48 \, kPa\), which is negligible compared to the pressure drop due to friction and acceleration for the range of the parameters considered in the present study.  
\end{enumerate}

\subsection{Properties determination}

Variation of fluid properties with pressure is not considered in the models for the bubble growth presented by \citet{He2016} and \citet{Gedupudi2011}. \citet{Thome2004} and \citet{Wang2010} have also not considered the fluid property variation with local transient pressure in their heat transfer models. In the case of water, there is a significant difference in the amplitudes of pressure fluctuation if the variable properties are not considered (presented in the current study). This is due to the fact that as the pressure drop increases across the channel, the local pressure changes, which leads to a change in vapor properties that defines the bubble growth rate. These, in turn, influence the local velocity and pressure drop. The fluid properties are taken from \citet{Bell2014}. The saturated properties corresponding to the local pressure are estimated at each time step. The pressure at the channel outlet is assumed to be 101 kPa. In the current study, the working fluid is water unless otherwise specified.

\subsection{Location of nucleation site}
The current transient model requires a nucleation site as an input parameter. This is one of the major challenges in the area of flow boiling as nucleation site is not deterministic and it depends on surface characteristics. Therefore, it is difficult to estimate the nucleation site accurately. Though, \citet{Liu2005} summarizes relationships available in the literature and proposes the condition for nucleation based on wall superheat. In the present study, Eq. (\ref{eq:1}) is used to determine nucleation site.

\begin{equation} \label{eq:1}
\sqrt{T_{wall}} - \sqrt{T_{sat}}  = \sqrt{\frac{2  \, (1+\cos(\theta))\, \sigma \,  q}{\rho_{v} \, k_{l} \, i_{lv}}}
\end{equation}

The location of nucleation site is determined iteratively. Initially for the specified inlet temperature and for the assumed channel inlet pressure equal to the exit pressure, the location where the wall temperature obtained from single-phase heat transfer coefficient correlation matches the wall temperature obtained from Eq. (\ref{eq:1}) is chosen as the initial nucleation site. Time-averaged two-phase pressure drop is then calculated. Saturation temperature corresponding to the obtained pressure at the assumed nucleation site is determined, and then the new wall temperature required for nucleation from Eq. (\ref{eq:1}) is obtained. Then the new location of the nucleation site is determined using single phase correlation based on the required wall temperature for nucleation. The estimation of the location of nucleation site and two-phase pressure drop is performed iteratively till the required wall temperature from Eq. (\ref{eq:1}) matches the obtained wall temperature from single phase correlation. Contact angle varies with the substrate and should be chosen accordingly. For example, it is around \ang{90} for copper.

\subsection{Nucleation frequency}

The nucleation time period consists of two components i.e., waiting time and growth time. Bubble growth and its departure from the nucleation site disturbs the local wall temperature. It takes a while for the reformation of the thermal boundary layer, i.e., for the wall to attain the superheat required for nucleation. This duration is called as waiting period. This has been determined with the help of transient single phase simulation carried out using ANSYS FLUENT 16.1. Another component is bubble growth period and is determined using the correlation given by \citet{Mikic1970}. The initial theoretical value obtained using this correlation is quite small and is of the order of \(100 \,\mu s\). The experimental data obtained from \citet{Kadam2018}, \citet{Lee2004} and \citet{Wang2010} shows that the nucleation time period varies in the range 1-500 ms for the water based on both heat flux and mass flux. In the present case, with the increase in heat flux, the time period decreases as both the waiting and bubble growth time decreases. As the heat flux increases the wall reaches the condition of superheat at a faster rate. Similarly, with the increase in mass flux two forces are opposing one is the inertial force which detaches the bubble from the nucleation site this decreases the time period as also observed by \citet{Jafari2016} and another is that the increases the time periods as due to large mass flux the thermal boundary layer
thickness decreases.

\begin{equation} \label{eq:2a}
R =  \sqrt{\frac{12 \, k_{l} \, t_{grow}}{C1 \,\rho_{l} \, C_{p,l}\pi}} \, Ja 
\end{equation}
where,
\begin{equation*} \label{eq:2b}
Ja =  \frac{C_{p,l} \, \Delta T_{sup} \, \rho_{l}}{\rho_{v} \, i_{lv}}
\end{equation*}
Here,

\(C1\) is the adjustable factor to determine the growth time period. In the present study \(C1 = 210\) is used.

\begin{figure}[H]
	\centering
	\includegraphics[width=\linewidth]{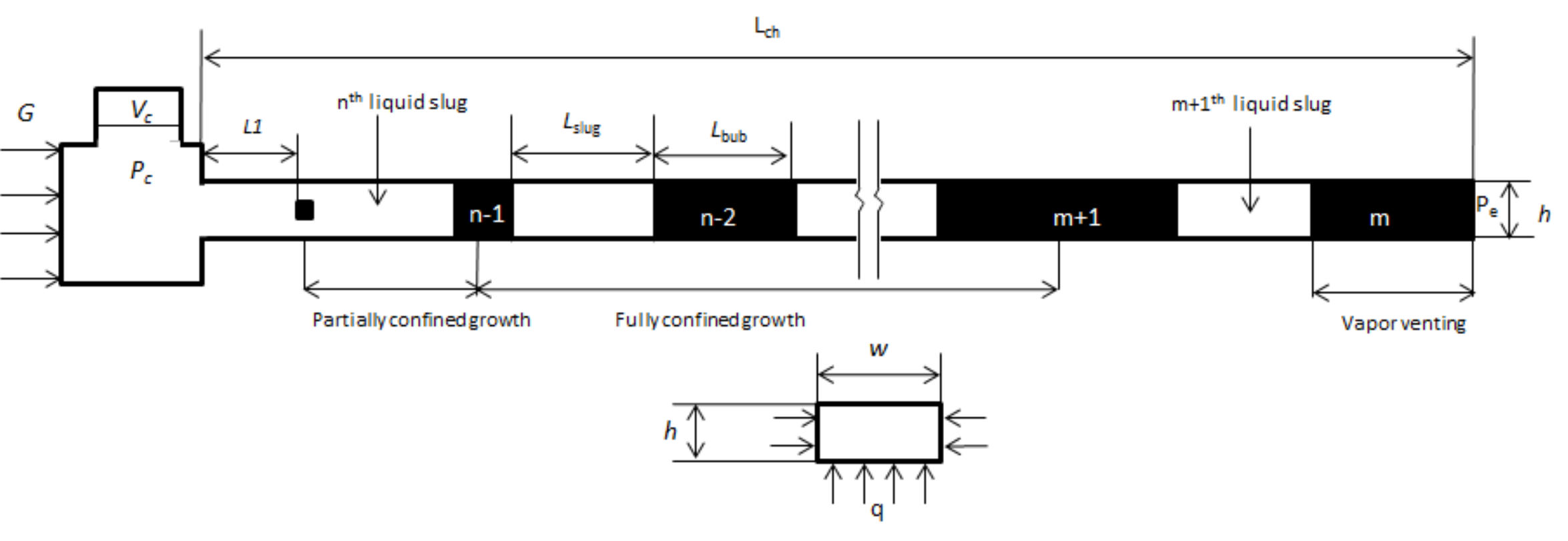}
	\caption{Stages of bubble growth in the one dimensional model.}
	\label{fig:1}
\end{figure}

\subsection{Bubble growth}

In the case of the flow boiling the bubble is initiated at the channel wall based on the wall superheat requirement. The initial stage consists of the free (unconfined) growth similar to the nucleate boiling. Once the bubble grows to the minor dimension of the channel the growth rate changes as observed by \citet{Yin2016}. \citet{Barber2010} and \citet{Gedupudi2011} also observed two different growth stages, namely partially confined growth where the bubble is confined by the minor dimension and grows widthwise and lengthwise  and fully confined growth where the bubble is confined by the minor and major dimensions of the cross-section and the bubble grows only along the length. The stage where the vapour bubble passes through the channel outlet is called vapour venting. 
Energy balance for the bubble growth is given in Eq. (\ref{eq:3}).

\begin{equation} \label{eq:3}
	q \; A_{heat} = \frac{d}{dt} \left(\rho_{v} \, A_{base} \, h \, i_{lv}\right)
\end{equation}

In the partially confined region, the area \(A_{base}\) differs from the partially confined condition and fully confined condition. \(A_{base} = L_{bub} \times L_{bub}\) for partially confined stage. \(A_{base} = L_{bub} \times w\) for fully confined stage. \(A_{heat} = L_{bub} \times L_{bub}\) for partially confined stage and \(A_{heat} = p_{h} \times L_{bub}\) for fully confined stage. \(p_{h}\) is the heated perimeter equal to \(w\) for  single-sided heating and equal to \(2h\, +\, w\) for three-sided heating.

\subsubsection*{Time constant}
Time constant is defined in Eq. (\ref{eq:4}).

\begin{equation}\label{eq:4}
\tau = \frac{w \, h\, i_{lv} \, \rho_{v}}{p_{h} \, q}
\end{equation}

where \(p_{h}\) is the heated perimeter. The value of $\tau$ is calculated for each time step and substituted in the equations to obtain the downstream end velocity, position and the bubble length.

\subsubsection{Partially confined bubble growth}

The growth rate is obtained by substituting the expression for \(A_{base}\) in Eq. (\ref{eq:3}) and is given as

\begin{equation} \label{eq:5a}
\frac{d L_{bub}}{dt} = \frac{L_{bub}}{2 \, \tau }
\end{equation}

Using Taylor series expansion, length of the bubble at t+\(\Delta\)t is estimated as in Eq. (\ref{eq:5}) where \textit{`t'} refers to the time from the start of the simulation.

\begin{equation} \label{eq:5}
(L_{bub})_{t + \Delta t} = (L_{bub})_{t} + \left( \frac{L_{bub}}{2 \, \tau} \right)_{t} \, \Delta t
\end{equation}

From continuity,  the velocity of the downstream end of the bubble is obtained as shown in Eq. (\ref{eq:6a}) and (\ref{eq:6}). 

\begin{equation} \label{eq:6a}
U_{d} = U_{u}  + \left(\frac{d A_{base}}{w \,dt}\right)
\end{equation}
\begin{equation} \label{eq:6}
U_{d} = U_{u}  + \left(\frac{L_{bub}^{2}}{w \,\tau}\right)
\end{equation}
Similarly, the location of the downstream end of the bubble is calculated by Eq. (\ref{eq:7}).
\begin{equation}{\label{eq:7}}
z_{d} = z_{u} + L_{bub}
\end{equation}

\subsubsection{Fully confined bubble growth}
Similar to the partially confined bubble growth, the length, the velocity and the position of the bubble in fully confined growth are determined using Eq. (\ref{eq:8}), (\ref{eq:9}) and (\ref{eq:10}) respectively.

\begin{equation} \label{eq:8}
(L_{bub})_{t + \Delta t} = (L_{bub})_{t} + \left( \frac{L_{bub}}{ \tau} \right)_{t} \, \Delta t
\end{equation}

\begin{equation} \label{eq:9}
U_{d} = U_{u}  + \left(\frac{L_{bub}}{\tau}\right)
\end{equation}

\begin{equation}{\label{eq:10}}
z_{d} = z_{u} + L_{bub}
\end{equation}

\subsubsection{Vapour venting}

During vapour venting stage, as the bubble passes through the channel outlet, the bubble is driven by the velocity on the upstream end and is given by
\begin{equation} \label{eq:11}
\frac{d \,L_{bub}}{dt} = -U_{u}
\end{equation}
The exit vapour velocity is determined using either Eq. (\ref{eq:6}) or (\ref{eq:9}) as the case may be.

\subsection{Upstream compressibility}

Due to the presence of sources of upstream compressibility, there can be flow reversal in the channel. In the present study, trapped non-condensable in the channel inlet plenum is considered, with  pressure \(P_{c}\)   and volume \(V_{c}\). Plenum is supplied with constant flow rate  \(\dot{V}_{in}\) . The channel inlet velocity can be calculated from the principle of continuity as.

\begin{equation}{\label{eq:fr1}}
\dot{V}_{in} + \frac{d V_{c}}{dt} = A_{cs} \, U_{i}
\end{equation}

For the current case, non condensable gas is assumed to undergo isothermal compression.

\begin{equation}{\label{eq:fr2}}
P_{c} \, V_{c}  = P_{0, ini} \, V_{0, ini} = constant 
\end{equation}

The rate of change of gas volume is determined using Eq. (\ref{eq:fr3}). 

\begin{equation}{\label{eq:fr3}}
\frac{d V_{c}}{dt} = - \frac{V_{c}}{P_{c}} \, \frac{d P_{c}}{dt}
\end{equation}
Stagnation pressure \((P_{0})\) is given by 

\begin{equation}{\label{eq:fr4}}
	P_{0} = P_{1} + (1.5K) \, \frac{\rho_{l} \, U^{2}_{i}}{2}
\end{equation}

Here \(P_{c}\) is equal to \(P_{0}\), and \(U_{i}\)  is channel inlet velocity and \(P_{1}\) is channel inlet pressure. \(K = 1\) for the forward flow and \(K = 0\) for the flow reversal.

Differentiating the pressure equation Eq. \ref{eq:fr4} gives

\begin{equation}{\label{eq:fr6}}
	\frac{d P_{0}}{dt} = \frac{d P_{1}}{dt} +  (1.5K ) \, U_{i} \, \frac{d U_{i}}{dt}
\end{equation}
Substituting the Eq. \ref{eq:fr3} in the Eq. \ref{eq:fr6}.
\begin{equation}{\label{eq:fr7}}
	\frac{d P_{1}}{dt} = - \frac{P_{c}}{V_{c}} \, \frac{d V_{c}}{dt} - (1.5K) \, U_{i} \, \frac{d U_{i}}{dt}
\end{equation}
Substituting the value of \(dV_{c}/dt\) from the Eq. \ref{eq:fr1}.

\begin{equation}{\label{eq:fr8}}
\frac{d P_{1}}{dt} =  -\frac{P_{0,ini}}{V_{0,ini}}(A_{cs}\, U_{i} \, - \, \dot{V}_{in}) - (1.5K) \, U_{i} \, \frac{d U_{i}}{dt}
\end{equation}
The initial compressible volume of the non-condensable gas is assumed to be a variable parameter and would be varied in order to study its effect.

\subsection{Pressure drop}
The channel pressure drop consists of the viscous pressure drop and the acceleration pressure drop across the liquid slugs. Pressure drops due to momentum change and friction across vapour bubble or vapour slug are assumed to be negligible \citet{Gedupudi2011} as \(\rho_{v} << \rho_{l}\). As explained in the \textit{Assumptions section}, pressure drop across the vapor bubble and pressure drop due to surface tension are neglected.  

\subsubsection{Viscous pressure drop}
The frictional pressure drop arises due to the motion of liquid slugs present in the channel. The total frictional pressure drop is the sum of the frictional pressure drops calculated for the individual liquid slugs. Apart from this, there will also be viscous pressure drop due to the liquid slug between the channel inlet and upstream end of the newest bubble. 

\begin{equation} \label{eq:13}
\Delta P_{vis} =  \frac{2 \, \rho_{l}\, f \, z_{n,u} \, U^{2}_{i}}{D_{h}} \, + \, \sum_{i = m}^{n} \left(\frac{2 \, \rho_{l} \, f \, L_{slug, i} \,U_{slug, i}^{2}}{D_{h}}\right)_{i}
\end{equation}
where,

where \textit{`m'} represents the liquid slug in between the exit and the last bubble, \textit{`n'} represents the liquid slug between the the newest bubble \textit{`n'} and \textit{`n-1'} bubble, as shown in Fig. \ref{fig:1}.

The Fanning friction factor \textit{f}  used in Eq. (\ref{eq:13}) is for a rectangular cross section with the aspect ratio \(\alpha \) and is obtained from Eq. (\ref{eq:14}). If the value \textit{f} goes below 0.01, then \textit{f} is taken as a constant equal to 0.01 to account for the turbulent flow, as considered in the previous study by \citet{Gedupudi2011}. 

\begin{equation}\label{eq:14}
\left(f \, Re\right) = 24 \left(1 -1.3553 \alpha + 1.9476 \alpha^{2}  -  1.7012 \alpha^{3}  + 0.9564 \alpha^{4} - 0.2537 \alpha^{5}\right)
\end{equation}




\subsubsection{Acceleration pressure drop}
The acceleration pressure drop is calculated using the momentum conservation applied for all the liquid slugs.
\begin{equation}\label{eq:15}
\Delta P_{acc} = \rho_{l} \, z_{n,u} \frac{d U_{i}}{dt} \, + \,\sum_{i = m}^{n} \left(\rho_{l} \, L_{slug ,i} \, \frac{d U_{slug, i}}{dt}\right)_{i}
\end{equation}
The first term in the above expression will be zero if there is no upstream compressibility.

\subsection{Heat transfer coefficient}
In the present work, the time-averaged heat transfer coefficient is calculated at each location. The heat transfer coefficient is calculated for each zone i.e. the liquid zone, partially confined bubble zone, elongated bubble zone, partial dry-out zone and full dry-out zone individually as they pass over a particular position cyclically. These are then time-averaged based on their residence time to obtain the local time-averaged heat transfer coefficient. The estimation of the heat transfer coefficients for different zones is described below.

\subsubsection{Liquid slug}

This is the liquid present between the two successive bubbles. The liquid slug moves with the velocity equal to the downstream end velocity of the bubble present on the upstream end or with the inlet velocity. In the literature various correlations have been used to estimate the heat transfer coefficient. \citet{Thome2004} estimates the liquid slug heat transfer coefficient (HTC) with the help of the laminar and turbulent correlations for the circular tube case. In order to avoid the jump from the laminar to turbulent region, asymptotic approach is used. This method averages the laminar and turbulent region HTC. In the case of four zone model developed by \citet{Wang2010}, single phase laminar fully developed region correlation is used. \citet{He2010} studied the heat transfer characteristics of liquid slug trapped between the two bubbles. Due to the movement of the liquid and no slip condition at the wall, internal vortex is seen. This vortex increases the convection and therefore increases the HTC. In the present study the fluid flow is considered laminar for \(Re < \, 1600\) and in the transition zone if \(Re\) is in between 1600 and 3000 and turbulent for  \(Re \, > \,3000\) as used by the \citet{Kandlikar2004}. For the transition range, based on \(Re\), linear average of the heat transfer coefficient is used. Eq. (\ref{eq:25}) is used for the calculation of HTC for the rectangular channel in laminar region with the aspect ratio \((\alpha = h/w)\).

For the laminar fully developed condition, the Nusselt number expressions for four-sided and three-sided heating configurations are as given in Eq. (\ref{eq:16}) and Eq. (\ref{eq:17}) respectively.

\begin{equation}{\label{eq:16}}
Nu_{4,\infty} = 8.235 (1-2.0421 \, \alpha + 3.0853 \alpha^{2} \, - 2.4765 \alpha^{3} +1.0578 \alpha^{4} -0.1861 \alpha^{5})
\end{equation}

\begin{equation}{\label{eq:17}}
Nu_{3,\infty} = 8.235 (1-1.883 \, \alpha + 3.767 \alpha^{2} \, - 5.814 \alpha^{3} +5.361 \alpha^{4} -2 \alpha^{5})
\end{equation}

\citet{Lee2006} suggested heat transfer correlation for laminar thermally developing region for the rectangular channel with the aspect ratio \((\alpha)\) for the constant heat flux condition for four sided heat condition which was further modified by \citet{phillips1988microchannel} for three side heating condition as given in Eq. (\ref{eq:25}).

\begin{equation}\label{eq:25}
Nu_{3} = \frac{Nu_{3,\infty}}{Nu_{4,\infty}} \, Nu_{4}
\end{equation}

In the case of turbulent flow the relation, Eq. (\ref{eq:26}), proposed by \citet{wu1984measurement} for a microchannel is used.

\begin{equation}\label{eq:26}
Nu = 0.00222 \, Re^{1.09} \, Pr^{0.4}
\end{equation}

\subsubsection{Thin film region}
As elucidated in the literature, evaporation of the thin film is the major contributor of the heat transfer coefficient. Hence it is very important to estimate the thin film thickness. Initial film thickness plays an important role in this process. Various researchers have proposed the relationship based on the experimental observations. \citet{Thome2004} used the basic relationship provided by \citet{moriyama1996thickness} and used an asymptotic method to estimate the film thickness. A factor was also used to correct the above relationship based on the experimental data available in the literature. \citet{Wang2010} also used the similar equation as by \citet{Thome2004} but the factor is based on the experiments conducted. \citet{Han2010} proposed the relationship for the initial film thickness for the bubble accelerating through the channel. The initial film thickness decreases with the increase in the acceleration due to the thinning of boundary layer. Film thickness at any location or at any instant of time is influenced by initial film thickness and rate of thinning. The latter depends on the rate of evaporation and shear stress. \citet{Thome2004} and \citet{Wang2010} considered only the evaporation of the thin liquid film, deducing the new film thickness with the help of the energy balance. On the contrary, \citet{consolini2008convective} considered the effect of shear stress on the film thickness for a bubble growing inside  a circular tube. \citet{sun2018transient} considered the thinning of the film due to evaporation and proposed an expression  for the additional thinning of the thin film due to the shear stress. The relationship was derived analytically and a constant was used in order to match the experimental film thickness. \citet{younes2012analytical} also employed the analytical approach by considering the shear stress at the liquid-vapour interface to calculate the heat transfer coefficient. The heat transfer coefficient is calculated by Eq. \ref{eq:27}.

\begin{equation}\label{eq:27}
HTC = \frac{k_{l}}{\delta_{film}}
\end{equation}
The initial film thickness Eq. (\ref{eq:28}) for a vapour bubble in the two phase region is:
\begin{equation}\label{eq:28}
\frac{\delta_{film}}{D_{h}} = C_{\delta} \, \left(3 \sqrt{\frac{\nu_{l}}{U \, D_{h}}}\right)^{0.84}  \, \left[\left(0.07 \, \left(\frac{\rho_{l} \, D_{h} \, U^{2}}{\sigma}\right)^{0.41}\right)^{-8} \,  +  \, 0.1^{-8} \right]^{-1/8} 
\end{equation}

\citet{Thome2004} obtained the value of the constant \(C_{\delta}\) equal to 0.29, by matching the model results with the experimental data for refrigerants boiling in a circular tube. The initial film thickness based on this constant is of the order of few microns.  \citet{Wang2010} used a value equal to 0.8117 by matching their model results with their experimental heat transfer data for water boiling in a rectangular channel (\(D_{h}\) = 0.137 mm). The initial film thickness based on this constant is in the range \(1-2 \, \mu m\) for the bubble velocity in the range \(0.2 - 0.5 \, m/s\) considered in their study. But in the present study with water as a working fluid, for the range of the parameters considered (\(D_{h}\) = 0.324 mm and 0.658 mm), the value of the constant  \(C_{\delta}\) equal to 3.48 results in a good match between the modeling results and the experimental heat transfer data from the literature.  The initial film thickness based on this constant is in the range \(10-30 \, \mu m\) for the bubble velocity in the range \(0.5 - 2 \, m/s\). The initial film thickness predictions made using this constant is in line with the experimental thin film thickness measured by \citet{sun2018transient} for water. Their experimental observations for flow boiling indicate initial film thickness of 25 and 31 \(\mu m\) for the bubble velocities 3.09 m/s and 3.63 m/s and tube diameter 0.94 mm. The corresponding values obtained using the proposed constant (equal to 3.48) is 17 and 16 \(\mu m\) respectively.  Further, their experimental observations show that from the point of initial film thickness, there is a gradual decrease in local film thickness until it reaches dryout thickness which is around 2-3 \(\mu m\). The film thickness estimated by \citet{Gedupudi2011} is in the range 20-30 \(\mu m\) and that observed by \citet{Balasubramanian2005} is near 50 \(\mu m\), all for water boiling in a mini/microchannel close to atmospheric pressure. 

In the present study, both the heat transfer and shear stress effect are taken into account to evaluate the depletion of the thin film thickness. Thin film exists in partially confined bubble region, elongated bubble region and partial dryout region.

\subsubsection*{Partially confined bubble region}
In this region, the bubble is partially confined by the minor channel dimension. The heat transfer coefficient is calculated considering the heat transfer through the thin film between the bottom wall and the bubble and the heat transfer through the liquid on either side of the bubble. The heat transfer coefficient is calculated using Eq. (\ref{eq:pc_1}) to Eq. (\ref{eq:pc_3}). The liquid heat transfer coefficient is estimated using the equations presented in \textit{Liquid slug} section considering the effective liquid flow cross-section obtained after subtracting the bubble cross-sectional area from the channel cross-section. Hydraulic diameter is calculated using the effective cross-sectional area and the corresponding wetted perimeter.

\begin{figure}[H]
	\centering
	\includegraphics[width=0.3\linewidth]{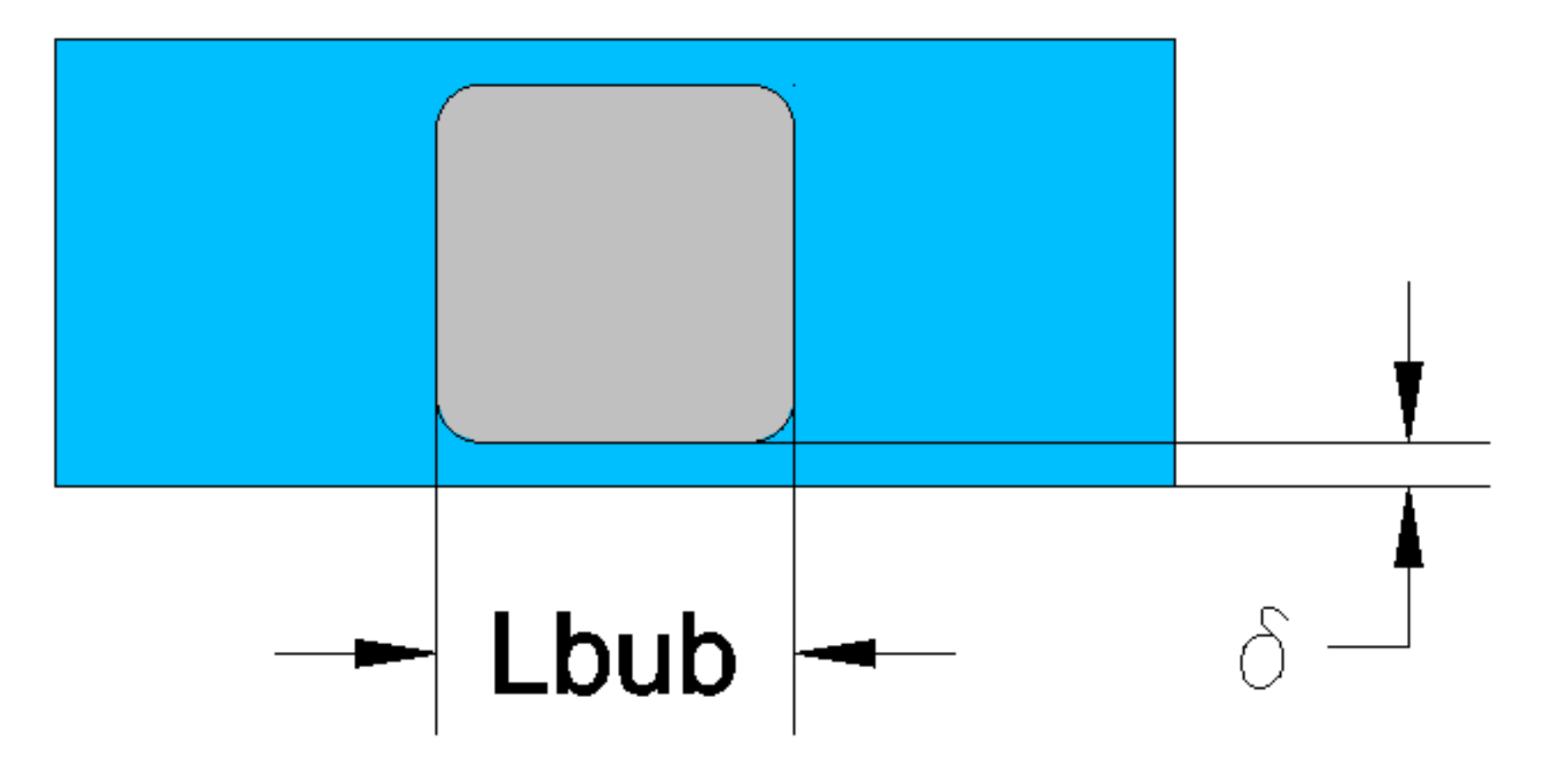}
	\caption{Paritally confined bubble zone.}
	\label{fig:zone_partial_conf}
\end{figure}

\begin{equation}\label{eq:pc_1}
(T_{w} - T_{sat}) \times (0.5 \times w \, + h) =  \int_{0}^{0.5 \times L_{bub}} (T_{w} - T_{sat}) ds  
+ \int_{0}^{h} (T_{w} - T_{sat}) ds 
+ \int_{0}^{0.5 \times ( w - L_{bub})} (T_{w} - T_{sat}) ds
\end{equation}

\begin{equation}
\frac{q \times (0.5 \times w  + h)}{HTC_{pc}} = \frac{q \times 0.5 \times L_{bub}}{HTC_{film}} +  \frac{q \times (0.5 \times w + h - 0.5 \times L_{bub} )}{HTC_{l}}
\end{equation}

\begin{equation}\label{eq:pc_3}
\frac{1}{HTC_{pc}} = \frac{0.5 \times w + h - 0.5 \times L_{bub}}{0.5 \times w + h} \times \frac{1}{HTC_{l}} + \frac{0.5 \times L_{bub}}{0.5 \times w + h} \times \frac{1}{HTC_{film}} 
\end{equation}

The \(HTC_{film}\) is given by the Eq. \ref{eq:pc_htc_1}. 

\begin{equation}\label{eq:pc_htc_1}
HTC_{film} = \frac{k_{l}}{\delta_{eq}}
\end{equation}

The \(\delta_{film}\) is the film thickness over the flat region and its depletion is given by Eq. \ref{eq:pc_htc_2}.
\begin{equation} \label{eq:pc_htc_2}
\left(\delta_{film}\right)_{t + \Delta t} = \left(\delta_{film}\right)_{t} - \frac{q \, L_{bub} \, \Delta t}{\rho_{l} \, i_{lv} \, L_{bub} }
\end{equation}

The \(\delta_{eq}\) is the equivalent thickness over the horizontal region which is given by Eq. \ref{eq:pc_htc_3}.

\begin{equation} \label{eq:pc_htc_3}
\delta_{eq} = \frac{h \times L_{bub} - A_{cs,bub}}{2 \times L_{bub}}
\end{equation}

\(A_{cs,bub}\) is calculated for hyperellipse with  \(n1 = 4\) , \(a = 0.5 \times L_{bub}\) and \(b = 0.5 \times h - \delta_{film}\) in Eq. \ref{eq:pc_htc_4} presented by \citet{Tamayol2010}.

\begin{equation}\label{eq:pc_htc_4}
A_{cs,bub} = 4 a \, b \frac{\sqrt{\pi} \: \Gamma \left(\frac{n1+1}{n1}\right)}{4^{1/n1} \, \Gamma \left(\frac{n1+2}{2n1}\right)}
\end{equation}

Here, \(\Gamma\) is the gamma function.

\subsubsection*{Elongated bubble region}

\begin{figure}[H]
	\centering
	\includegraphics[width=0.3\linewidth]{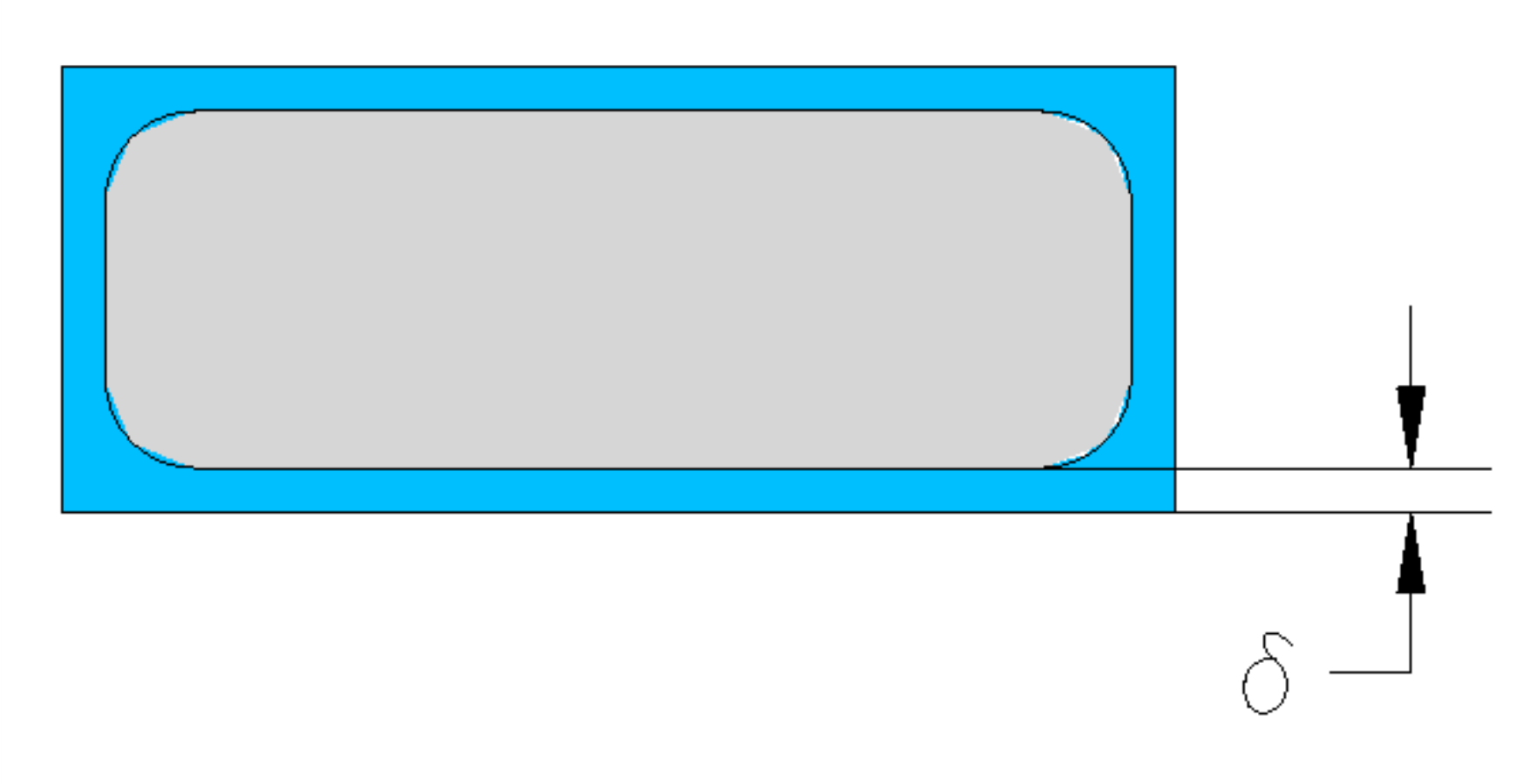}
	\caption{Elongated bubble zone.}
	\label{fig:zone_elong}
\end{figure}

This is the region where the cross- section has liquid film over the entire perimeter of the channel as shown in Fig. \ref{fig:zone_elong}. Vapour bubble is assumed to be of rectangular shape with rounded corners, obtained with \(n1 = 4\) in hyperellipse formula \citet{Tamayol2010}.   The region between the curved portion at the corner and the channel walls form the bulk liquid region.

Here \(\delta_{film}\) is the film thickness at the flat region of the channel. 


\begin{equation} \label{eq:29a}
\left(\delta_{film}\right)_{t + \Delta t} = \left(\delta_{film}\right)_{t} - \frac{q \, p_{h} \, \Delta t}{\rho_{l} \, i_{lv} \, p_{lv} }
\end{equation}

Here \(p_{lv}\) is the perimeter of liquid and vapor interface at any cross-section used by \citet{Tamayol2010}. \(p_{ch}\) is the channel perimeter and \(p_h\) is the heated perimeter.

\begin{equation}\label{eq:29b}
\delta_{eq} = \frac{A_{l}}{p_{ch}} =  \frac{w \times h - A_{cs,bub}}{2 \, \times (w + h)}
\end{equation}
The cross-sectional area for a rectangle with rounded corner is given by following equation and used by \citet{Tamayol2010}. Here \(n1 = 4 \) for rectangle with rounded corners. 
\(a = 0.5 \times w - \delta_{film}\) and \(b = 0.5 \times h - \delta_{film}\)

\begin{equation}\label{eq:29c}
A_{cs,bub} = 4 a \, b \frac{\sqrt{\pi} \: \Gamma \left(\frac{n1+1}{n1}\right)}{4^{1/n1} \, \Gamma \left(\frac{n1+2}{2n1}\right)}
\end{equation}

Here, \(\Gamma\) is the gamma function.


The Eq. \ref{eq:31} is used to determine the heat transfer coefficient in the elongated bubble region.

\begin{equation}\label{eq:31}
HTC_{elb} = \frac{k_{l}}{\delta_{eq}}
\end{equation}

\subsubsection*{Partial dryout region}

\begin{figure}[H]
	\centering
	\includegraphics[width=0.3\linewidth]{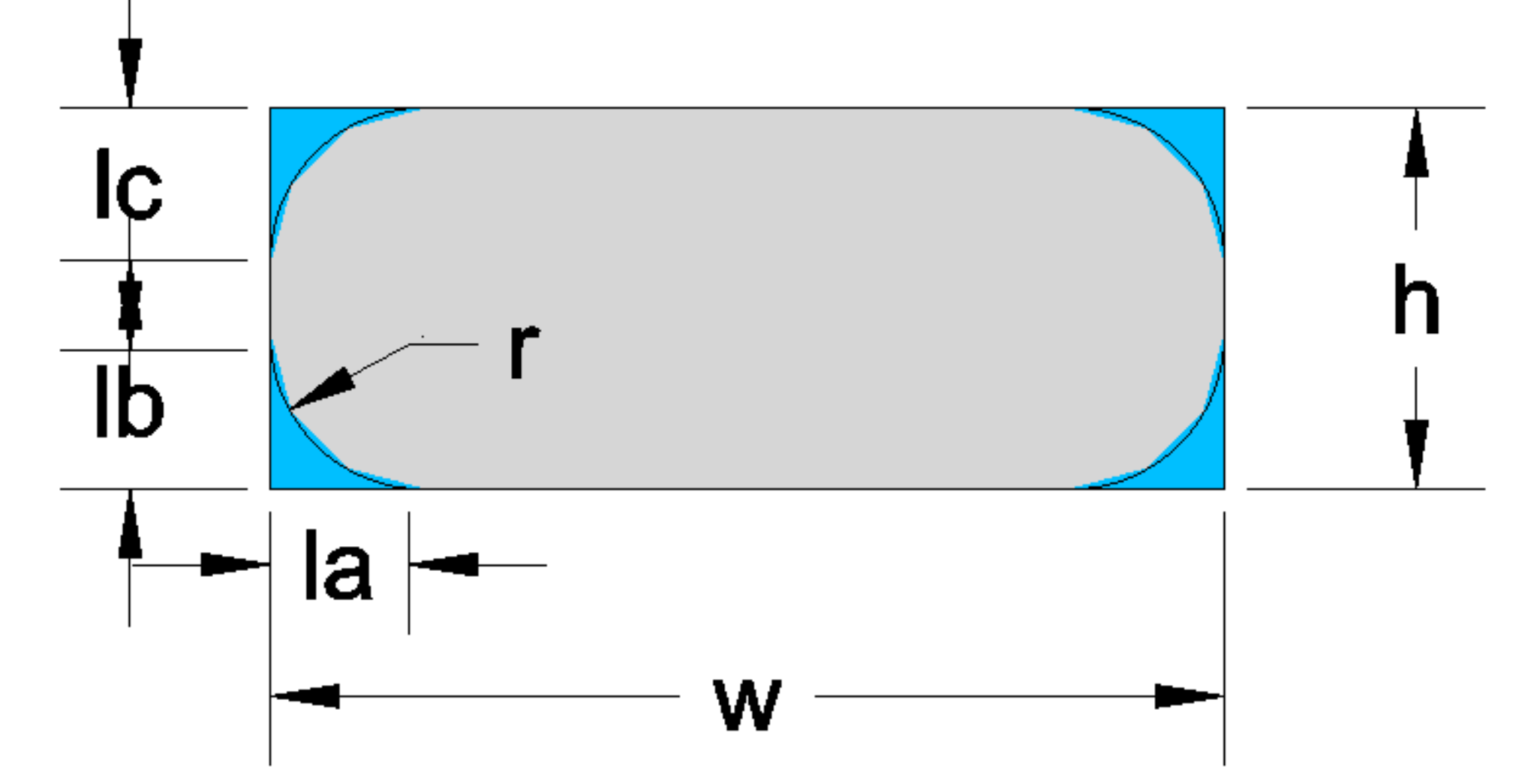}
	\caption{Partial dryout zone.}
	\label{fig:2part}
\end{figure}

Once the thin film in the flat region of the elongated bubble is evaporated fully, i.e., \(\delta_{film}  = 0 \), it leads to partial dryout as shown in Fig. \ref{fig:2part}. The corners of the channel still have the liquid which is not evaporated. Thus the equivalent heat transfer coefficient is the average of the heat transfer coefficient of the dryout region i.e. over the wall in contact with vapour and liquid heat transfer coefficient due to liquid in the corners of the channel. In the present case, the cross-sectional area occupied by vapour bubble is of rectangular in shape with rounded corners. In partial-dryout region, perimeter of bubble cross-section intersects with the channel wall.  and the corner liquid area and lengths are calculated.

\begin{equation}\label{eq:32_1}
(T_{w} - T_{sat}) \times (0.5 \times w \, + h) =  \int_{0}^{l_{a}} (T_{w} - T_{sat}) ds  
+ \int_{0}^{l_{b}} (T_{w} - T_{sat}) ds 
+ \int_{0}^{l_{c}} (T_{w} - T_{sat}) ds
+ \int_{0}^{0.5 \times w + h - l_{l}} (T_{w} - T_{sat}) ds
  \end{equation}

\begin{equation}
\frac{q \times (0.5 \times w + h)}{HTC_{pdr}} = \frac{q \times l_{a}}{HTC_{l}} + \frac{2 \,q \times l_{b}}{HTC_{l}}  + \frac{q \times (0.5 \times w + h - l_{l} )}{HTC_{v}}
\end{equation}

\begin{equation}\label{eq:32}
\frac{1}{HTC_{pdr}} = \frac{0.5 \times w + h - l_{l}}{0.5 \times w + h} \times \frac{1}{HTC_{v}} + \frac{l_{l}}{0.5 \times w + h} \times \frac{1}{HTC_{l}} 
\end{equation}

Here \(l_{l}\)  refers to the length of the wall in contact with the liquid over the heated perimeter. \(l_{l} \, = \, l_{a} \, + \,l_{b} \, + \, l_{c} = l_{a} \, + \, 2 l_{b}\).

\begin{equation}\label{eq:32a}
\left(\delta_{corner}\right)_{t + \Delta t} = \left(\delta_{corner}\right)_{t} - \frac{q \, p_{h} \, \Delta t}{ \rho_{l} \, i_{lv} \, p_{lv}} 
\end{equation}

\(A_{l}\) is the area enclosed by the sides of the channel and the curve formed by the liquid-vapour interface at the corner.
This curve is obtained by assuming a hyperellipse (\citet{Tamayol2010}) with  \((0.5 \times w + \delta_{corner})\) and \((0.5 \times h + \delta_{corner})\) as major axis (a) and minor axis (b), as described in \citet{Tamayol2010}. The enclosed area is calculated by numerical integration.

\begin{equation}\label{eq:32c}
\delta_{eq} = \frac{A_{l}}{p_{lv}}
\end{equation}

And the corner bulk liquid heat transfer coefficient is calculated by the Eq. (\ref{eq:32d}).

\begin{equation}\label{eq:32d}
HTC_{l} = \frac{k_{l}}{\delta_{eq}}
\end{equation}

\subsubsection{Full dry-out region}
This is the region developed due to the complete evaporation of the thin film. The HTC can be calculated by means of the Eq. (\ref{eq:16}) to (\ref{eq:26}) using vapor properties.

\subsubsection{Shear stress effect}
\citet{sun2018transient} experimentally investigated the evolution of thin film with time, for adiabatic and diabatic cases.   The depletion of the film thickness was observed to be faster in diabatic case than that in adiabatic case. They also derived the depletion rate of thin film thickness for circular tube. In the present study, effect of shear stress is extended to a rectangular channel.

From continuity,

\begin{equation}
\rho_{l} \, \delta_{eq, z} \, p_{lv} \, U_{l, mean, z} \, = \,\rho_{l} \, \delta_{eq, z+\Delta z} \, p_{lv} \, U_{l, mean, z+\Delta z}
\end{equation}

\begin{equation}
\delta_{eq} \, U_{l,mean} = constant
\end{equation}

Differentiating the above equation with time and using \(dz = U_{l,mean}dt\)  results in Eq. (\ref{eq:33}).
The rate of thin film depletion is defined by the Eq. (\ref{eq:33}).

\begin{equation}\label{eq:33}
\frac{d \delta_{eq}}{dt} = -\delta_{eq} \, \frac{d U_{l,mean}}{dz}
\end{equation}
This equation requires the variation of the mean liquid velocity with the axial position. In order to estimate the value, the shear stress on the vapour side is equated to the shear stress on the liquid side at liquid-vapour interface, which is given by Eq. \ref{eq:34}.
\begin{equation}\label{eq:34}
\tau_{l} = \tau_{v}
\end{equation}

The shear stress on the vapor side is given by the following Eq. (\ref{eq:35}).

\begin{equation}\label{eq:35}
\tau_{v} = \mu_{v} \frac{d U_{v}}{dr}
\end{equation}
Similarly the shear stress due on the liquid is given by the following Eq. (\ref{eq:36}).
\begin{equation}\label{eq:36}
\tau_{l} = \mu_{l} \, \frac{U_{l}}{\delta_{eq}}
\end{equation}
Equating the shear stress relationship mentioned in Eq. (\ref{eq:35}) and (\ref{eq:36}) to obtain Eq. (\ref{eq:37}).

\begin{equation}\label{eq:37}
U_{l, mean} = \frac{\mu_{v}}{\mu_{l}} \delta_{eq} \frac{d U_{v}}{dr}
\end{equation}
In the Eq. \ref{eq:37}, the spatial variation of the vapor velocity needs to be determined.

\begin{equation}\label{eq:38}
\frac{d U_{v}}{dr} = \frac{d(u^{*} U_{max})}{dr} = U_{max} \frac{du^{*}}{dr}
\end{equation}

In the Eq. (\ref{eq:38}), \(u^{*}\) is a non-dimensional velocity calculated as described by \citet{Tamayol2010}.
The equation below shows the relationship between the mean velocity and maximum velocity with the help of the velocity distribution along the cross-section.

\begin{equation}\label{eq:39}
U_{max} = \frac{U_{v,mean} \times A_{cs,bub}}{\iint u^{*} r \, dr \, d\theta}
\end{equation}

Differentiating Eq. (\ref{eq:37}) w.r.t  dz results in the Eq. (\ref{eq:39a}).  

\begin{equation}\label{eq:39a}
\frac{dU_{l, mean}}{dz} = \frac{\mu_{v}}{\mu_{l}} \delta_{eq} \, \frac{dU_{max}}{dz} \frac{du^{*}}{dr}
\end{equation}

Substituting the value of \(U_{max}\) from Eq. (\ref{eq:39}) into Eq. (\ref{eq:39a}) results in Eq. \ref{eq:39b}.

\begin{equation}\label{eq:39b}
\frac{dU_{l, mean}}{dz} = \frac{\mu_{v}}{\mu_{l}} \delta_{eq} \, \frac{A_{cs,bub}}{\iint u^{*} r \, dr \, d\theta} \times \frac{dU_{v, mean}}{dz} \frac{du^{*}}{dr}
\end{equation}

From the energy balance the mean velocity variation vapor velocity is given by the Eq. (\ref{eq:40}).

\begin{equation}\label{eq:40}
\frac{d U_{v, mean}}{dz} = \frac{p_{h} \, q}{A_{cs,bub} \, i_{lv} \, \rho_{v}}
\end{equation}

 \(dU_{v,mean} / dz\) from Eq. (\ref{eq:40}) is substituted into Eq. (\ref{eq:39b}) and the resulting  value of  \(dU_{v,mean} / dz\)  into Eq. (\ref{eq:33}) to obtain the rate of depletion of the thin film due to shear stress.

In the case of partial dryout region , liquid is present in the corners and the effect of the shear stress on the depletion of the thin film is modeled as described below.  

\begin{equation}\label{eq:40a}
\frac{d \delta_{corner}}{dt} = -\delta_{corner} \, \frac{d U_{l,mean}}{dz}
\end{equation}

\(dU_{l}/dz\) is calculated using the Eq. (\ref{eq:37}) to (\ref{eq:39a}).  \(dU^{*}/dr\) is calculated using hyperellipse with exponent as 4 and major and minor axes as mentioned in the section on partial dryout region. The limits of \(\theta\) and \(r\) in Eq. (\ref{eq:39}) and (\ref{eq:39b}) correspond to intersection made by the hyperellipse with the rectangle.

\vspace{5mm}
In the present study, for pressure drop module, a single vapour bubble is assumed that comprises of all stages-partially confined, elongated, partial dryout and full dryout-considered in the heat transfer module. This will not affect pressure drop calculation as pressure drop across vapour bubbles is negligible, as mentioned in \textit{Pressure drop} section.  But these stages do influence the heat transfer coefficient and hence are considered in heat transfer module. 

\section{Solution procedure}
In order to solve the equations presented in the bubble growth and heat transfer coefficient sections the channel is divided into several discretized elements with the length as \textit{dz}. Forward time step marching is used with time step \textit{dt}. Sequence of steps followed to obtain the solution is given below.
\begin{enumerate}
	\item The first step involves the determination of the location of nucleation site as explained in section 2.3. 
	\item	The nucleation time period is calculated by the procedure described in section 2.4.
	\item	Based on the time period and nucleation site, a partially confined bubble of dimension \(h \times h \times h\) is placed at the nucleation site, as mention in \textit{Assumptions} section. 
	\item	Velocity at the downstream end of the bubble is calculated using Eq. (\ref{eq:6}) or (\ref{eq:9}) and bubble length based on Eq. (\ref{eq:5}) or Eq. (\ref{eq:8}) depending on the on the  bubble growth stage.
	\item	Then frictional and acceleration pressure drops are calculated using Eq. (\ref{eq:13}) and Eq. (\ref{eq:15}) respectively.
	\item	Inlet stagnation pressure is calculated and then the inlet velocity using Eq. (\ref{eq:fr4}) if inlet compressibility is present.  
	\item	Local heat transfer coefficients are calculated based on the zones as explained in section 2.8, considering the fluid property variation with the local transient pressure.
	\item	Time step is then incremented and the steps (4)-(7) are repeated until periodic behavior is achieved.
	\item	Time averaged and channel averaged pressure drop and heat transfer coefficient are calculated.
\end{enumerate}

\section{Results and discussion}
\subsection{Validation and verification}
\subsubsection{Time step and grid size independence study}

\begin{figure}[H]
	\centering
	\subfloat[Transient pressure drop.\label{fig:2a}]{\includegraphics[width=0.5\textwidth]{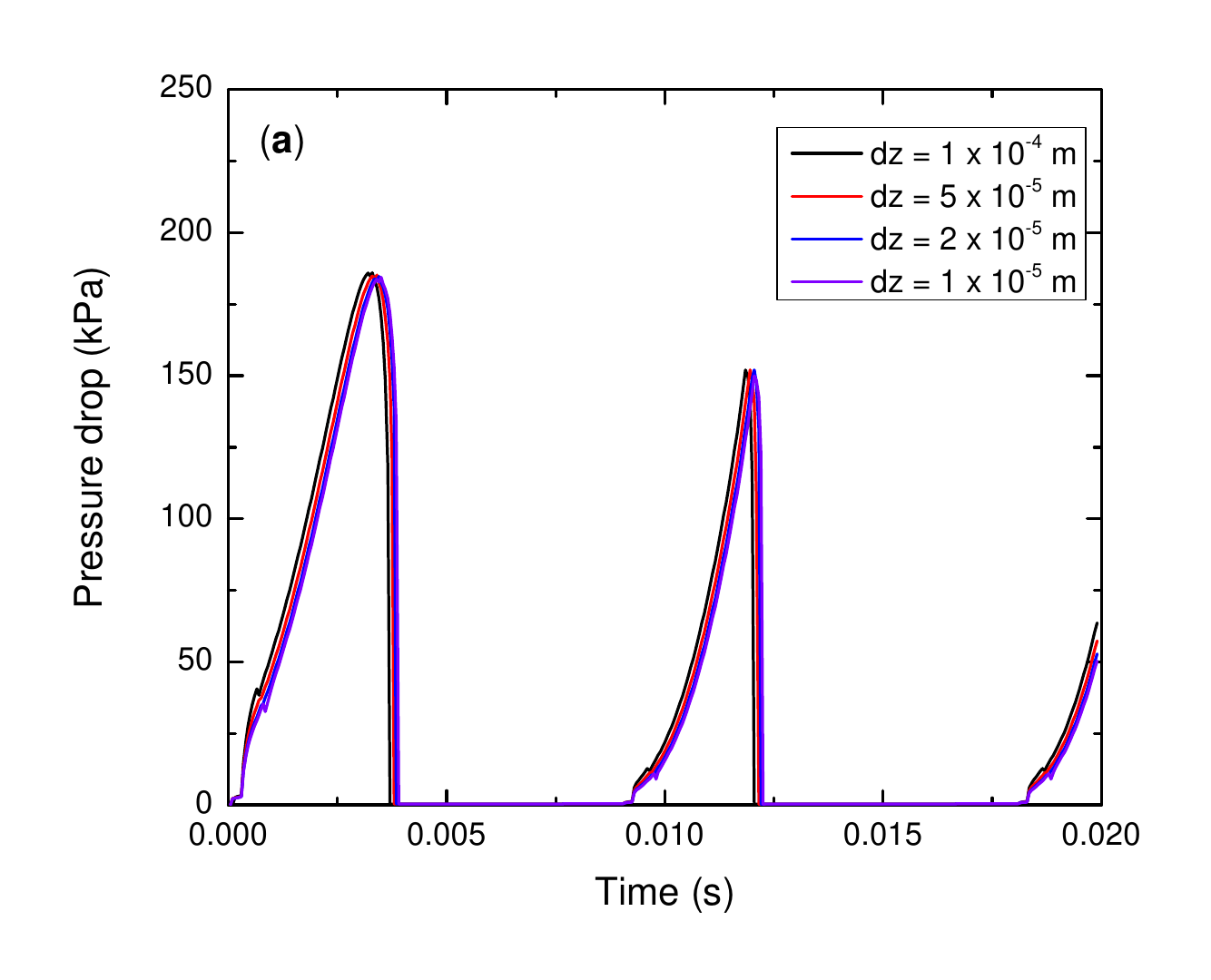}}\hfill
	\subfloat[Transient heat transfer coefficient.\label{fig:2b}] {\includegraphics[width=0.5\textwidth]{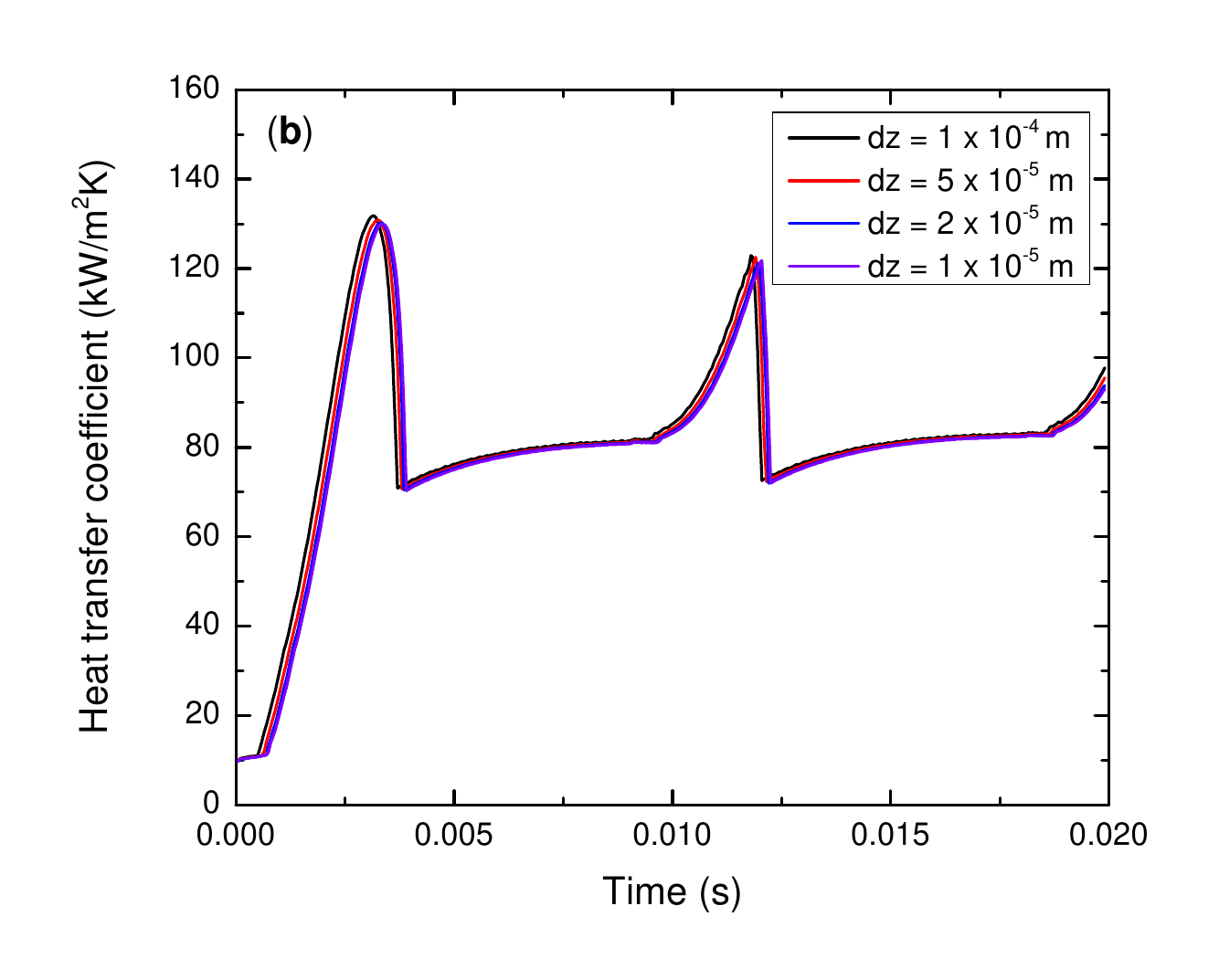}}
	
	\caption{Grid size independence study for channel \( 0.24 \times 0.50 \times 25\) mm, \(q = 400 \, kW/m^{2} \), \(G = 500 \, kg/m^{2}s\),  \(T_{in} = 373 \, K\) and \(P_{e} = 101 \, kPa\).} \label{fig:2}
\end{figure}

\begin{figure}[H]
	\centering
	\subfloat[Transient pressure drop.\label{fig:3a}]{\includegraphics[width=0.5\textwidth]{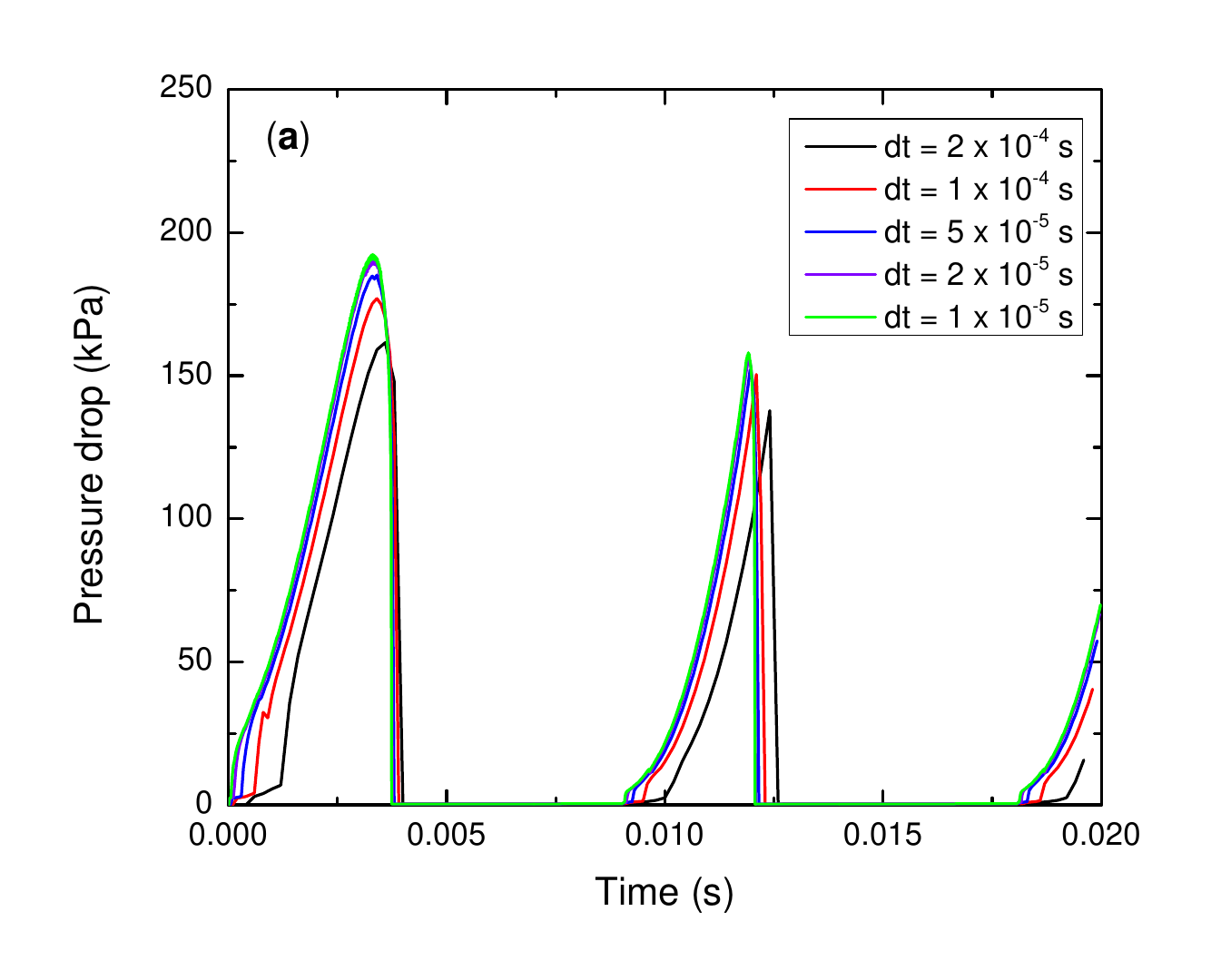}}\hfill
	\subfloat[Transient heat transfer coefficient.\label{fig:3b}] {\includegraphics[width=0.5\textwidth]{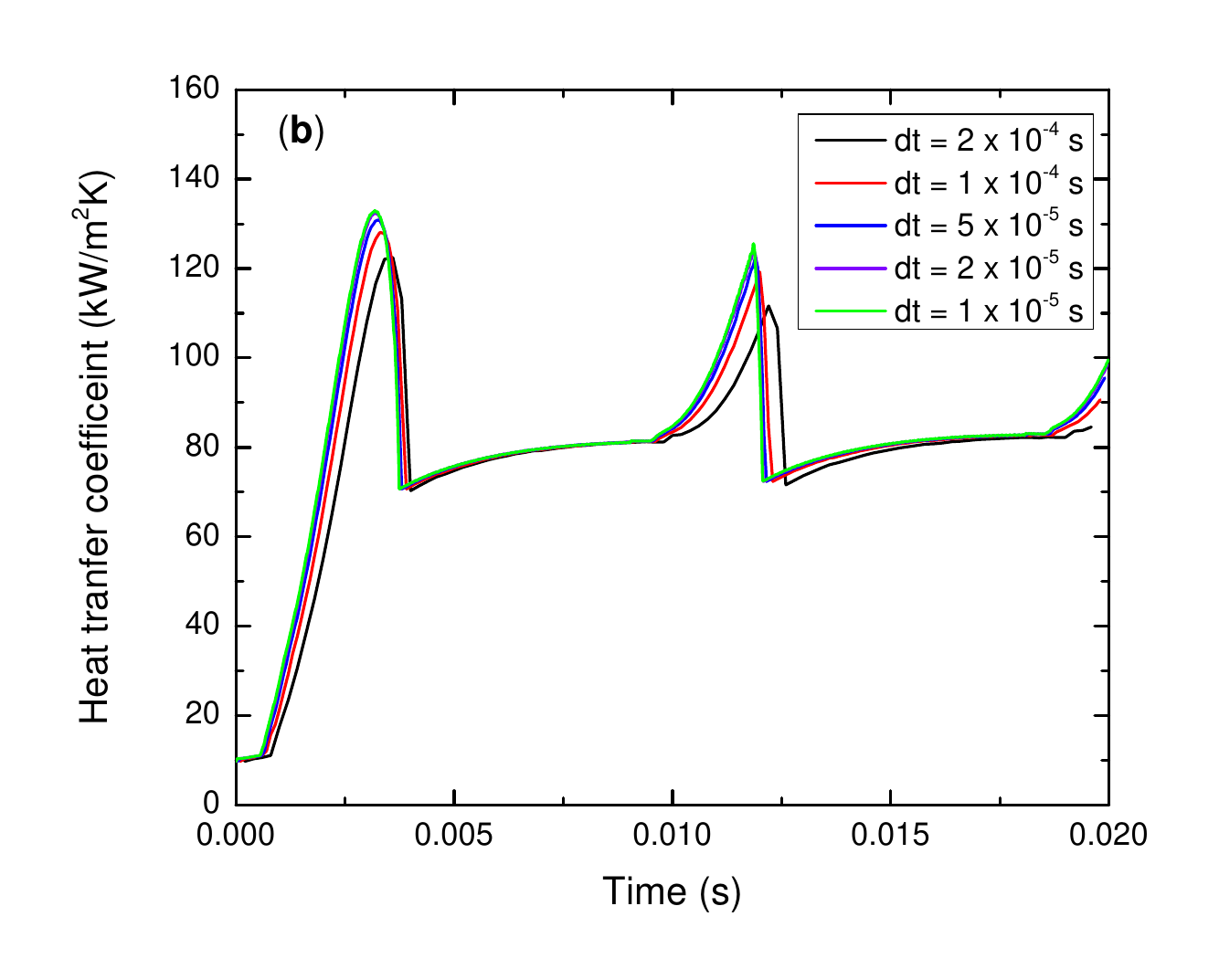}}
	
	\caption{Time step independence study for channel \( 0.24 \times 0.50 \times 25\) mm, \(q = 400 \, kW/m^{2} \), \(G = 500 \, kg/m^{2}s\),  \(T_{in} = 373 \, K\) and \(P_{e} = 101 \, kPa\).} \label{fig:3}
\end{figure}

Grid independence and time-step independence studies are carried out to make sure that the numerical errors do not affect the solution.  Transient heat transfer coefficient and pressure drop variations are observed and it can be seen from Fig. \ref{fig:2} and \ref{fig:3} that the change is negligible for the grid size and time step smaller than \(2 \times 10^{-5}\) m and  \(2 \times 10^{-5}\) s respectively.

\subsubsection{Transient pressure and heat transfer coefficient fluctuations}

\begin{figure}[H]
	\centering
	\subfloat[Comparison with  \citet{Brutin2004}. \label{fig:transp_a}]{\includegraphics[width=0.5\textwidth]{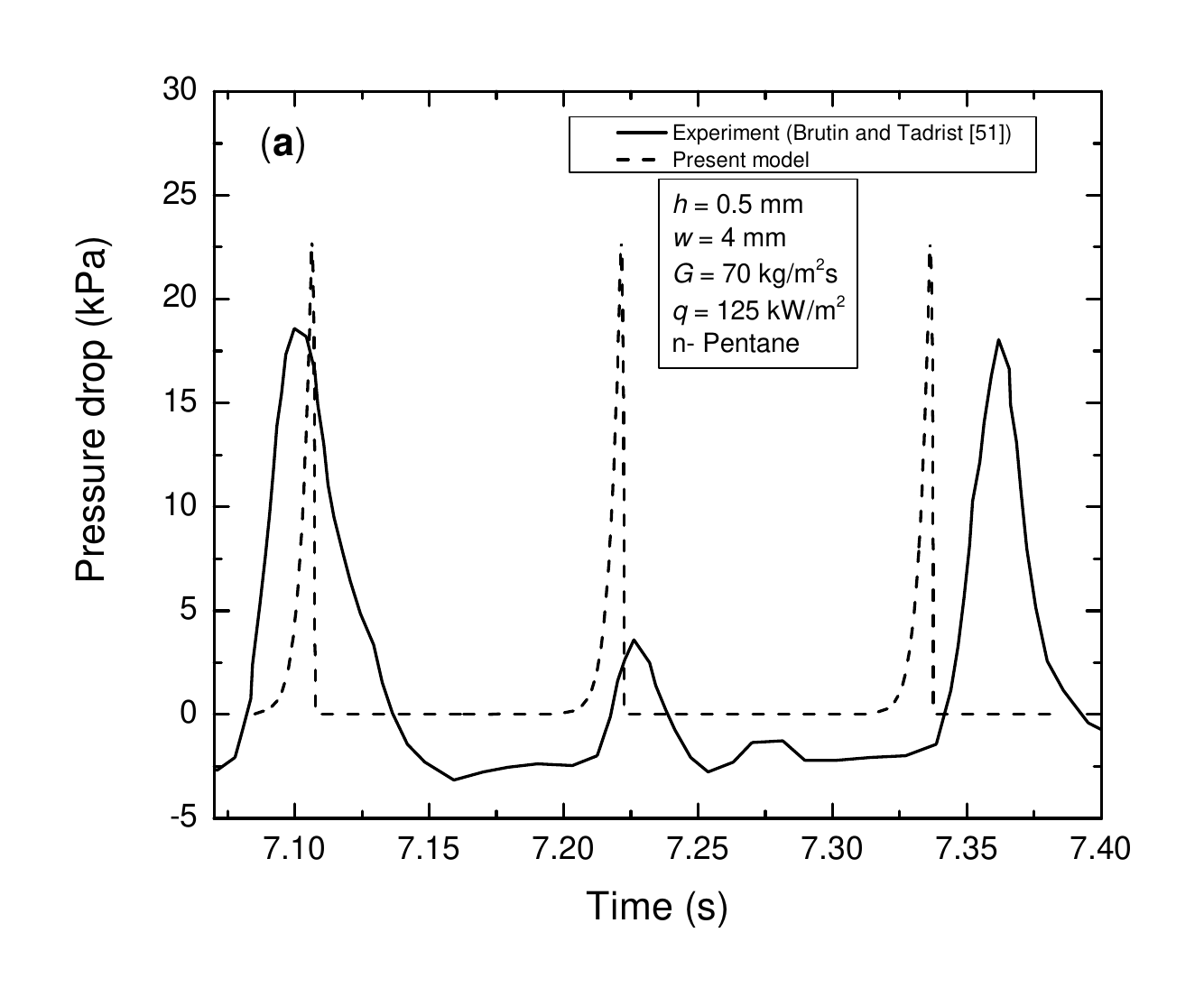}}\hfill
	\subfloat[Comparison with \citet{Zu2011}.\label{fig:transp_b}] {\includegraphics[width=0.5\textwidth]{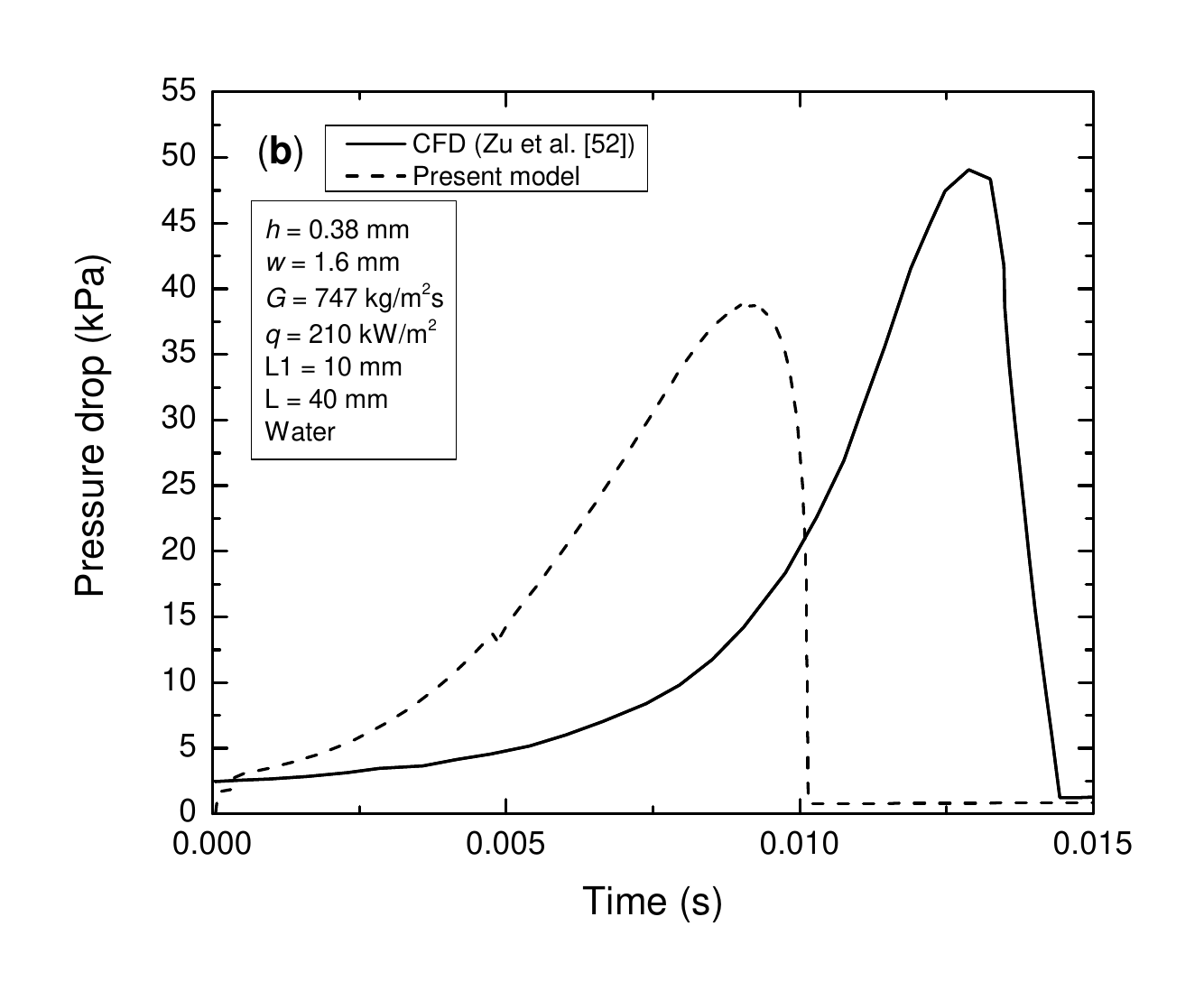}}
	
	\caption{Transient pressure fluctuation variation comparison. }   \label{fig:transp}
\end{figure}

Comparison of the modeled pressure fluctuation with the experimental data reported by \citet{Brutin2004} for n-Pentane is shown in Fig. \ref{fig:transp_a}.  The amplitude predicted by the model closely matches the experimental observation. The nucleation time period used in the model is the same as the observed fluctuation period. The nucleation time period may also depend on the surface characteristics which has not been taken into account in the present study. The location of the nucleation site is calculated as described in section 2.3. The modeled transient pressure has also been compared with the CFD prediction made by \citet{Zu2011} for water and is shown in Fig. \ref{fig:transp_b}. The trend matches with the 3-dimensional CFD result, but the amplitude obtained from model is lower than that predicted by CFD study due to the fact that the modeling considers fluid property variation with local pressure, but the CFD study assumes constant fluid property. The increase in pressure drop is due to the acceleration of the liquid slugs and the corresponding frictional  pressure drop. As the liquid slug leaves the channel, the amount of accelerated liquid slug present in the channel decreases and hence the drop in pressure.

\begin{figure}[H]
	\centering
	\subfloat[Comparison with  \(q = 172 \, kW/m^{2}\) and \(G = 200 \, kg/m^{2}s\). \label{fig:transhtc_a}]{\includegraphics[width=0.5\textwidth]{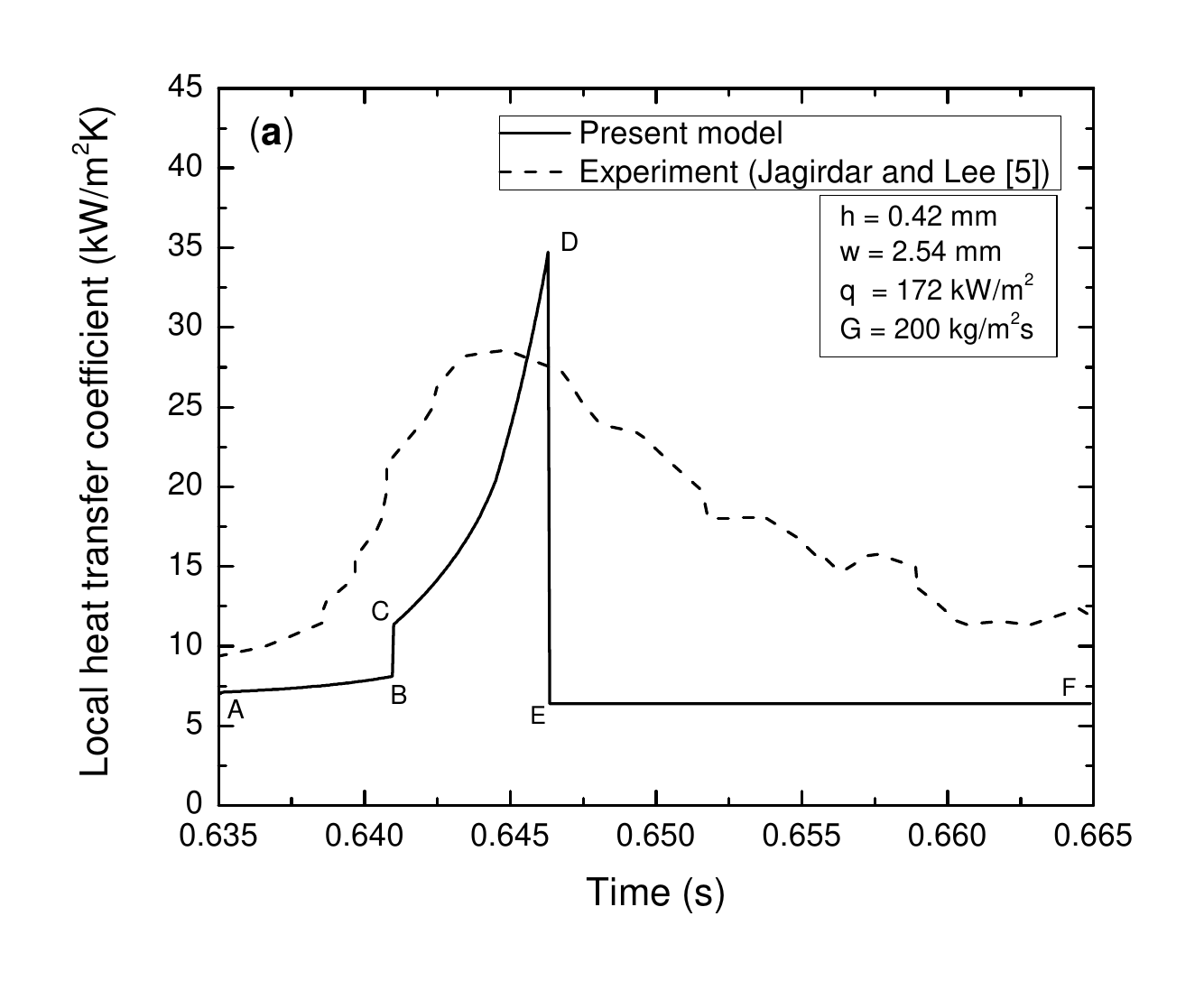}}\hfill
	\subfloat[Comparison with \(q = 320 \, kW/m^{2}\) and \(G = 400 \, kg/m^{2}s\).\label{fig:transhtc_b}] {\includegraphics[width=0.5\textwidth]{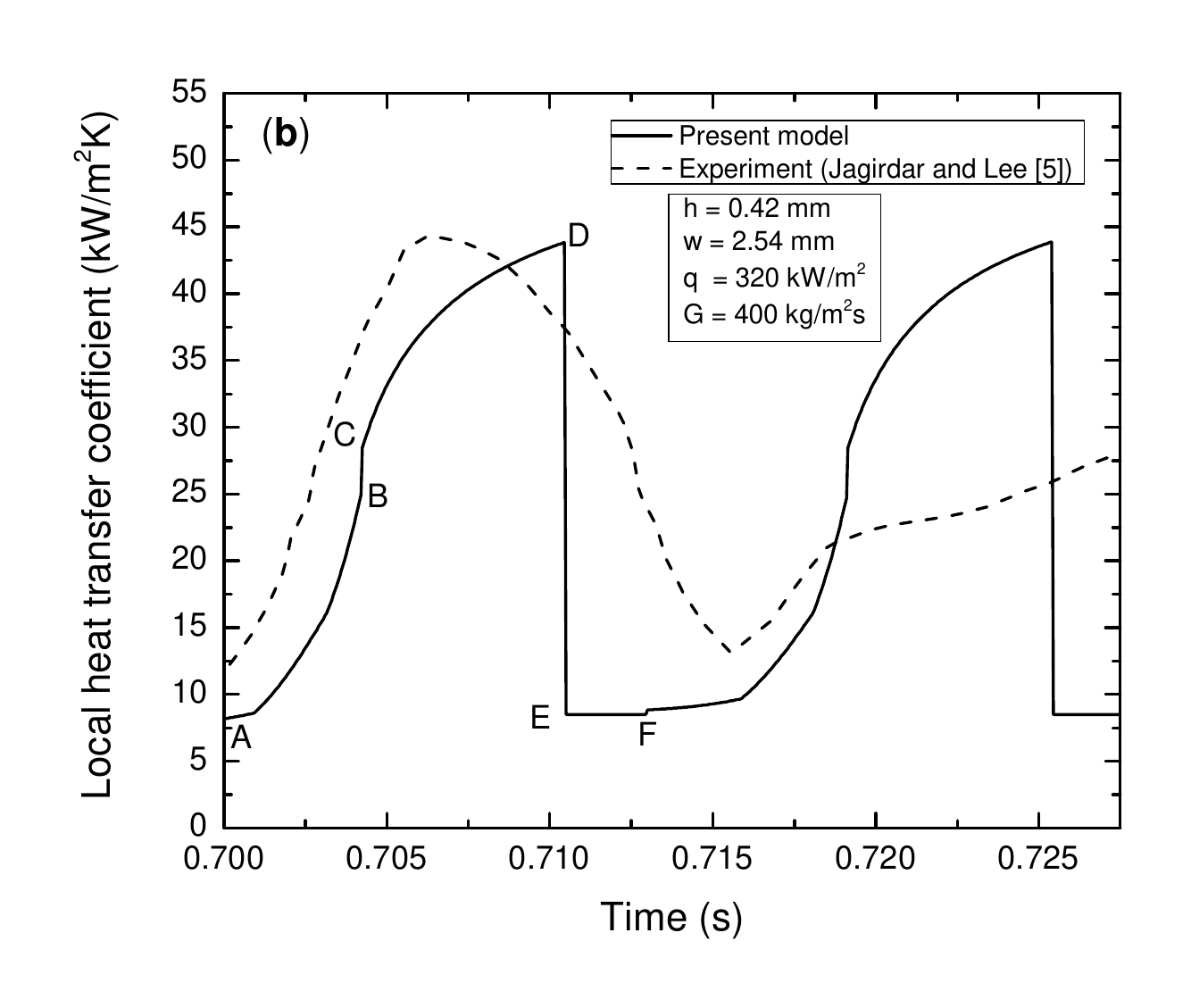}}
	
	\caption{Comparison of transient heat transfer coefficient variation with \citet{Jagirdar2016} at 19.05 mm from inlet.}   \label{fig:transhtc}
\end{figure}

Fig. \ref{fig:transhtc} shows the comparison of the modeled local transient heat transfer coefficient with the experimental data of \citet{Jagirdar2016} for water, under different operating conditions. The modeled heat transfer coefficient is in reasonably good agreement with the experimental data, both in trend and magnitude. 

The local heat transfer coefficient variation from A to B is due to the acceleration of the liquid slug caused by the bubble growth on the upstream side. The jump from B to C is due to the change in zone from liquid slug to vapour bubble that causes a sudden increase in heat transfer coefficient. The gradual increase in heat transfer coefficient from C to D is attributed to the reduction in the thin film thickness caused by evaporation and shear stress. The drop from D to E is due to the change in zone from vapour bubble to liquid slug at that location. The heat transfer coefficient remains constant from E to F, as the liquid slug with constant velocity passes over the location. This continues until a bubble nucleates, which then causes an increase in the liquid velocity and hence the heat transfer coefficient (from A to B).

\subsubsection{Time averaged pressure drop and heat transfer coefficient }
\begin{figure}[H]
	\centering
	\subfloat[Time average pressure drop.\label{fig:4a}]{\includegraphics[width=0.5\textwidth]{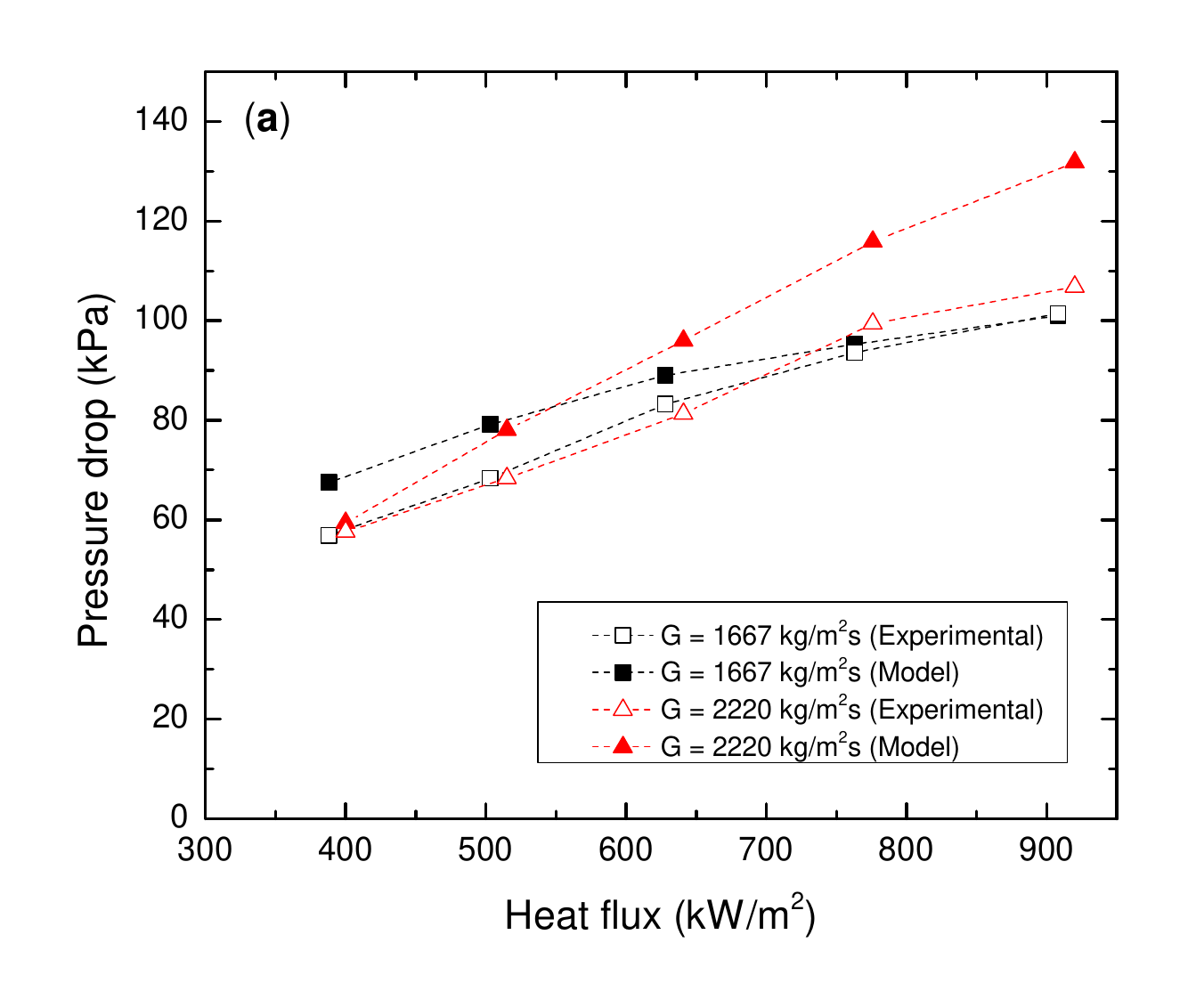}}\hfill
	\subfloat[Time and spactially heat transfer coefficient\label{fig:4b}.] {\includegraphics[width=0.5\textwidth]{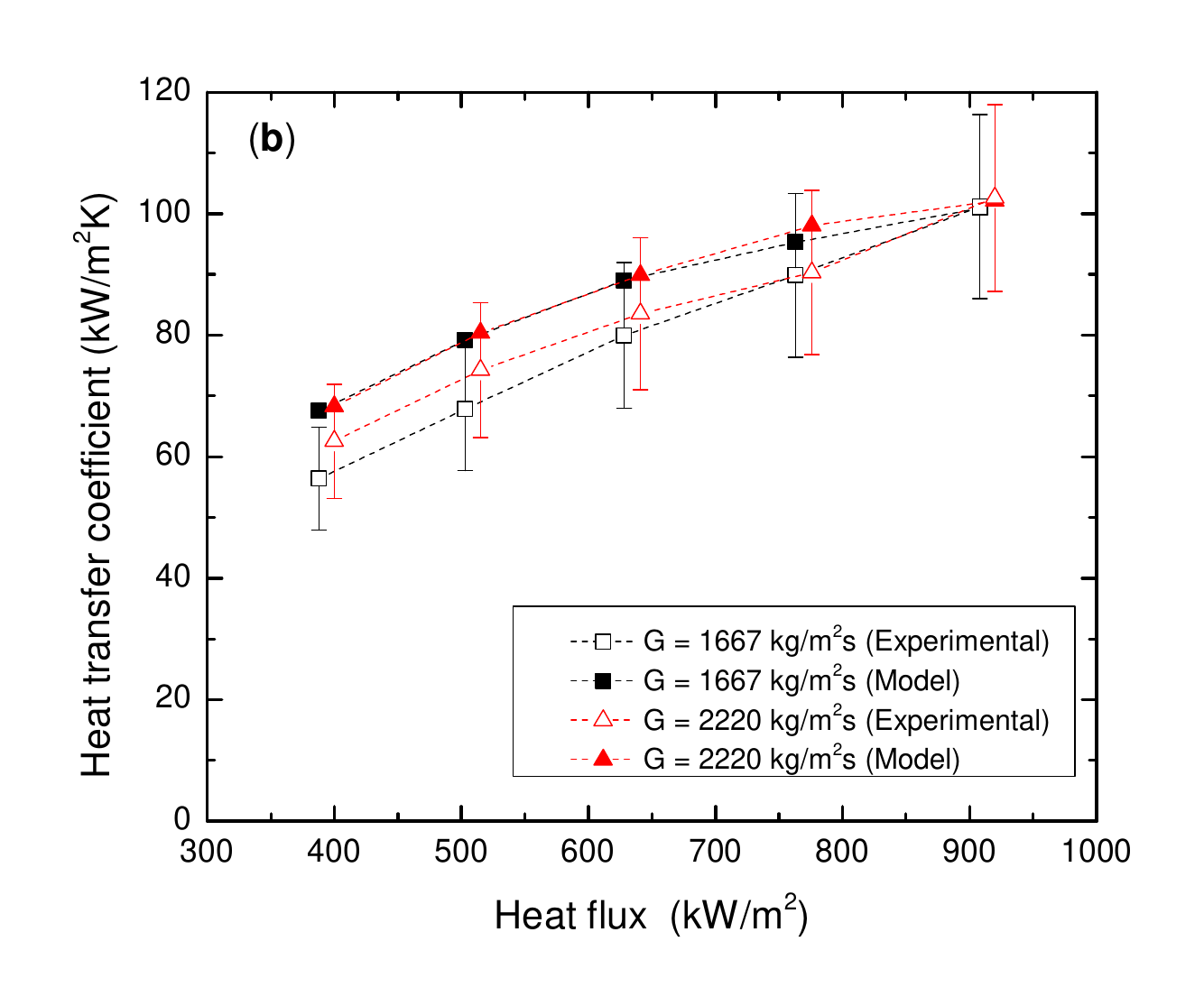}}
	
	\caption{Comparison of pressure drop and heat transfer coefficient with the experimental data \citet{jayaramu2018experimental}, channel dimension 0.24 \(\times\) 0.50 \(\times\) 40 mm.} \label{fig:4}
\end{figure}

In this section, time-averaged pressure drop and heat transfer coefficients are compared with the experimental data reported by \citet{jayaramu2018experimental}. Fig. \ref{fig:4} shows the comparison for a channel of dimension \(0.24  \times 0.50  \times 40 \) mm, for the specified operating conditions. The modeled time-averaged pressure drop is within 25\% of the experimental average pressure drop, as shown in Fig. \ref{fig:4a}, and the time-averaged heat transfer coefficient within 24\%, as in Fig. \ref{fig:4b}. The maximum deviations for a channel of dimension \(0.49 \times 1.00  \times 40 \) mm are 20\% and 5\% for pressure drop and heat transfer coefficient respectively, as shown in Fig. \ref{fig:4}.  The deviations may perhaps be attributed to the change in flow patterns. 

\begin{figure}[H]
	\centering
	\subfloat[{Time average pressure} drop.\label{fig:4a6}]{\includegraphics[width=0.5\textwidth]{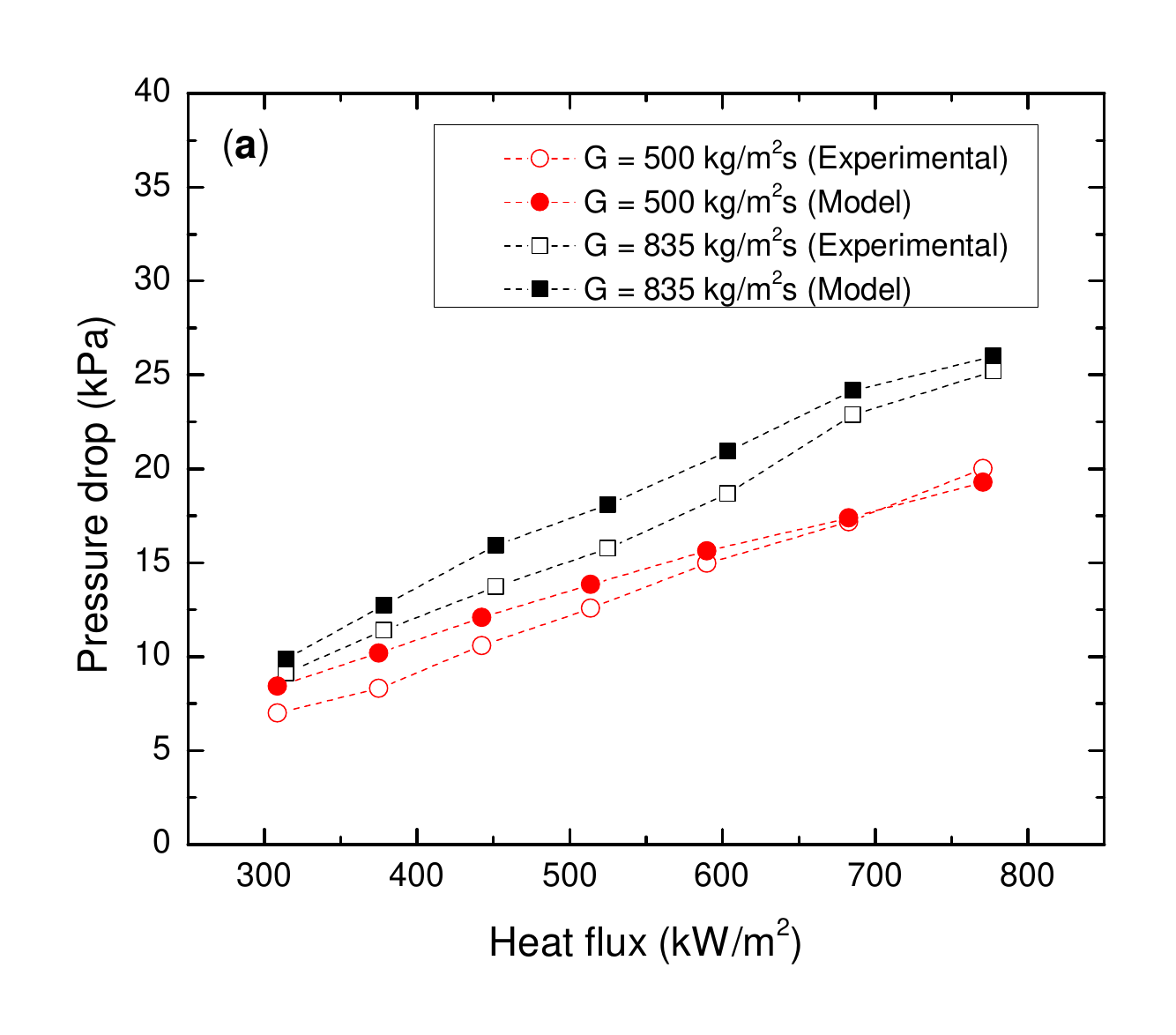}}\hfill
	\subfloat[Time and spatially heat transfer coefficient\label{fig:4b6}.] {\includegraphics[width=0.5\textwidth]{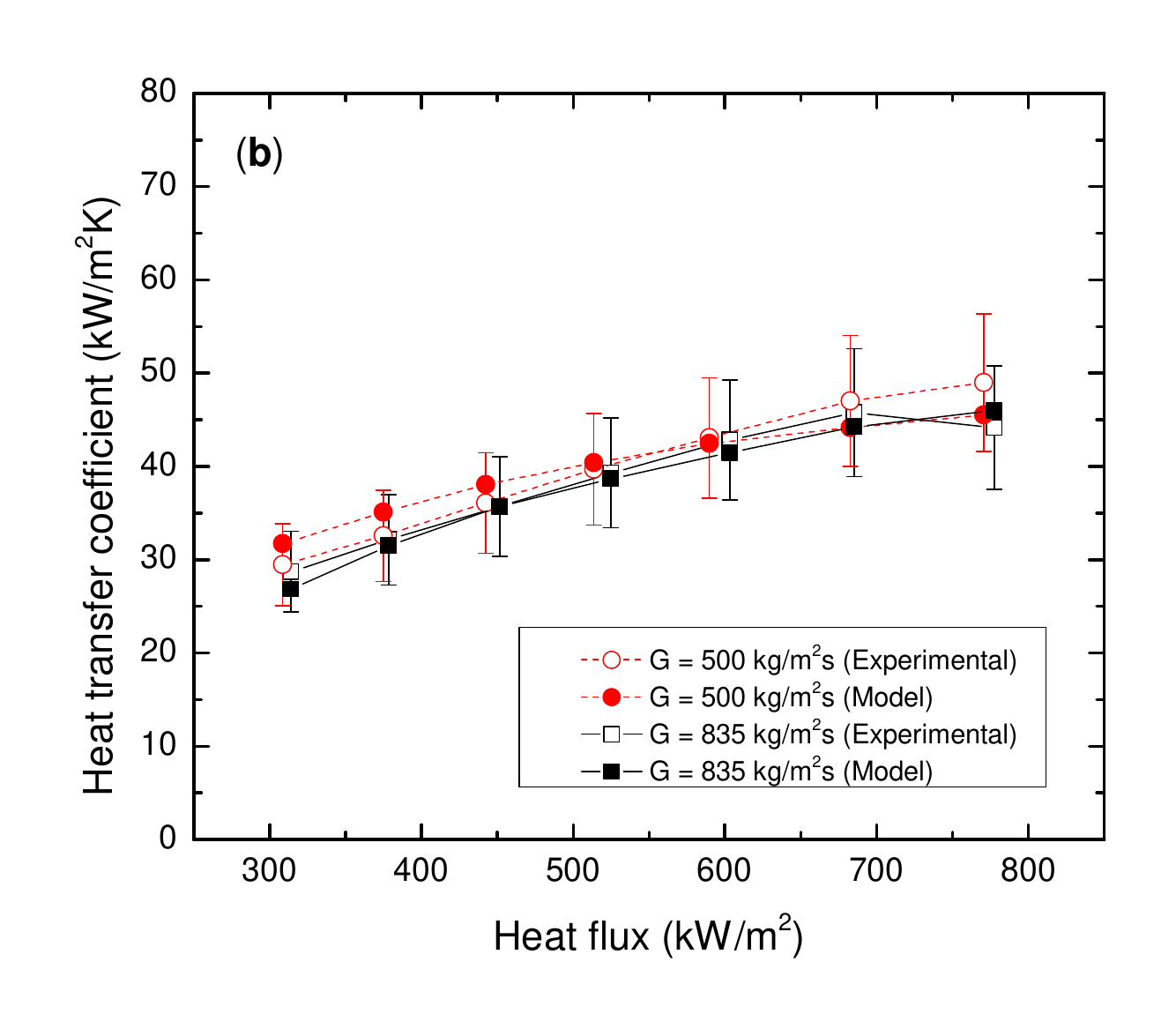}}
	
	\caption{Comparison of pressure drop and heat transfer coefficient with the experimental data \citet{jayaramu2018experimental}, channel dimension 0.49 \(\times\) 1.0 \(\times\) 40 mm.} \label{fig:46}
\end{figure}

In the present model, the calculated quality is different from the thermodynamic quality. Liquid slugs are assumed to be superheated due to the heat transfer from the wall and the heat conducted between the liquid slug and the vapour bubble is neglected as the area of contact between the liquid slug and vapour bubble is assumed to be small compared to the total heat transfer area of liquid slug.   The quality is estimated by the Eq. \ref{eq:q_est}.

\begin{equation}\label{eq:q_est}
x_{est} = \frac{\sum \rho_{v} U_{v}}{\sum \rho_{v} U_{v} +  \sum \rho_{l} U_{l}}
\end{equation}

\begin{figure}[H]
	\centering
	\subfloat[Channel dimension \(0.49 \times 1.0 \times 40 \) mm. \label{fig:qa}]{\includegraphics[width=0.5\textwidth]{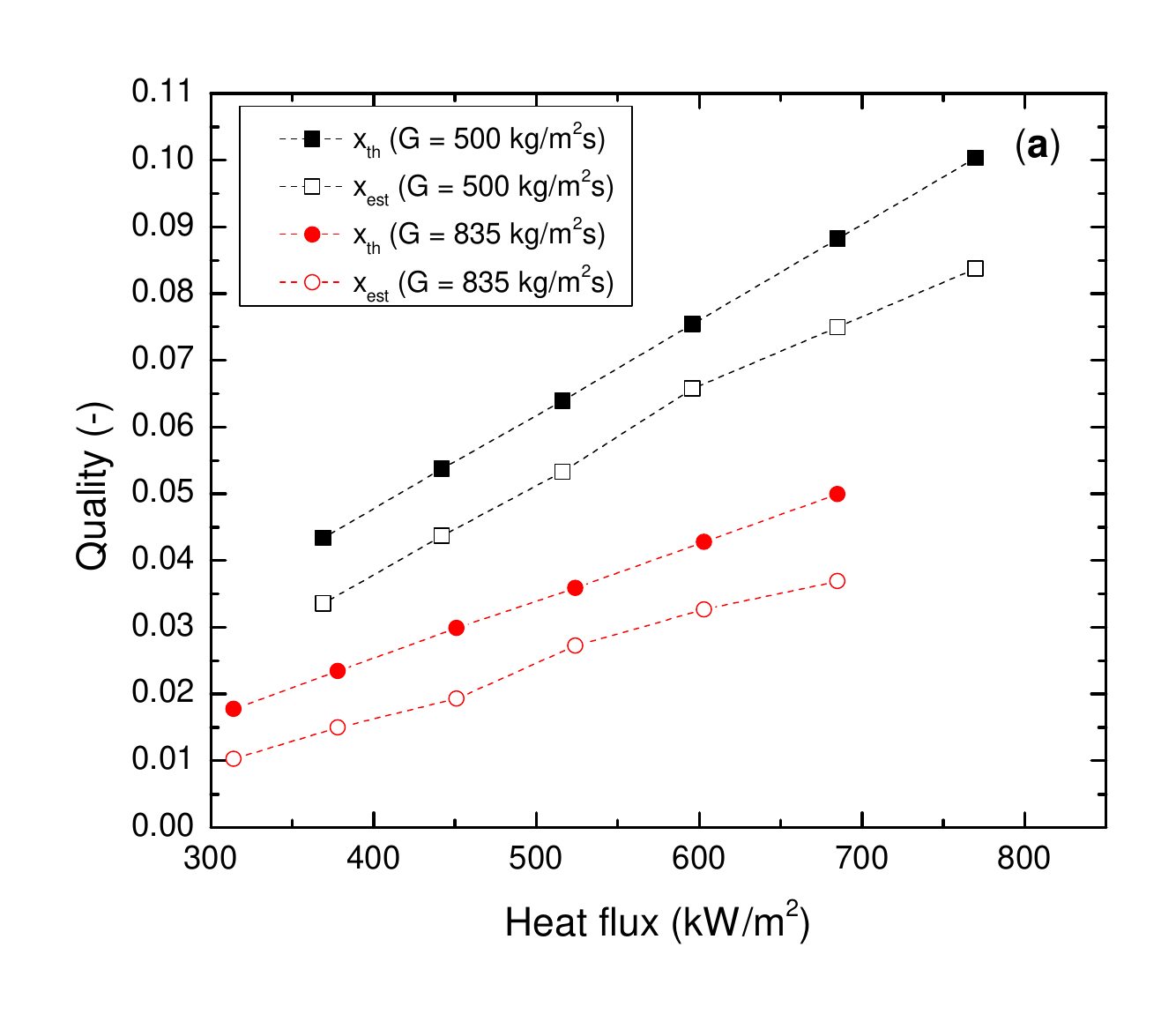}}\hfill
	\subfloat[Channel dimension \(0.24 \times 0.5 \times 40 \) mm.\label{fig:qb}] {\includegraphics[width=0.5\textwidth]{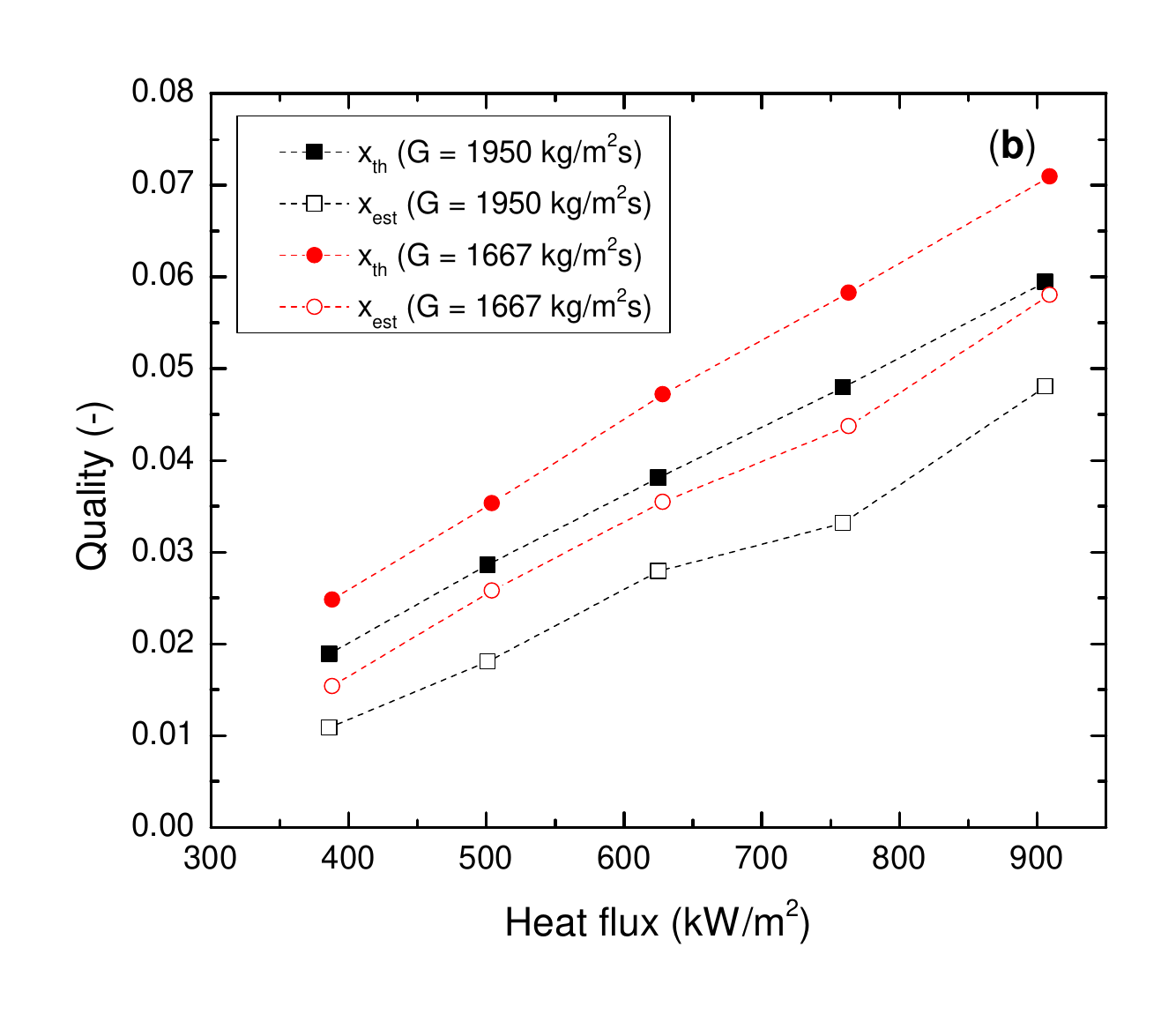}}\hfill
	\caption{Variation of estimated quality and thermodynamic quality.} \label{fig:q}
\end{figure}

The summation is over the time period at the channel exit. The variation in quality with heat flux is presented in Fig. \ref{fig:q}.  There is a reduction in the percentage of deviation between the thermodynamic quality and the actual quality as the heat flux increases. This is due to the decrease in the amount of liquid slug with the increase in frequency caused by the increase in heat flux.  There is also a reduction in the percentage of deviation with the decrease in mass flux, due to the decrease in the amount of liquid slugs. \citet{Magnini2015} performed CFD simulations for confined bubble growth in a microtube under constant heat flux condition and reported 82\% utilization of the heat flux for the bubble growth, which perhaps indicates the presence of superheated liquid and hence the deviation between the actual and thermodynamic quality. The observations from the present study are in line with their results.

\begin{figure}[H]
	\centering
	\subfloat[Correlation comparison for \(G = 600 \,kg/m^{2}s.\) \label{fig:corra}]{\includegraphics[width=0.5\textwidth]{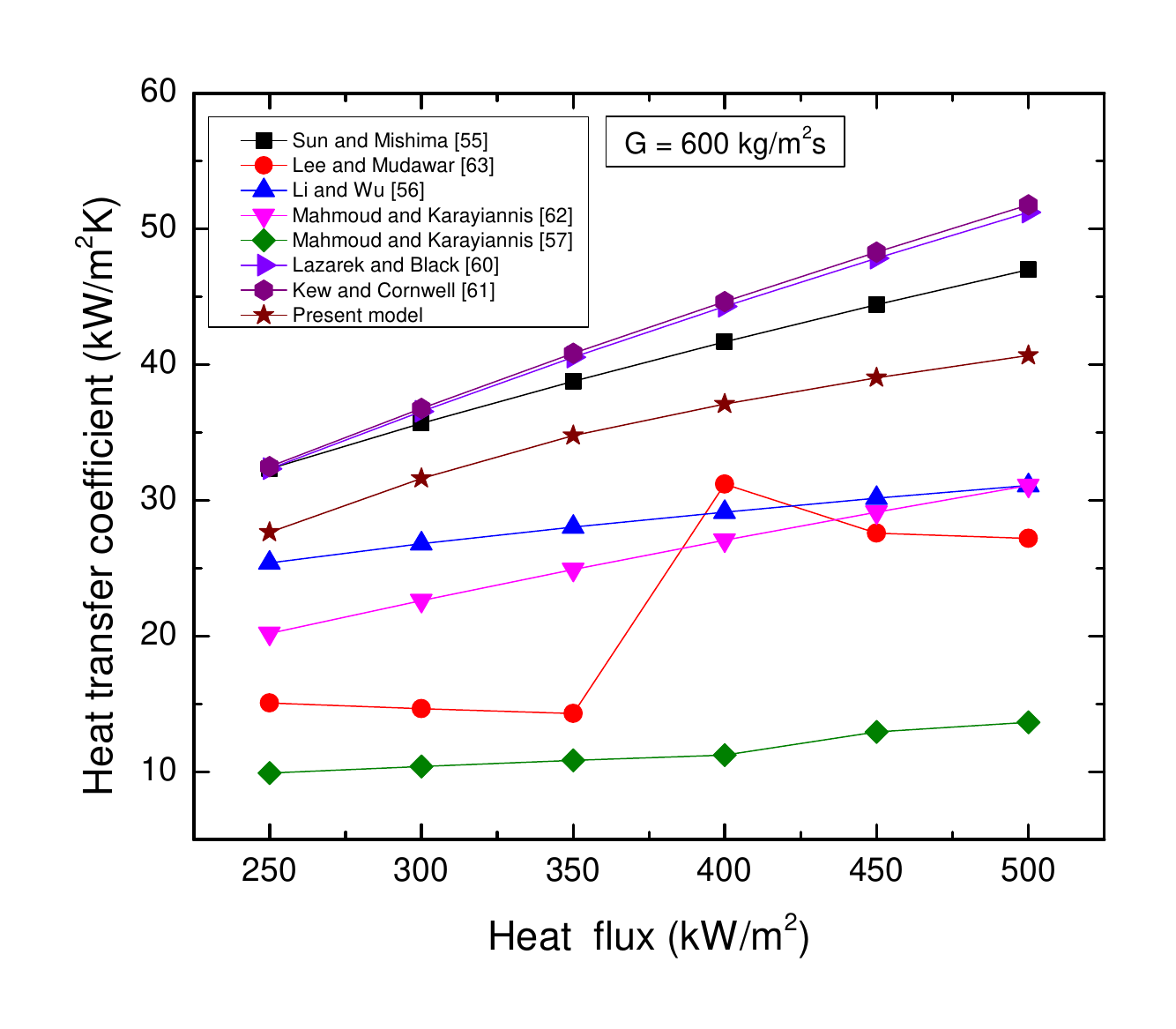}}\hfill
	\subfloat[Correlation comparison for \(G = 1000 \,kg/m^{2}s.\)\label{fig:corrb}] {\includegraphics[width=0.5\textwidth]{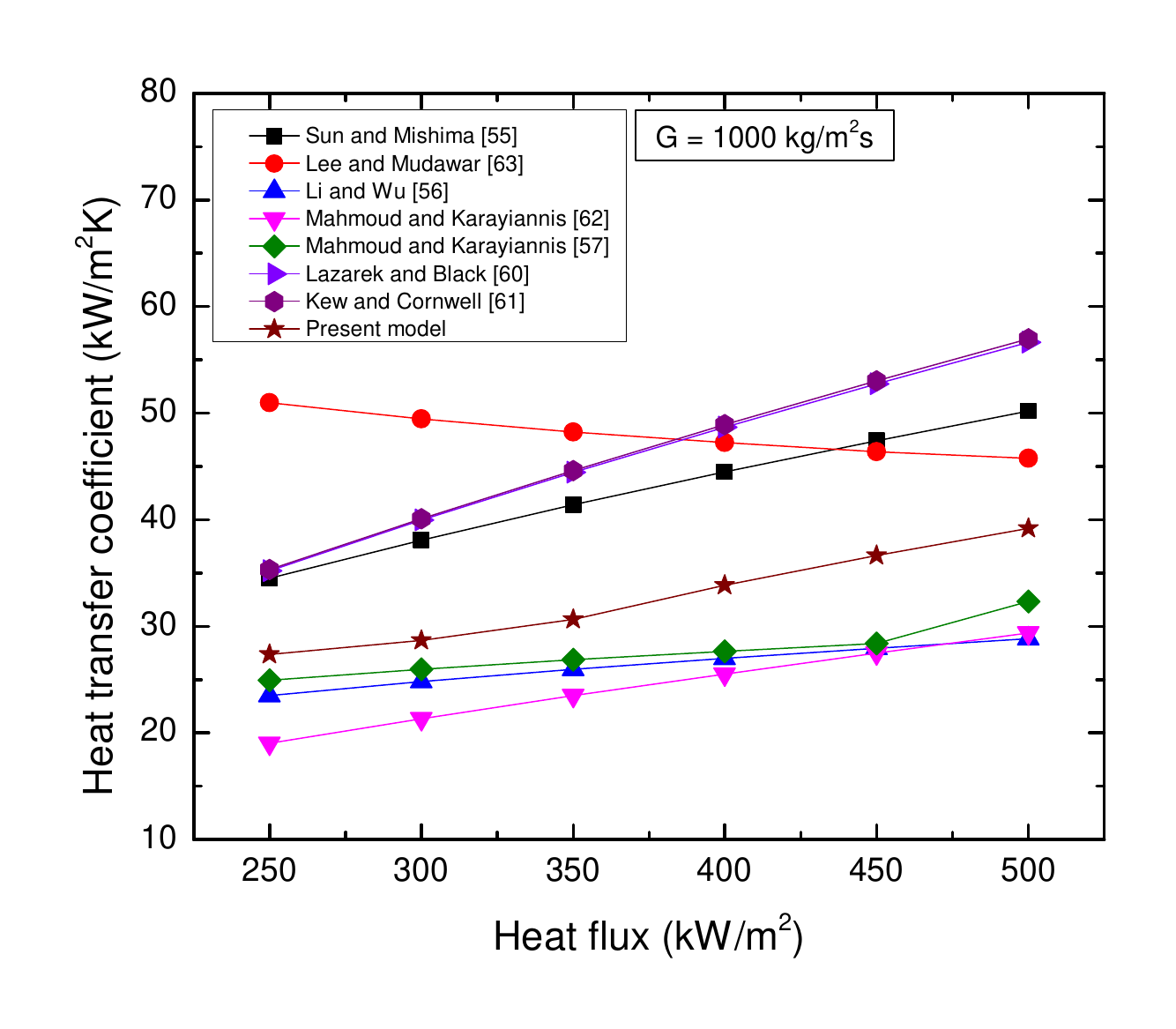}}\hfill
	\caption{Comparison of estimated heat transfer coefficient with the existing correlations for channel \(0.49 \times 1.0 \times 40 \, mm \), \(P_{e} = 101 \, kPa\) and \(T_{in} = 373 \,K\).} \label{fig:corr}
\end{figure}

Fig. \ref{fig:corr} makes a comparison between the spatio-temporally averaged heat transfer from the model and that obtained from the existing steady flow correlations (Table \ref{table}) in the literature.  For lower mass flux, the model prediction lies in between the ones obtained using \citet{Sun2009} model and \citet{li2010general} model, and for higher it is in between \citet{Sun2009} and \citet{mahmoud2013heat}. The possible reasons for deviation are the conditions used for the development of correlations and the influence of heating surface characteristics which has not been taken into account in the present model and also in the development of the correlations.    

\subsection{Different zones}
\begin{figure}[H]
	
	\subfloat[) Local heat transfer coefficient variation with time. \label{fig:zonea}] {\includegraphics[width=0.5\textwidth]{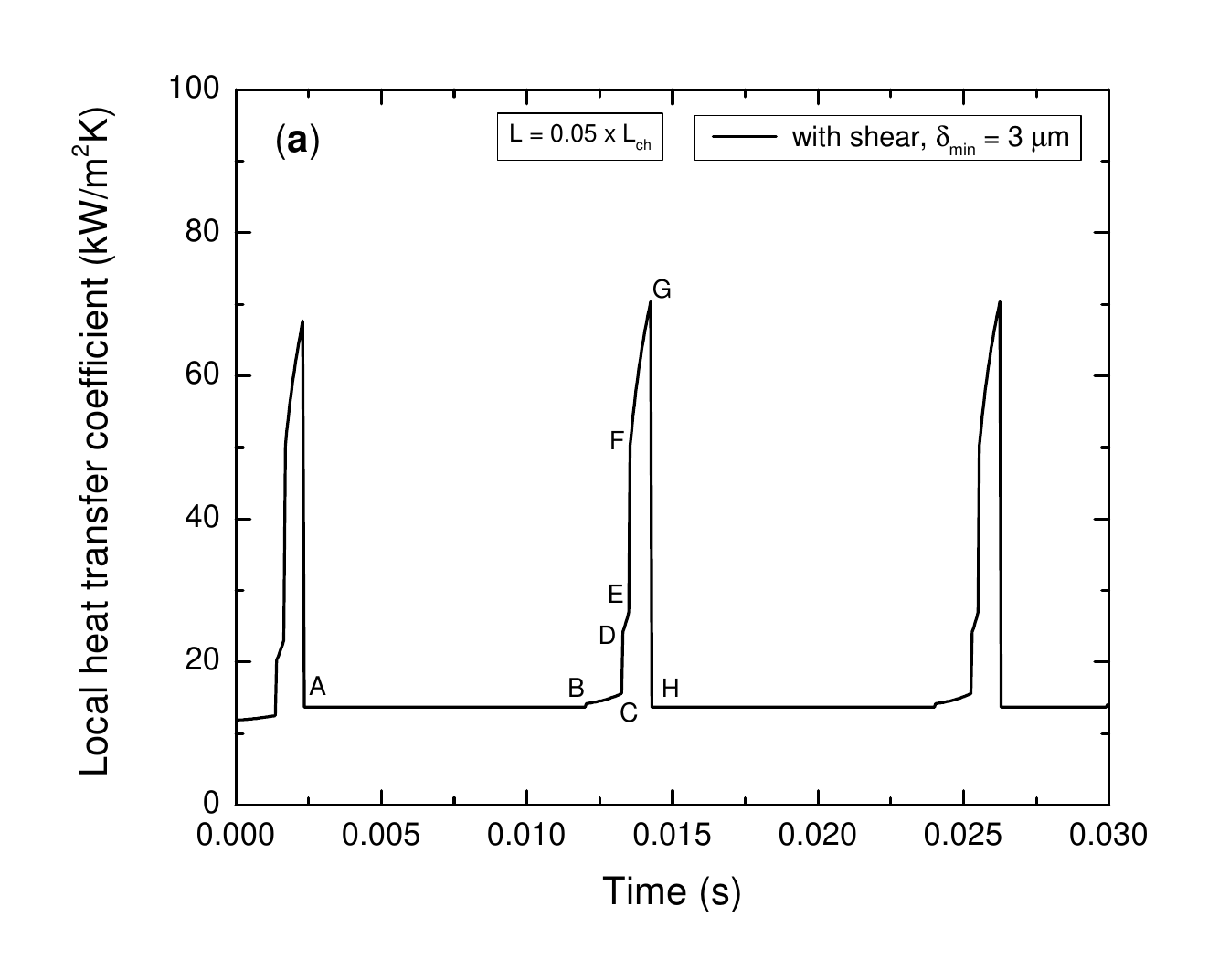}}\hfill
	\subfloat[Local heat transfer coefficient variation with time.\label{fig:zoneb}] {\includegraphics[width=0.5\textwidth]{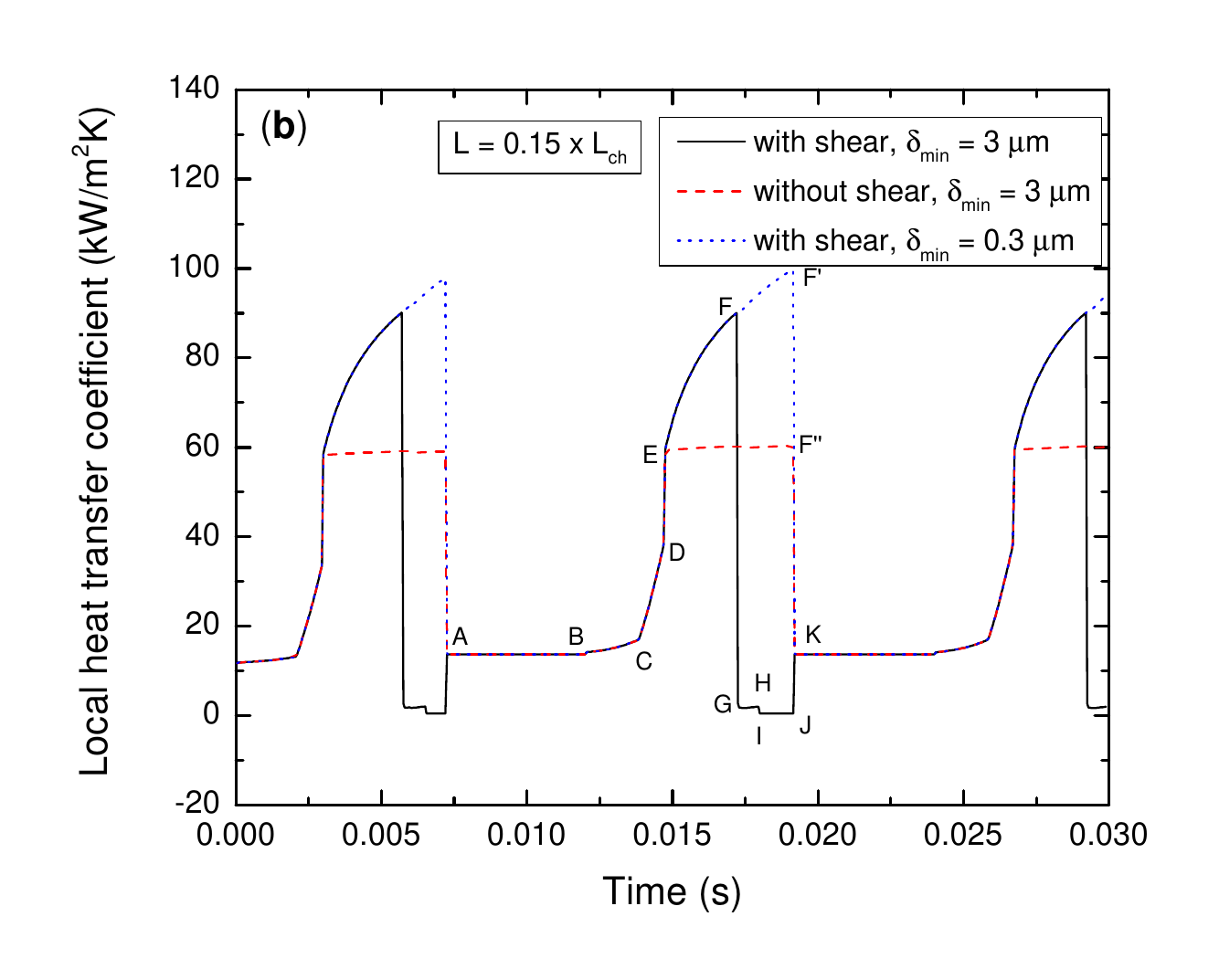}}
	\caption{Various zones during flow boiling, \( 0.20 \times 0.60 \times 25\) mm with \(G = 500 \, kg/m^{2}s\) , \(q = 200 \, kW/m^{2}\) , \(P_{e} = 101 \, kPa\) and \(T_{in} = 373 \, K\).} \label{fig:zone}
\end{figure}

The occurrence of five different zones during flow boiling depends on the channel dimension, inlet condition, mass flux and heat flux.  Fig. \ref{fig:zonea} shows different zones passing cyclically over the location that is 1.25 mm away from the inlet. Initially the transients are affected by the initial condition during simulation and later the cycle repeats itself. Zone A-B is the liquid slug zone (un-accelerated),  B-C is the liquid slug region (accelerated), the jump C-D represents transition from liquid to partial confinement, D-E the partially confined bubble region, E-F the jump from partial confinement to full confinement, F-G the fully confined or elongated bubble stage and G-H the transition from the bubble to the liquid slug zone. Fig. \ref{fig:zoneb} shows different zones over a location that is 3.75 mm from inlet, for with and without shear and for different minimum liquid film thickness values for dryout.  Zones A-B, B-C, C-D and D-E indicate the liquid slug zone (un-accelerated), the liquid slug zone (accelerated by partially confined bubble), the liquid slug zone (accelerated by fully confined bubble) and the transition from the liquid slug to the elongated bubble region, respectively. The zones E-F, E-F' and E-F'' represent elongated bubble region for the case with shear (3 \(\mu m\)), with shear (0.3 \(\mu m\)) and without shear (3   \(\mu m\)) respectively. The minimum liquid film thickness values considered in the present study are based on the experimental observations by \citet{sun2018transient}. Zones F-G, G-H, H-I, I-J and J-K represent the transition to partial dryout, the partial dryout region, the transition to full dryout, the full dryout region and the transition to the liquid slug region, respectively.  Zones F'-K and F''-K indicate the transition from elongated bubble to liquid slug region. The differences between the two figures are the presence of partially confined bubble region and the absence of partial and full dryout in Fig. \ref{fig:zonea} as against Fig. \ref{fig:zoneb}. The effect of shear stress is discussed in detail in the section 4.3.

\subsection{Shear stress effect}
\begin{figure}[H]
	\centering
	\subfloat[Variation of spatially averaged heat transfer coefficient with time. \label{fig:14a}]{\includegraphics[width=0.5\textwidth]{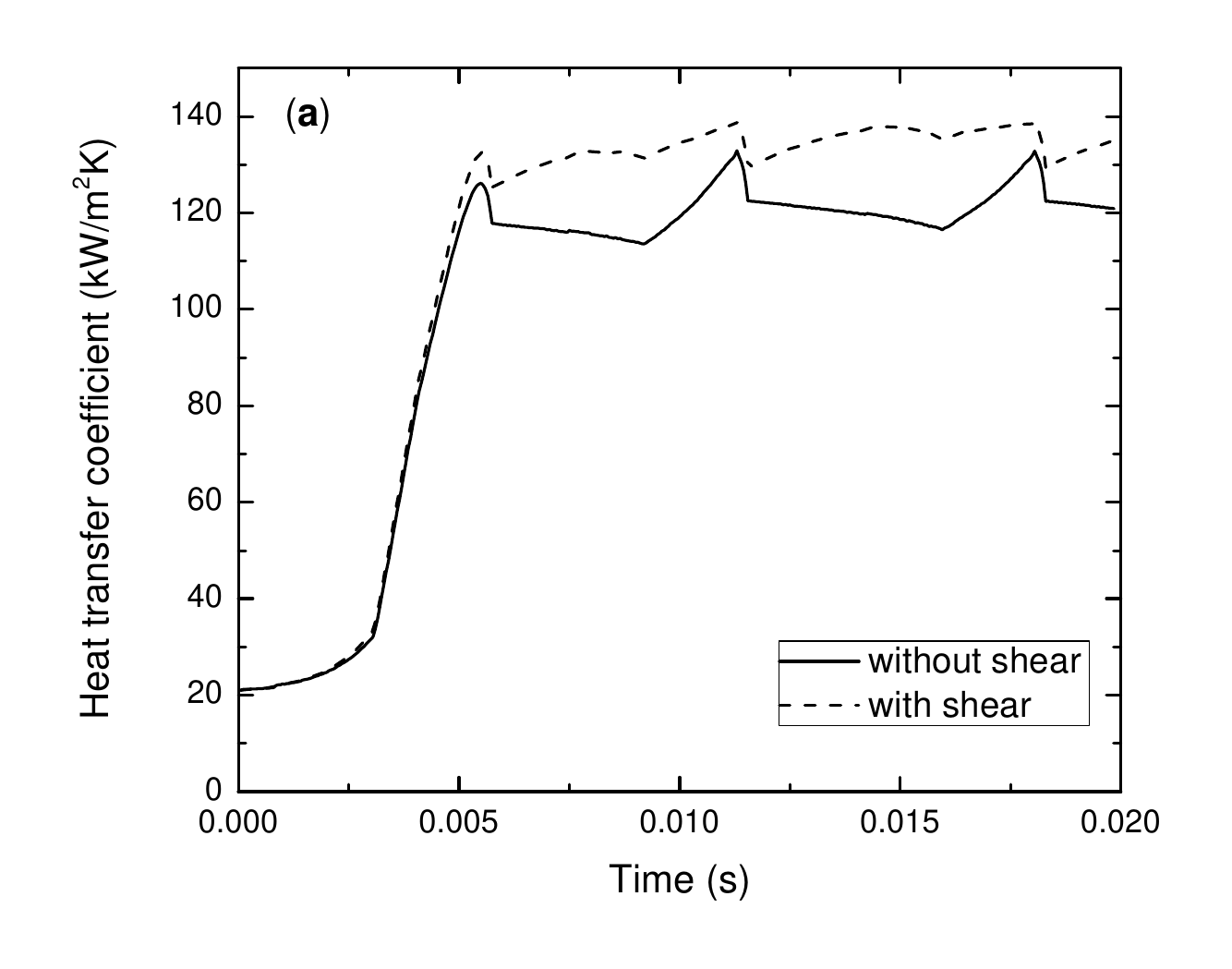}}\hfill
	\subfloat[Variation of time-averaged heat transfer coefficient with location. \label{fig:14b}] {\includegraphics[width=0.5\textwidth]{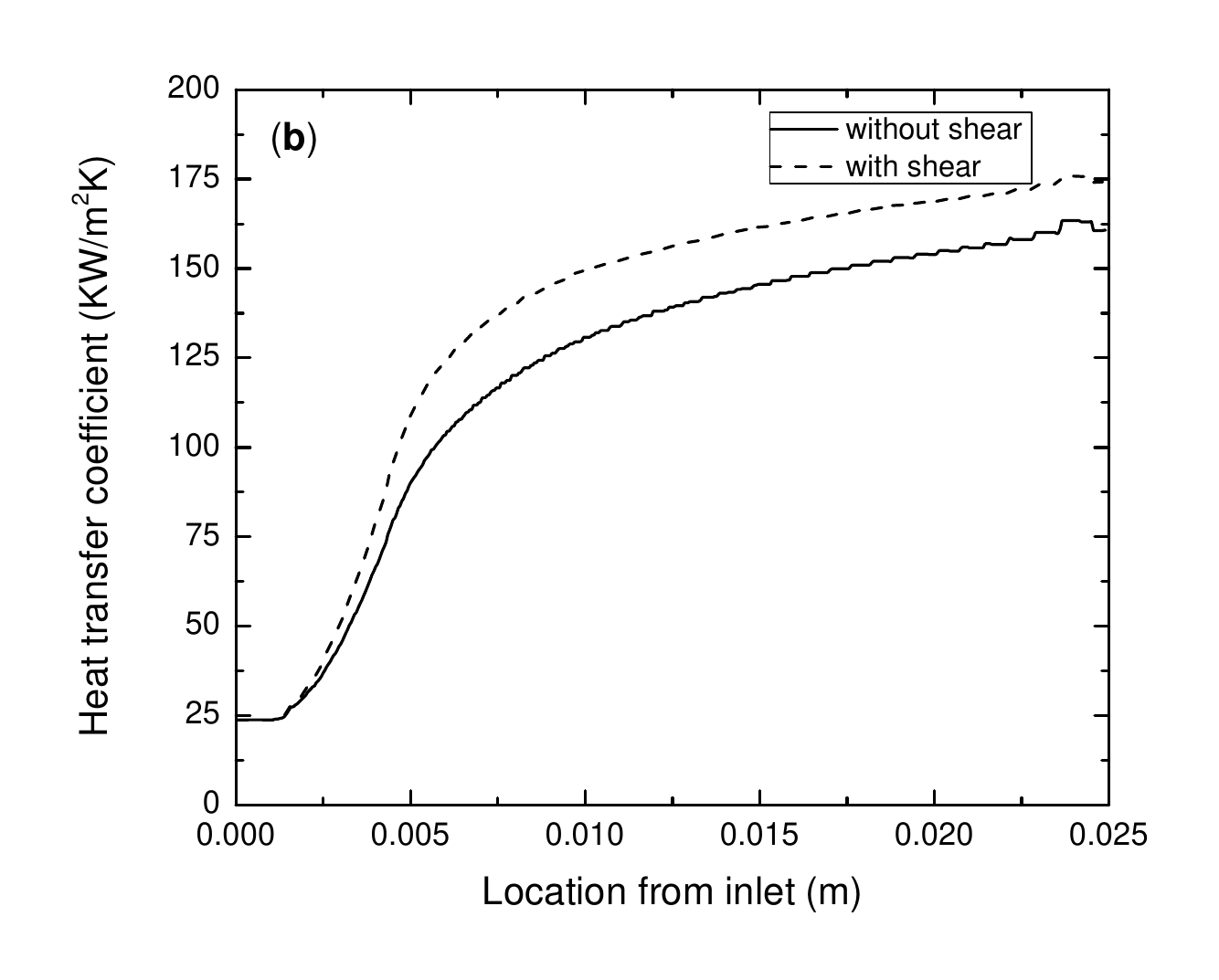}}\hfill
\end{figure}

\begin{figure}[H]
	
	\subfloat[ Local heat transfer coefficient variation with time. \label{fig:14c}] {\includegraphics[width=0.5\textwidth]{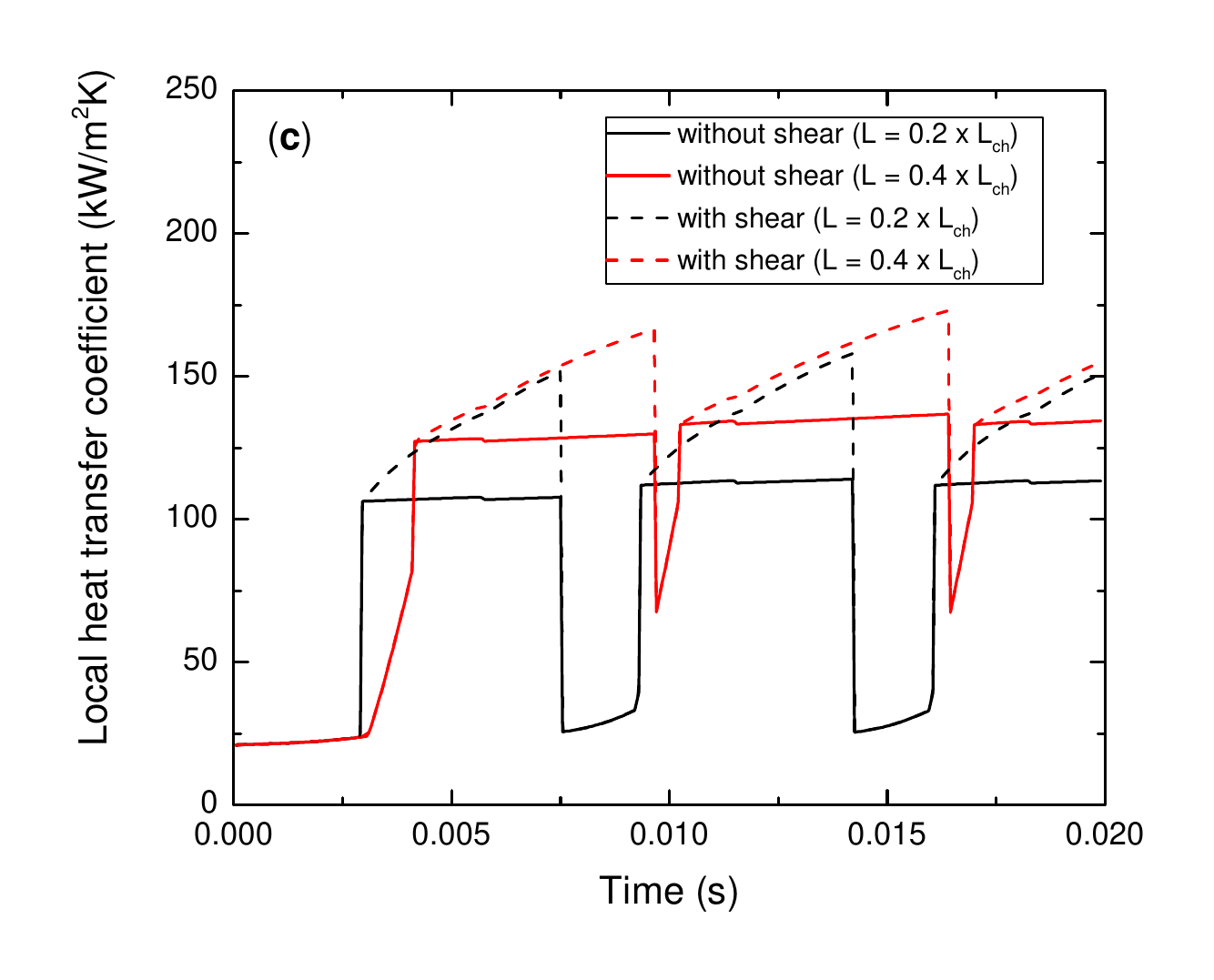}}\hfill
	\subfloat[Local thin film thickness variation with time.\label{fig:14d}] {\includegraphics[width=0.5\textwidth]{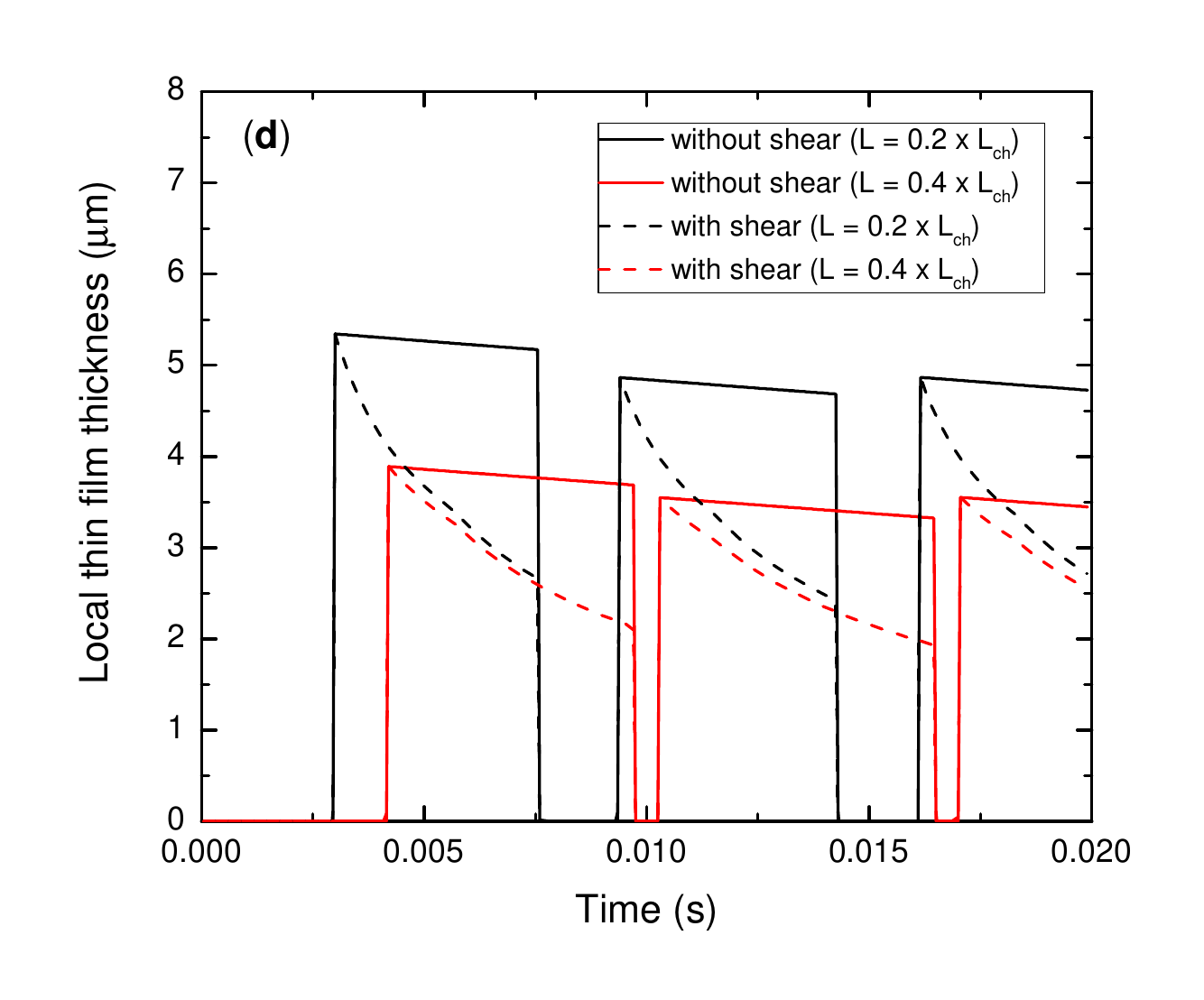}}
	\caption{Effect of shear stress, \( 0.10 \times 0.20 \times 25\) mm with \(G = 500 \, kg/m^{2}s\), \(q = 100 \, kW/m^{2}\), \(P_{e} = 101 \, kPa\) and \(T_{in} = 373 \, K\).} \label{fig:14}
\end{figure}

In the process of bubble growth, the thin liquid film gets depleted due to the evaporation of film and also due to the shear stress acting on the liquid-vapor interface. The models presented by \citet{Thome2004} and \citet{Magnini2017} for circular channel and \citet{Wang2010} for rectangular channel, do not take into account the influence of shear. Fig. \ref{fig:14} shows the effect of shear stress on the variations of heat transfer coefficient and thin film thickness. Dashed lines indicate the heat transfer coefficient with shear and solid lines without shear. The spatially averaged and the time-averaged heat transfer coefficients with shear stress are higher than that without shear, as shown in Fig. \ref{fig:14a} and Fig. \ref{fig:14b}, respectively. This can be explained from Fig. \ref{fig:14c} that shows the variation of local heat transfer coefficient with time for two different locations (\(0.2 \, \times \, L_{ch}\) and \(0.4 \, \times \, L_{ch}\) from the inlet). With shear, the heat transfer coefficient during the passage of elongated bubble  is higher compared to that without shear due to the higher depletion rate of thin film with shear as shown in Fig. \ref{fig:14d}. The reduction in the thin film thickness is larger for the upstream location as the rate of depletion of thin film is directly proportional to the film thickness as indicated by the Eq. (\ref{eq:33}). This leads to a larger increase in the heat transfer coefficient for the upstream location.


\subsection{Effect of inlet compressibility}

\begin{figure}[H]
	\centering

	\subfloat[Comparison of position of upstream and downstream end of bubble.\label{fig:17a}] {\includegraphics[width=0.5\textwidth]{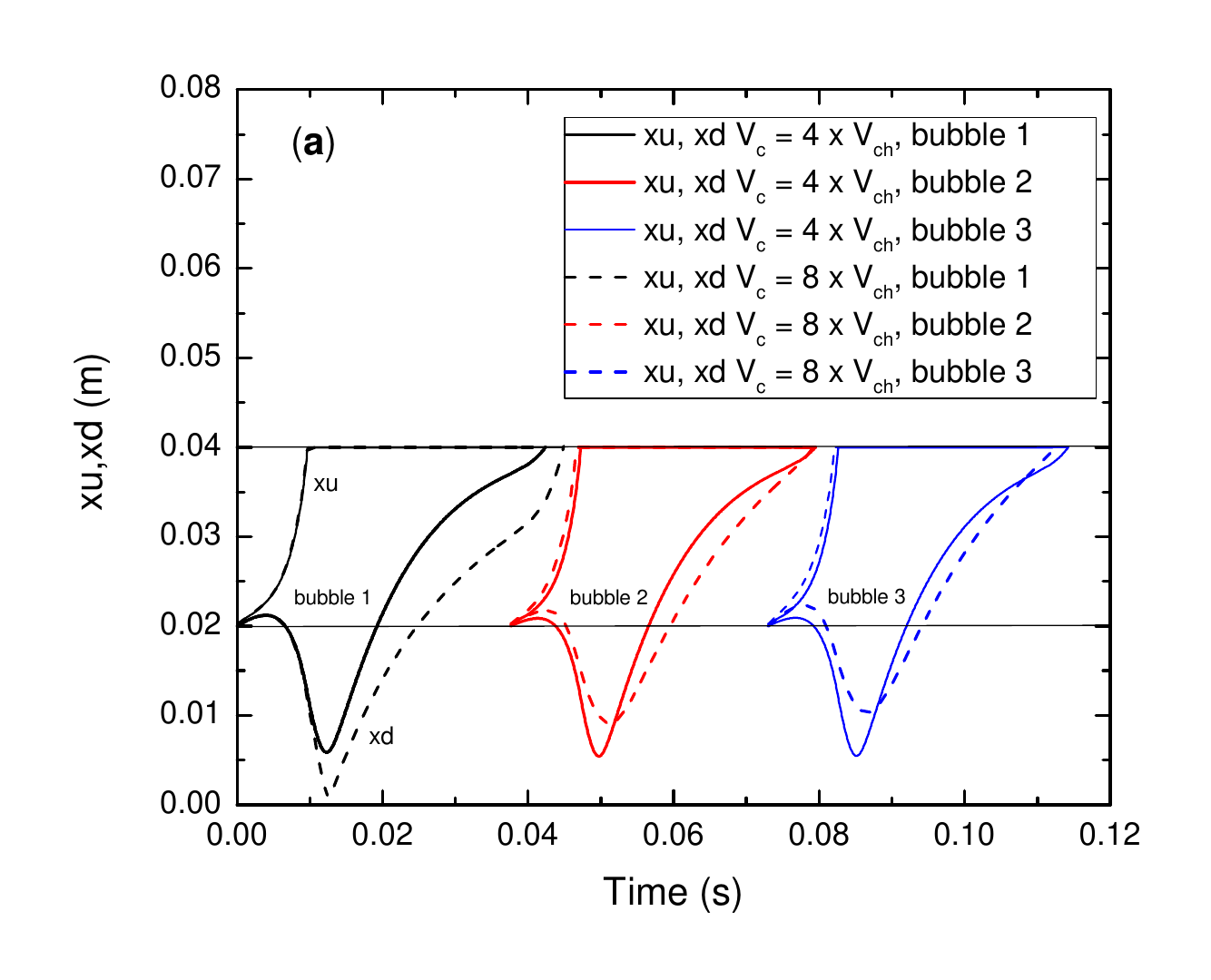}}\hfill
	\subfloat[Variation of pressure drop with time. \label{fig:17b}]{\includegraphics[width=0.5\textwidth]{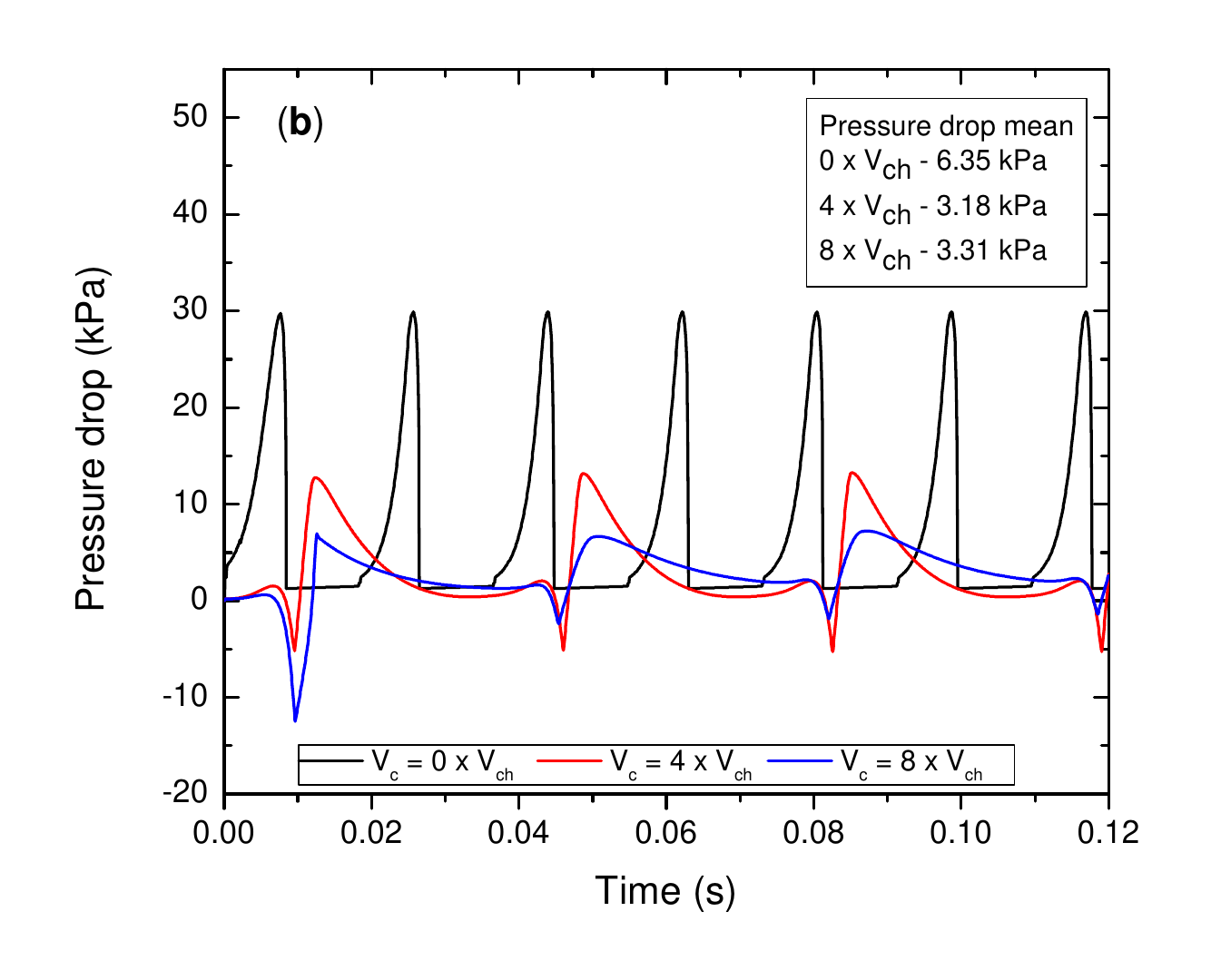}}\hfill
	\subfloat[Variation of spatial-averaged heat transfer coefficient with time. \label{fig:17c}] {\includegraphics[width=0.5\textwidth]{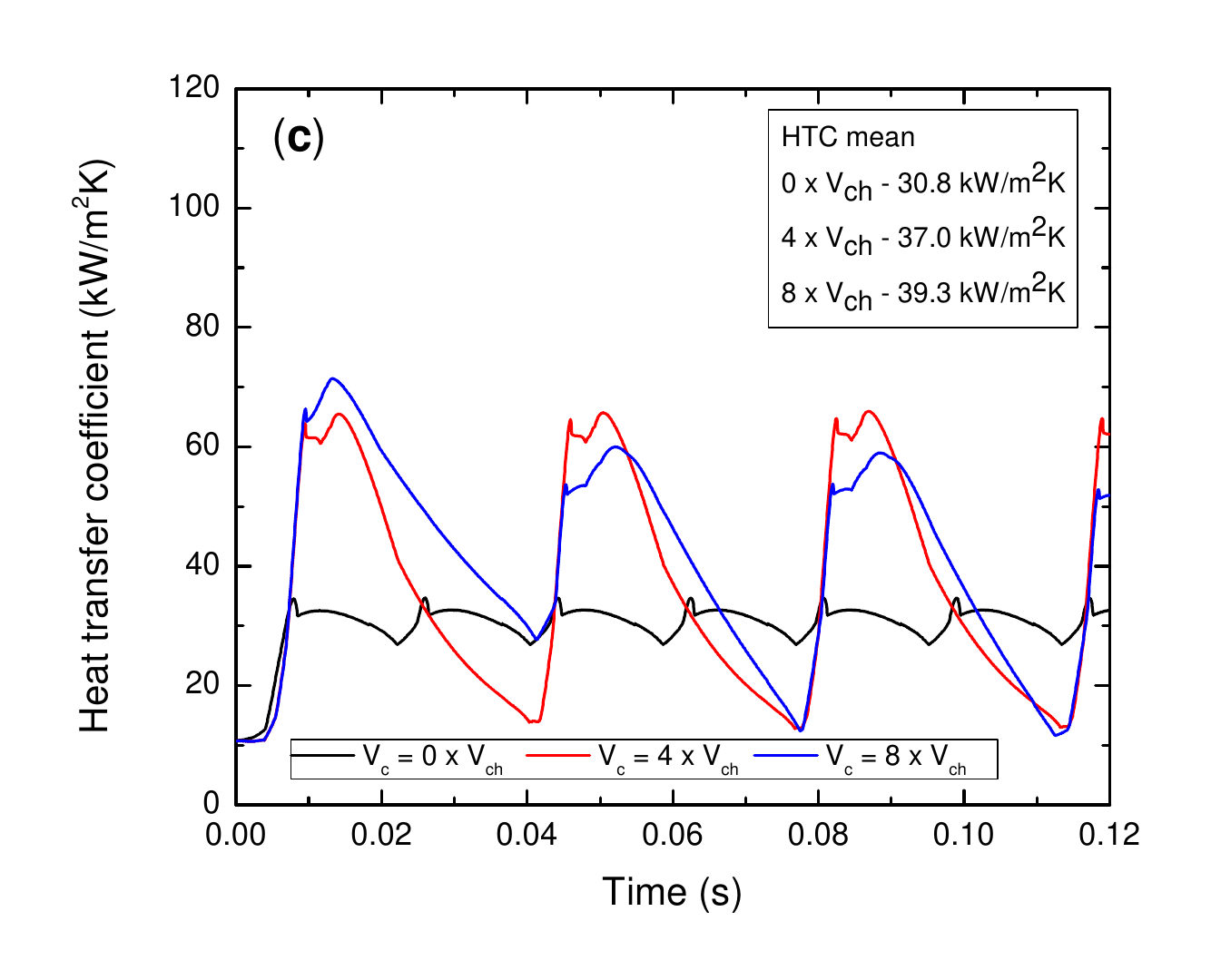}}\hfill
	
	\caption{Effect of inlet compressibility on bubble position, pressure drop and heat transfer coefficient, 0.24 x 0.50 x 40 mm with \(G = 500 \, kg/m^{2}s\), \(q = 200 \, kW/m^{2}\), \(L1 = 0.5 \, L_{ch}\), \(P_{e} = 101 \, kPa\) and \(T_{in} = 373 \,K\).}\label{fig:17}
\end{figure}

\begin{figure}[H]
	\centering
	\subfloat[At \(L = 0.2 \times L_{ch}\). \label{fig:9a}]{\includegraphics[width=0.5\textwidth]{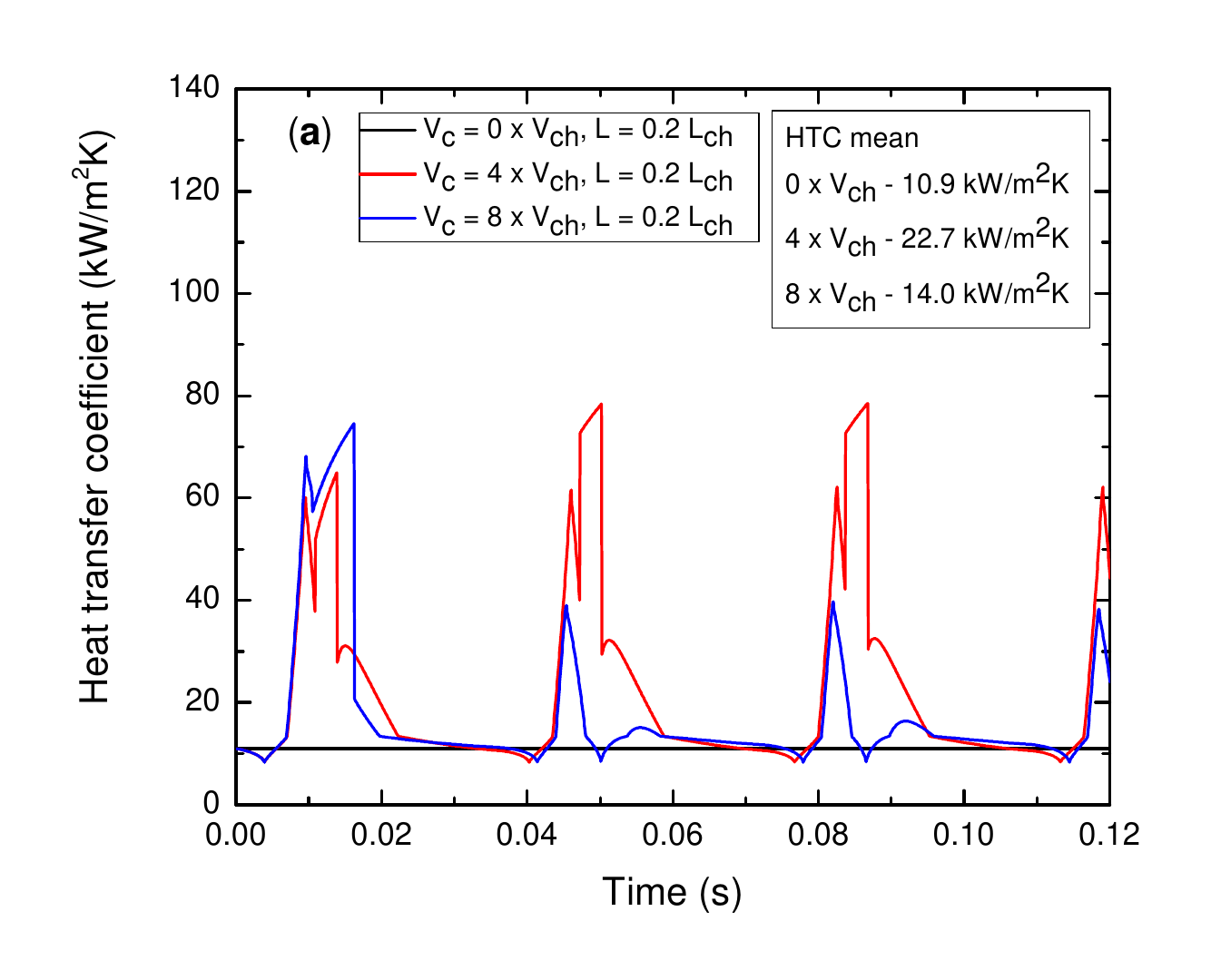}}\hfill
	\subfloat[At \(L = 0.4 \times L_{ch}\).\label{fig:9b}] {\includegraphics[width=0.5\textwidth]{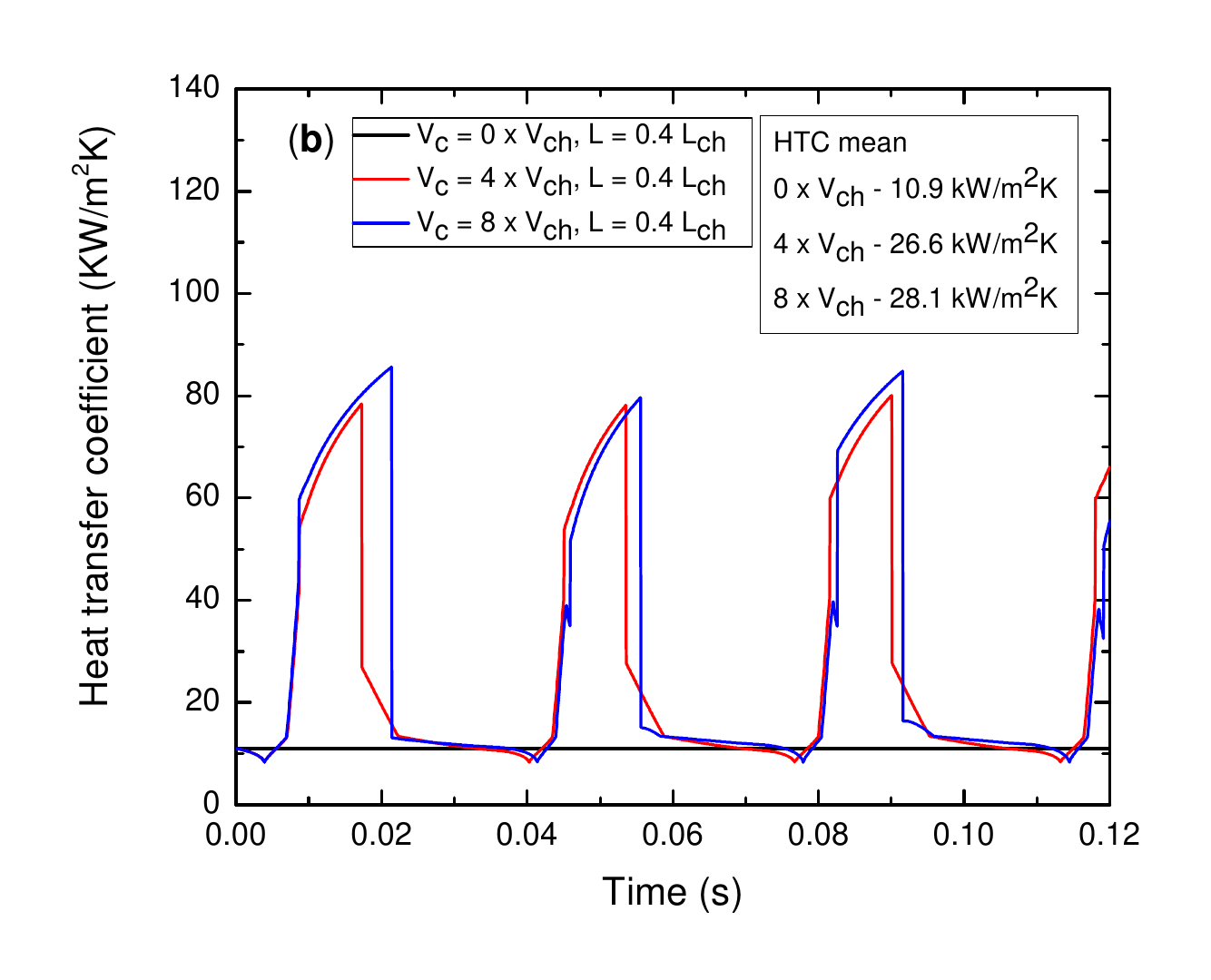}}\hfill
\end{figure}

\begin{figure}[H]
	
	\subfloat[At \(L = 0.6 \times L_{ch}\).\label{fig:9c}] {\includegraphics[width=0.5\textwidth]{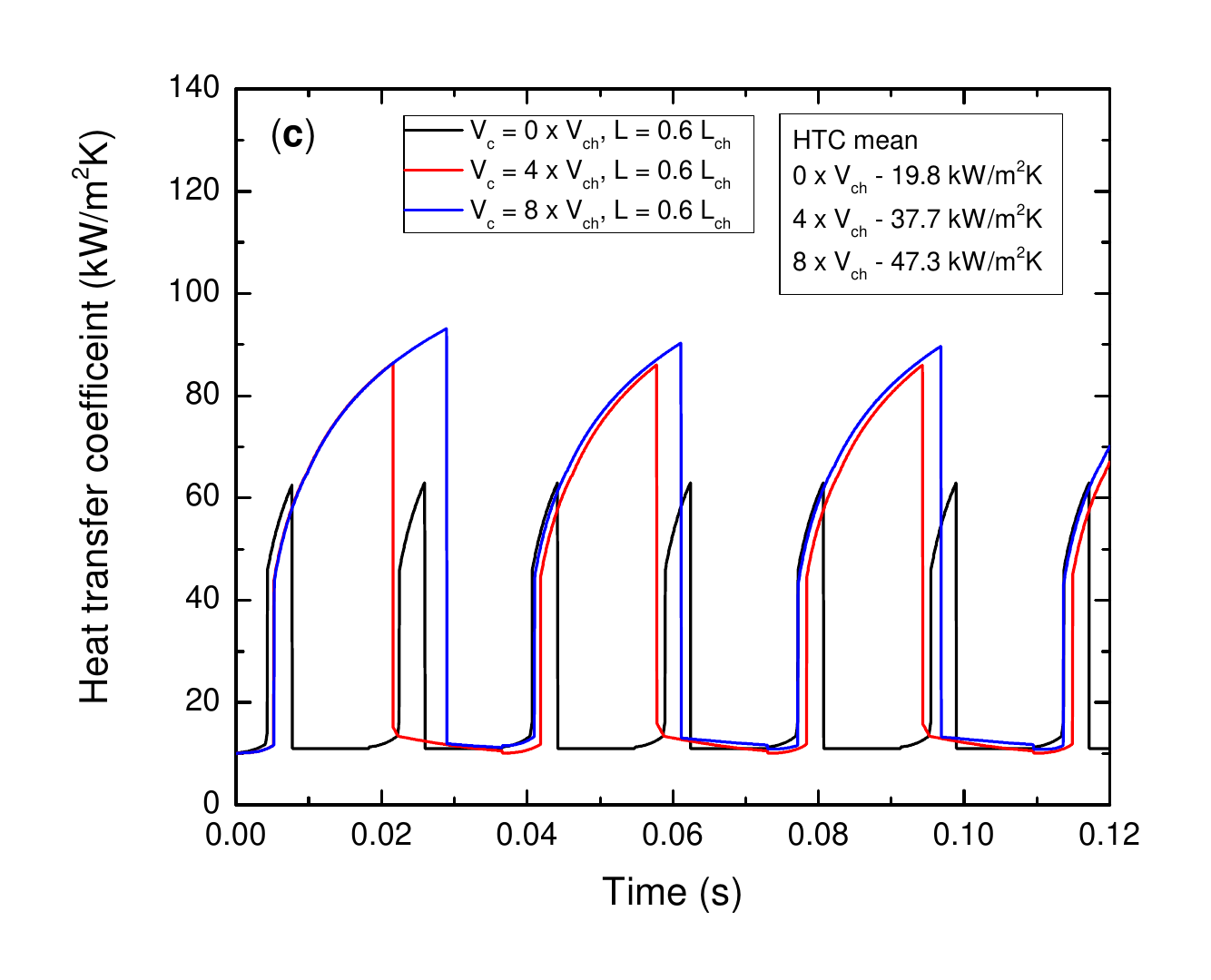}}\hfill
	\subfloat[At \(L = 0.8 \times L_{ch}\).\label{fig:9d}] {\includegraphics[width=0.5\textwidth]{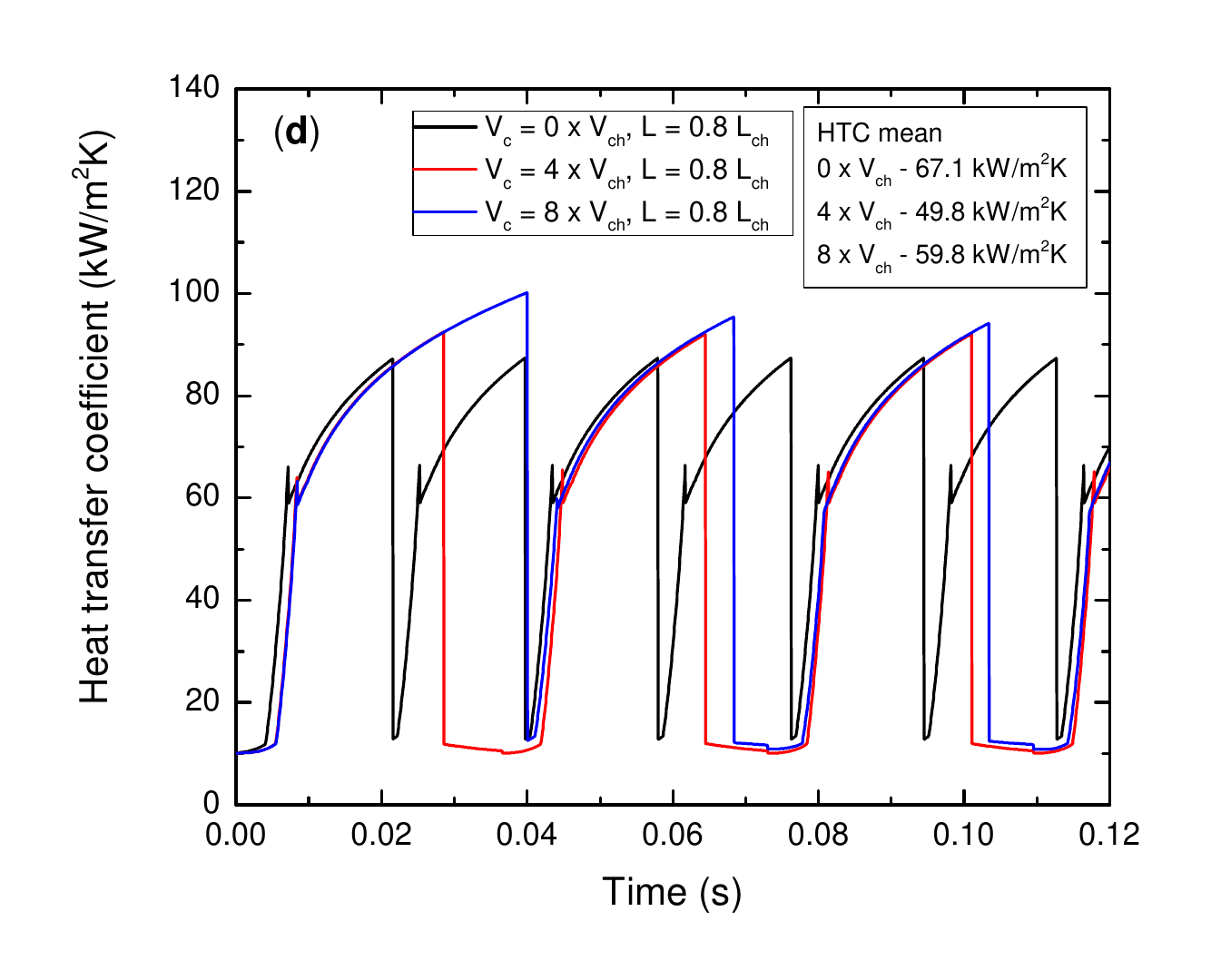}}
	\caption{Variation of heat transfer coefficient at various locations under different inlet compressibility, \(0.24 \times 0.50 \times 40\) mm with \(G = 500 \, kg/m^{2}s\) , \(q = 200 \, kW/m^{2}\),\(L1 = 0.5 \times L_{ch}\), \(P_{e} = 101 \, kPa\) and \(T_{in} = 373 \,K\).} \label{fig:9}
\end{figure}

This section presents the influence of inlet compressibility caused by trapped non-condensable gas on pressure fluctuations, flow reversals and heat transfer coefficients. Modeling the location of nucleation site and nucleation frequency with flow reversals is difficult due to the complex thermo-fluidics and hence the approaches presented in Sections 2.3 and 2.4 are not valid for the cases with flow reversals. Here, for simplicity, the nucleation site is assumed to be in the middle of the channel to account for the bubble growth in the upstream direction during flow reversal. Bubble is assumed to nucleate at regular intervals provided there is no bubble passing over the nucleation site at that instant of time. With inlet compressibility, there is a flow reversal, i.e., the bubble grows in both the upstream direction and the downstream direction as shown in Fig. \ref{fig:17a}.  The flow reversal can be explained using Eq. (\ref{eq:fr1}) and (\ref{eq:fr8}). The flow, after reversing for certain duration, changes its direction and moves forward. The extent of flow reversal is coupled with the change in the inlet (plenum or stagnation) pressure. Fig. \ref{fig:17b} shows that there is a reduction in the amplitude of pressure fluctuation with the increase in compressible volume, which is in line with the experimental observations made by \citet{Liu2013}. This can be attributed to the decrease in the net acceleration (and hence velocities) due to flow reversal. Fig. (\ref{fig:17c}) indicates higher amplitudes of heat transfer coefficient fluctuation and enhanced average heat transfer coefficient in the presence of instabilities. The spatial and temporal averaged values of heat transfer coefficient are 30.8, 37.0 and 39.3 \(kW/m^{2}K\) for no compressible volume, 4 \(V_{ch}\) and 8 \(V_{ch}\) respectively. The maximum heat transfer coefficient is higher for the case with flow reversal due to larger residence period of the confined bubble leading to larger contribution from the thin film evaporation as shown in Fig. \ref{fig:9a} to \ref{fig:9d}. The minimum heat transfer coefficient with flow reversal is lower than that without flow reversal, due to lower velocities (caused by reduction in the net acceleration) of liquid slugs with flow reversals. Fig. \ref{fig:9a} and \ref{fig:9b} show the variation of local heat transfer coefficients with time on the upstream side of the nucleation site and Fig. \ref{fig:9c} and \ref{fig:9d} show the variation  on downstream side of the nucleation site. It is clear that the thin film evaporation exists for a longer duration over a larger length for the cases with flow reversals, resulting in higher heat transfer coefficients.  Higher heat transfer coefficients associated with the flow reversal cases are in line with the experimental observations made by \citet{Gedupudi2011}. \citet{Kenning2001} reported no appreciable change in the heat transfer coefficient due to compressible volume for the range of parameters and dimensions considered in their experimental study. \citet{Liu2013}  reported higher heat transfer coefficients with compressible volume for saturated boiling of water for certain heat flux and mass flux and enhanced heat transfer coefficient only at thermodynamic quality greater than 0.05 for a different heat flux and mass flux. \citet{li2017effect} carried out the experimental investigation of the influence of periodic flow reversal on the performance of microchannel heat exchanger and reported higher heat transfer coefficients, mainly in the upstream part. The results obtained from the present model, shown in Fig. \ref{fig:9a} and \ref{fig:9b}, are very similar to their experimental observations. 


\begin{figure}[H]
	\subfloat[Local heat transfer coefficient variation with time.\label{fig:zone_reva}]
	{\includegraphics[width=0.5\linewidth]{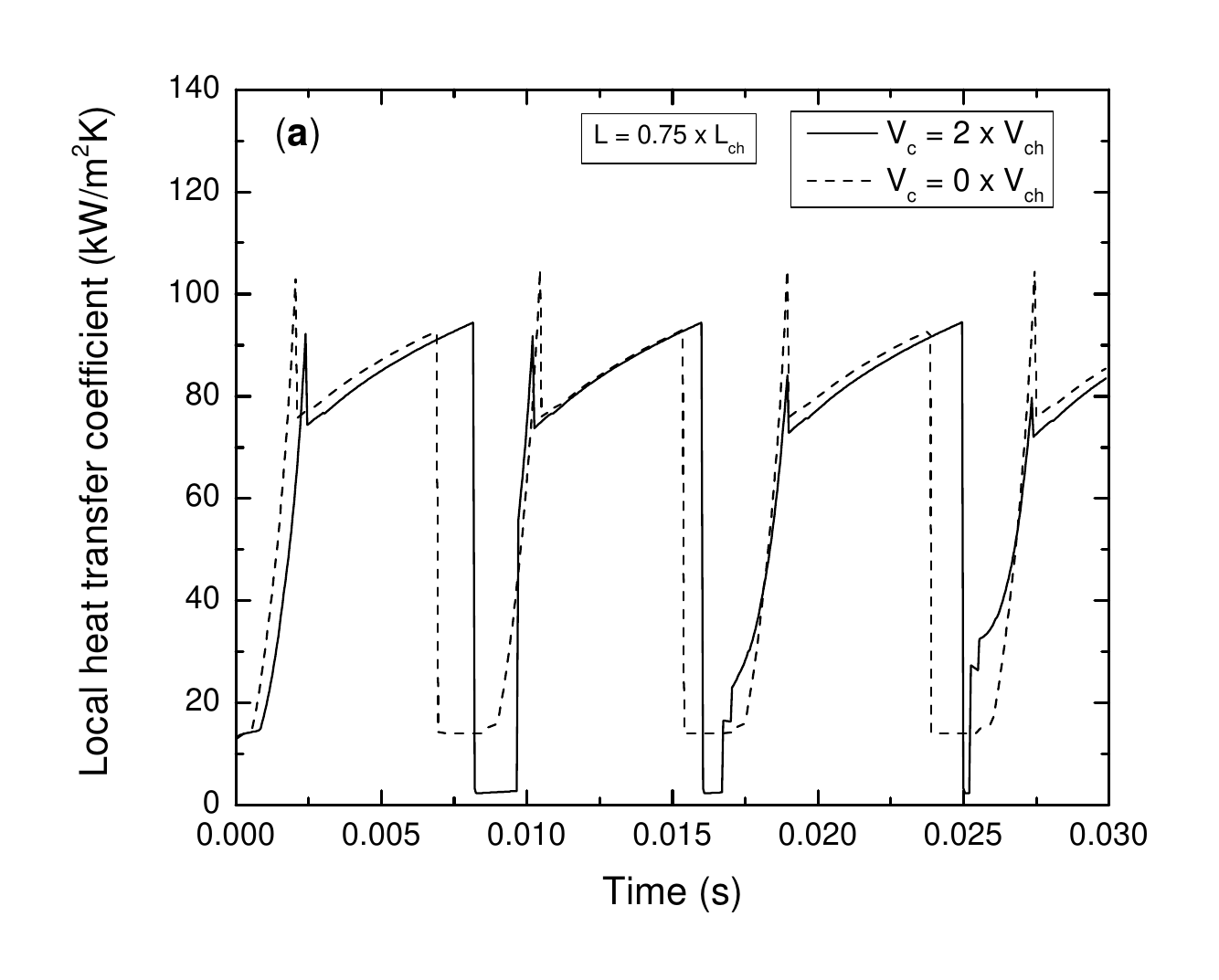}}\hfill
	\subfloat[Local thin film thickness variation with time.\label{fig:zone_revb}]
	{\includegraphics[width=0.5\linewidth]{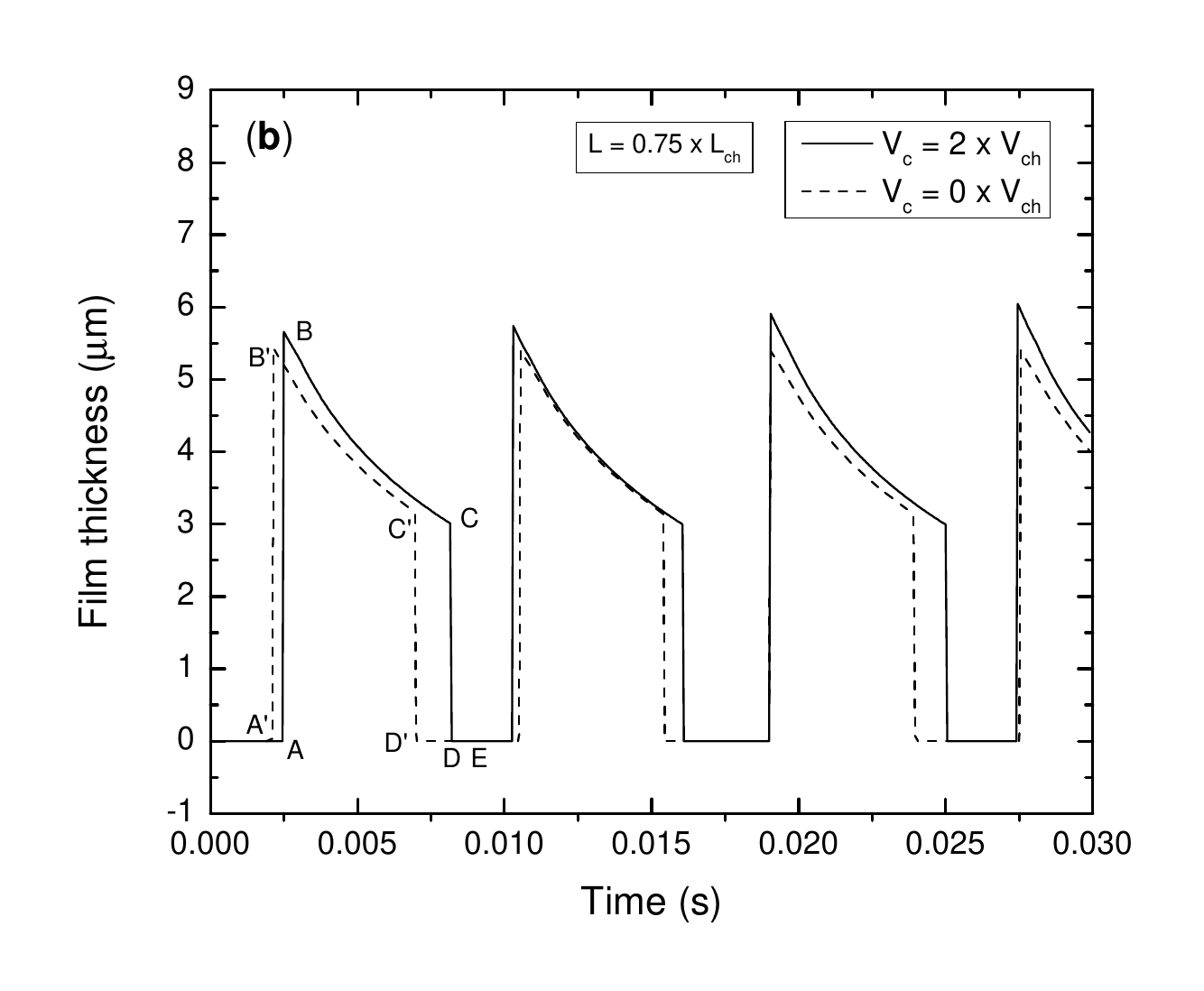}}
	\caption{Local heat transfer coefficient under flow reversal and no flow reversal case, \( 0.20 \times 0.40 \times 25\) mm with \(G = 900 \, kg/m^{2}s\), \(q = 300 \, kW/m^{2}\), \(L1 = 0.5 \times L_{ch}\) and \(P_{e} = 101 \, kPa\).}
	\label{fig:zone_rev}
\end{figure}

Fig. \ref{fig:zone_reva} and Fig. \ref{fig:zone_revb} show the local transient variation of heat transfer coefficient and thin film thickness
respectively. Larger residence time of the bubble for the case of flow reversal, as shown in Fig. \ref{fig:zone_revb}, causes complete dryout on the flat surfaces  leaving the liquid film only at the corners, termed as partial dryout that results in lower local heat  transfer coefficient as shown in Fig. \ref{fig:zone_reva}.  In Fig. \ref{fig:zone_revb}, A-B indicates transition from liquid slug to elongated bubble, B-C depletion of thin film on flat face in elongated bubble region, C-D transition to partial dryout and D-E partial dryout. \citet{tuo2013periodic} observed decrease in heat transfer coefficient due to the emergence of dryout during flow reversal. C'-D' denotes transition from vapour bubble to liquid slug. The film thickness at C reaches minimum film thickness whereas at C' it is higher than minimum film thickness. A-E and A'-D' indicate the bubble residence period for flow reversal case and no flow reversal case respectively.

\subsection{Effect of local pressure}

\begin{figure}[H]
	\centering
	\subfloat[Variation of pressure drop with time. \label{fig:6a}]{\includegraphics[width=0.5\textwidth]{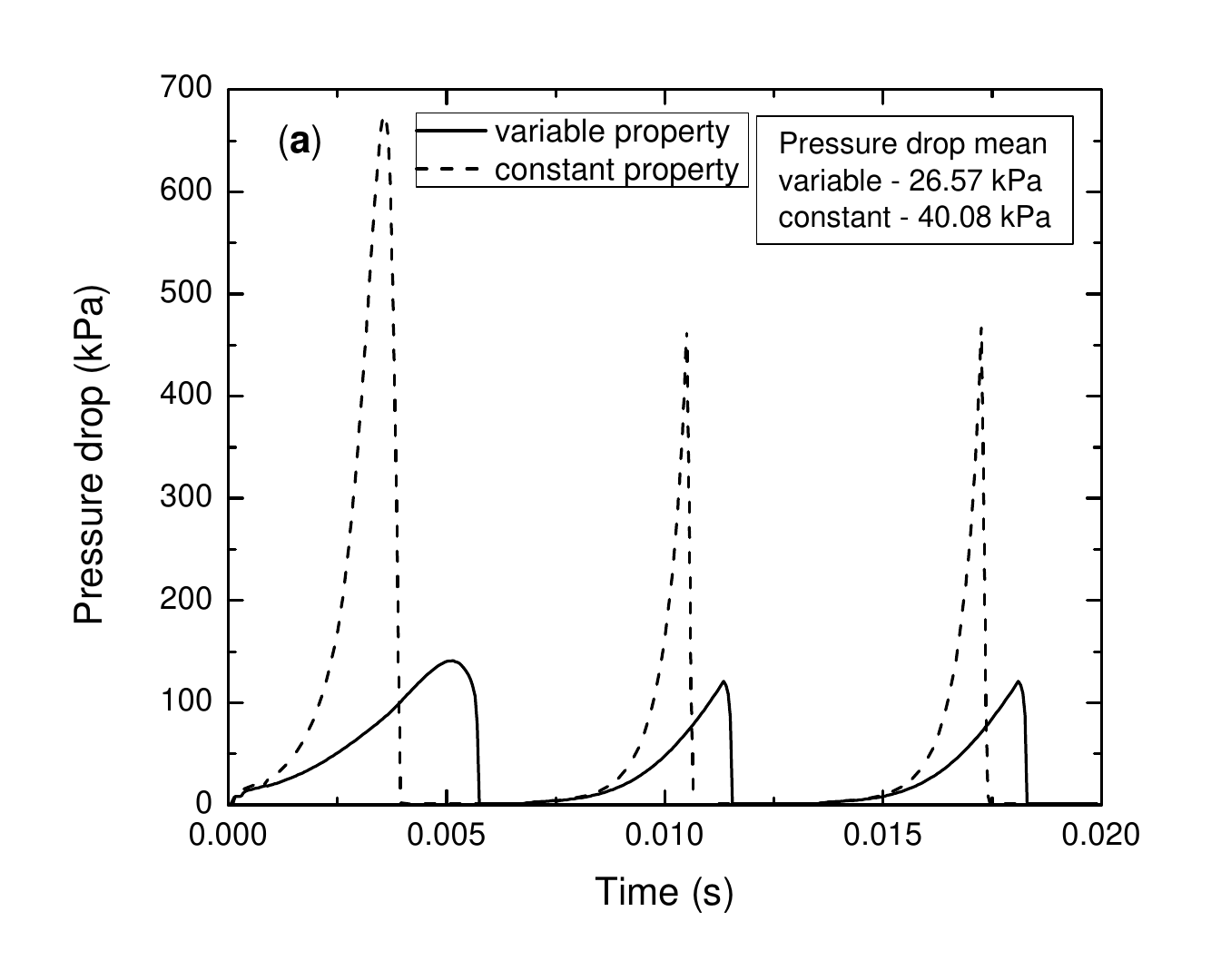}}\hfill
	\subfloat[Variation of spatial averaged heat transfer coefficient with time.\label{fig:6b}] {\includegraphics[width=0.5\textwidth]{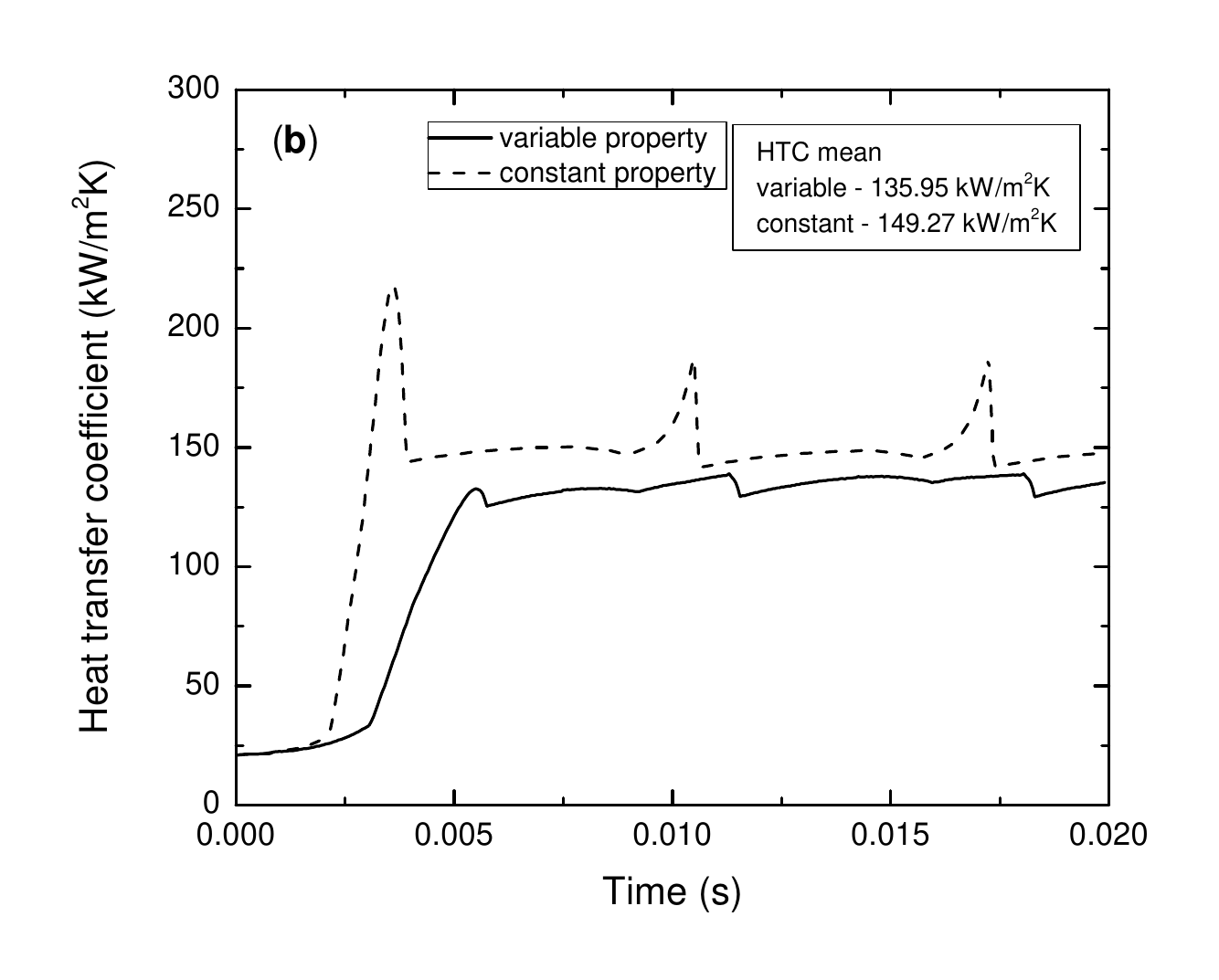}}\hfill
	\caption{Effect of local pressure condition on pressure drop and heat transfer coefficient, \(0.10 \times 0.20 \times 25 \) mm with \(G = 500 \, kg/m^{2}s\), \(q = 100 \, kW/m^{2}\), \(P_{e} = 101 \, kPa\) and \(T_{in} = 373 \, K\).} \label{fig:6}
\end{figure}

\begin{figure}[H]
	\centering
	\subfloat[At \(L = 0.2 L_{ch}\). \label{fig:7a}]{\includegraphics[width=0.5\textwidth]{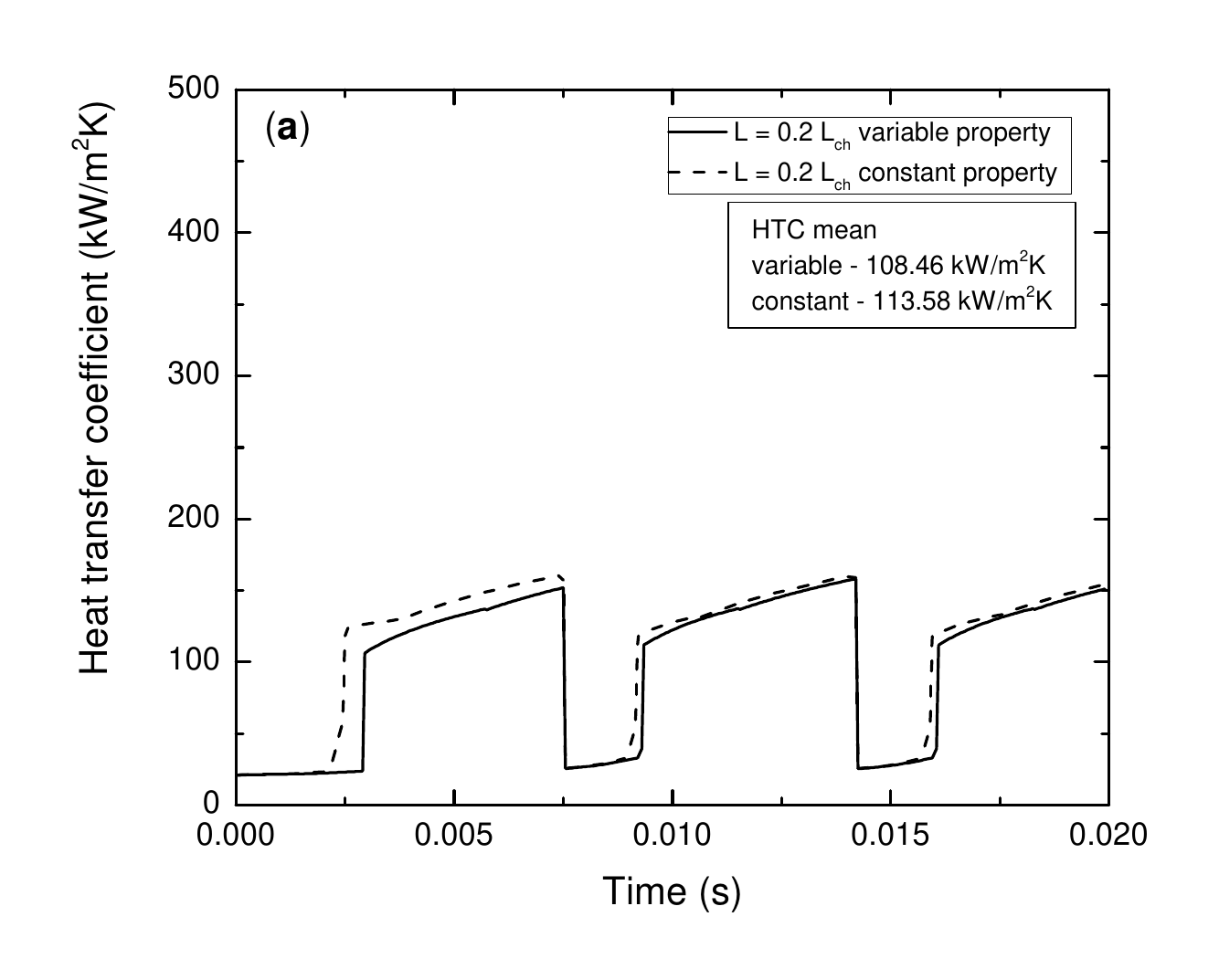}}\hfill
	\subfloat[At \(L = 0.4 L_{ch}\).\label{fig:7b}] {\includegraphics[width=0.5\textwidth]{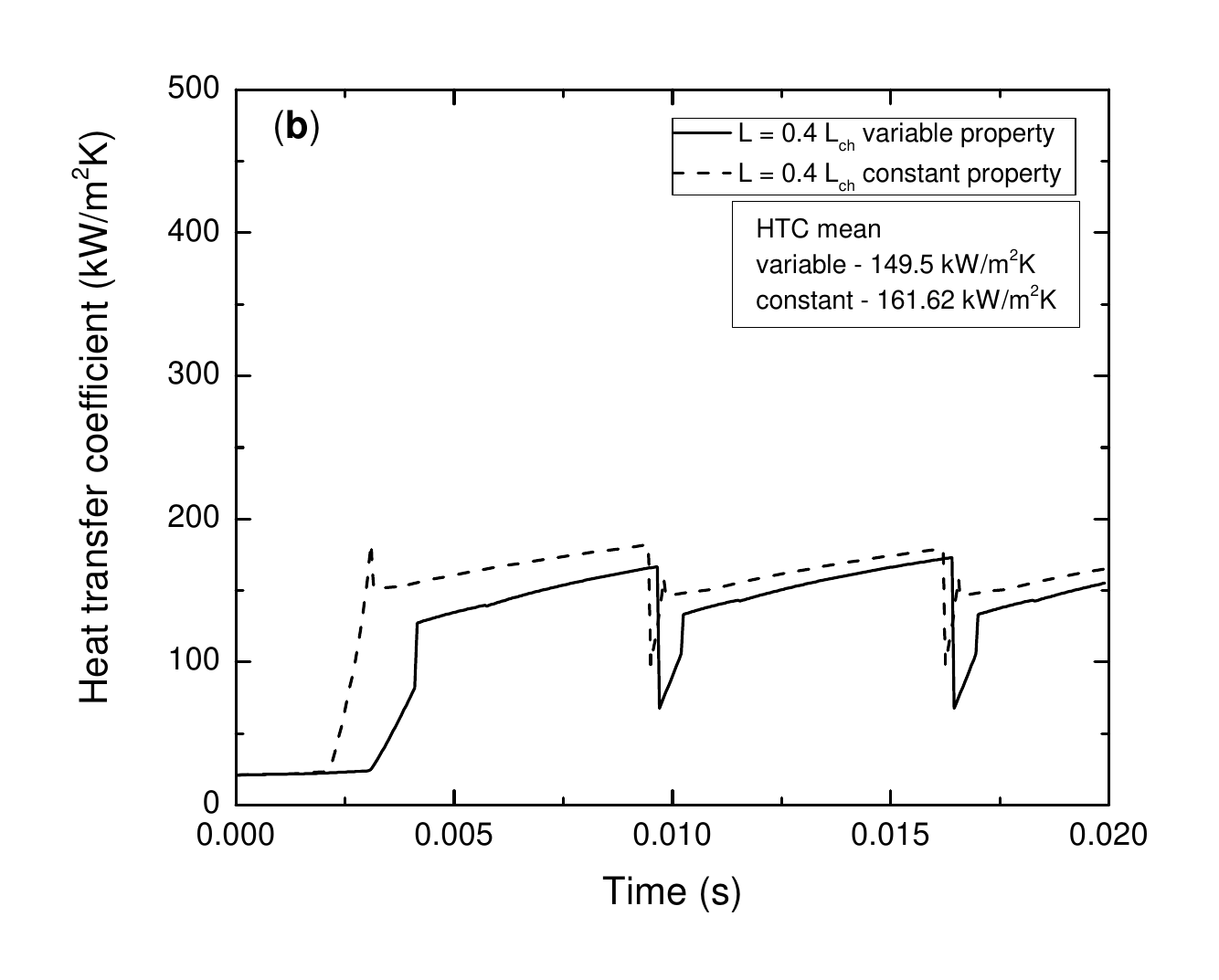}}\hfill
\end{figure}

\begin{figure}[H]
	
	\subfloat[At \(L = 0.6 L_{ch}\).\label{fig:7c}] {\includegraphics[width=0.5\textwidth]{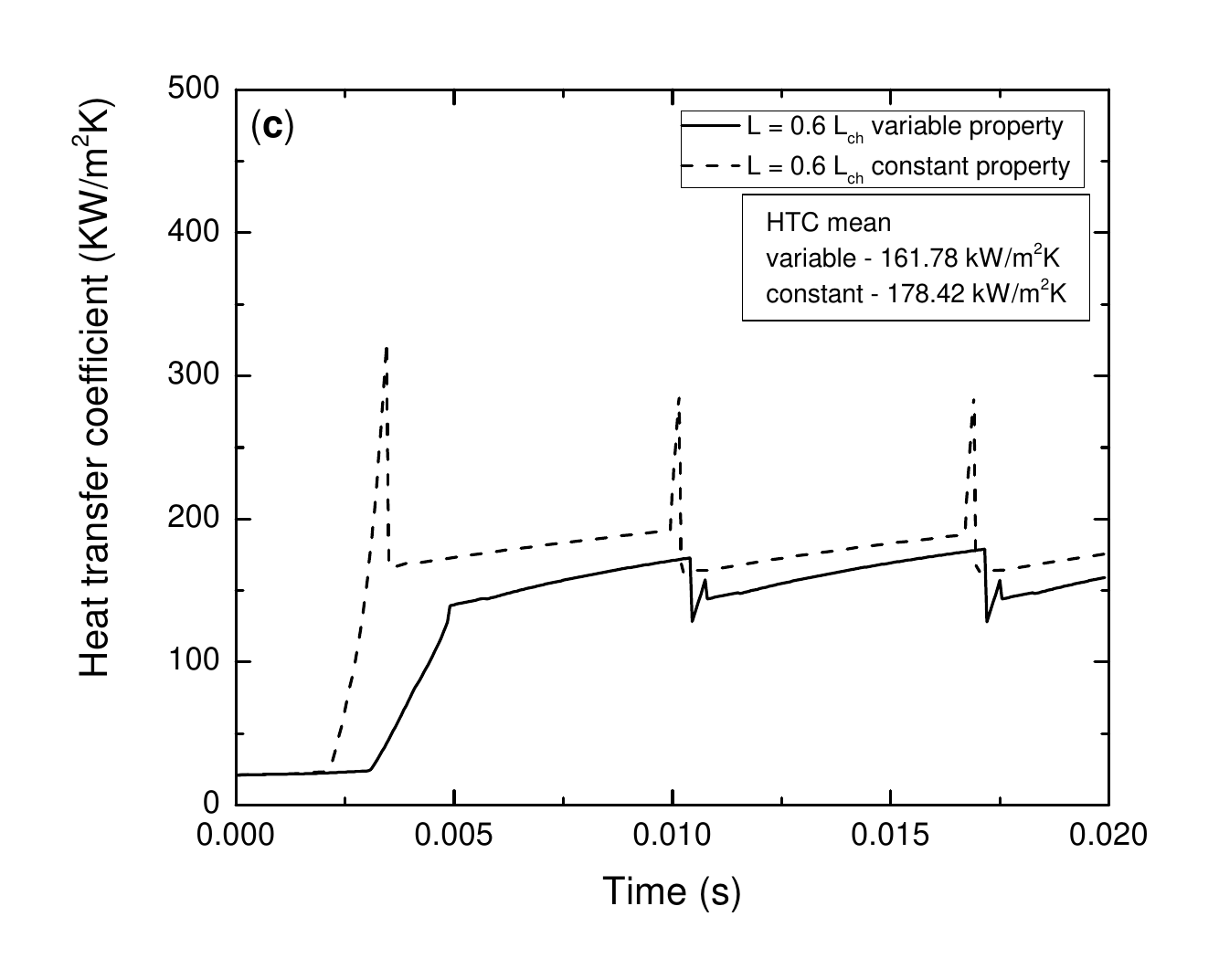}}\hfill
	\subfloat[At \(L = 0.8 L_{ch}\).\label{fig:7d}] {\includegraphics[width=0.5\textwidth]{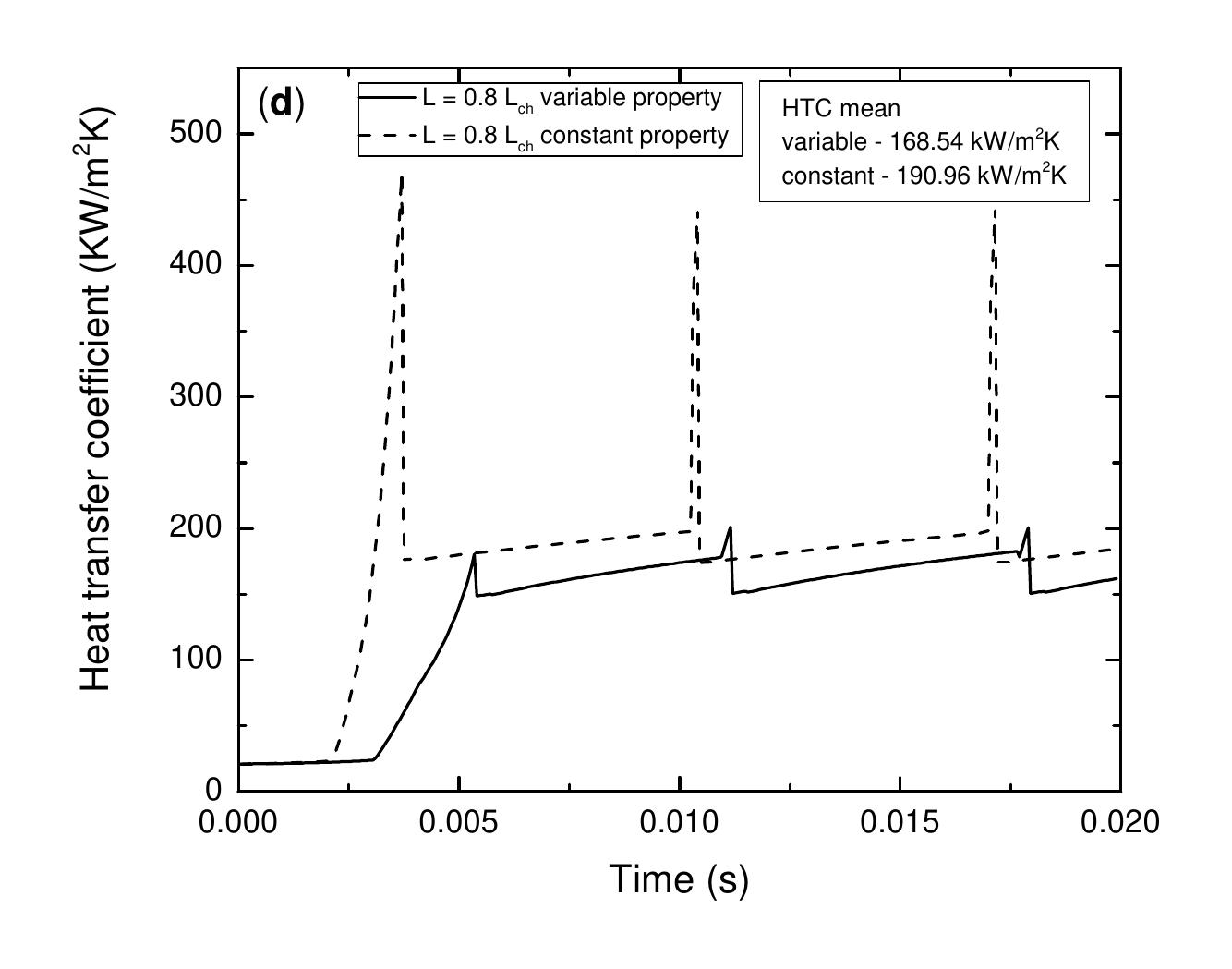}}
	\caption{Effect of local pressure condition on heat transfer coefficient, \(0.10 \times 0.20 \times 25\) mm with \(G = 500 \, kg/m^{2}s\), \(q = 100 \, kW/m^{2}\), \(T_{in} = 373 \,K\) and \(P_{e} = 101 \, kPa\).} \label{fig:7}
\end{figure}

In this section, the effect of local pressure on the pressure drop and the heat transfer coefficient is studied. In the models previously developed by \citet{kew1996pressure}, \citet{Wang2010}, \citet{Thome2004} and \citet{He2016}, fluid property variation with local transient pressure was not considered. Fig. \ref{fig:6a} makes a comparison between the transient variation of channel pressure drop calculated considering the property variation with local pressure and assuming constant pressure equal to exit pressure.  The peak pressure obtained considering the property (vapour density, latent heat, liquid density and liquid viscosity) variation with local pressure is much lower and the bubble residence time is higher than that with constant property case. This is due to the reduction in the bubble acceleration due to property variation with local pressure.  Similarly, Fig. \ref{fig:6b}  the transient variation of channel heat transfer coefficient. The heat transfer coefficient obtained with variable property is lower than that with constant property, due to the changes in initial film thickness and the velocities of accelerated liquid slugs.  Fig. \ref{fig:7a} to \ref{fig:7d} shows the transient variation of local heat transfer coefficient at different axial locations. The difference between the constant property case and variable property case, especially the peak values, increases along the flow direction, due to the increase in the difference between the accelerations (which influence thin film depletion rate and liquid slug velocity) obtained with constant property case and variable property case.



\subsection{Effect of fluid properties}
\begin{figure}[H]
	\centering
	\subfloat[Variation of pressure drop with time. \label{fig:13a}]{\includegraphics[width=0.5\textwidth]{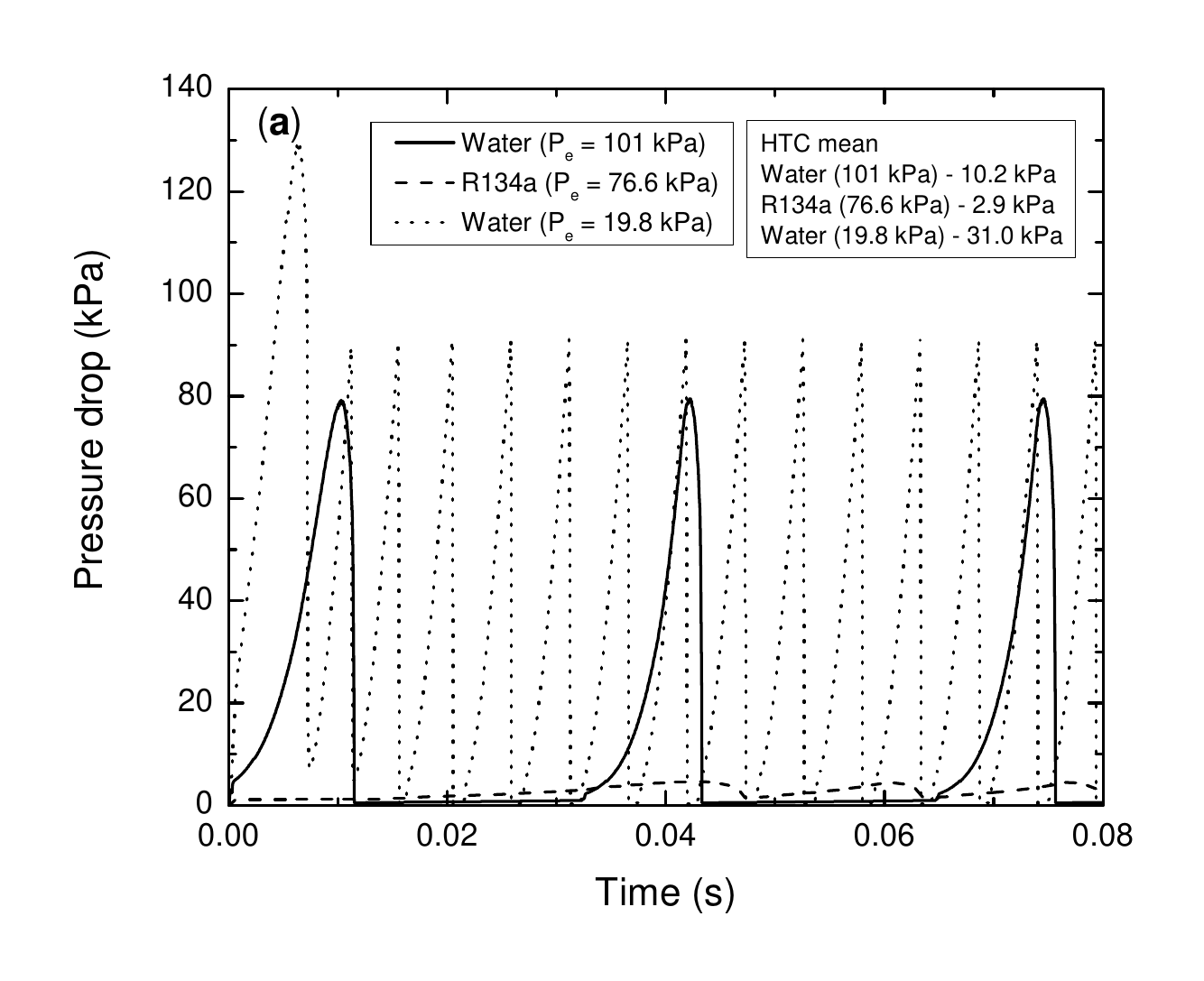}}\hfill
	\subfloat[Variation of heat transfer coefficient with time.\label{fig:13b}] {\includegraphics[width=0.5\textwidth]{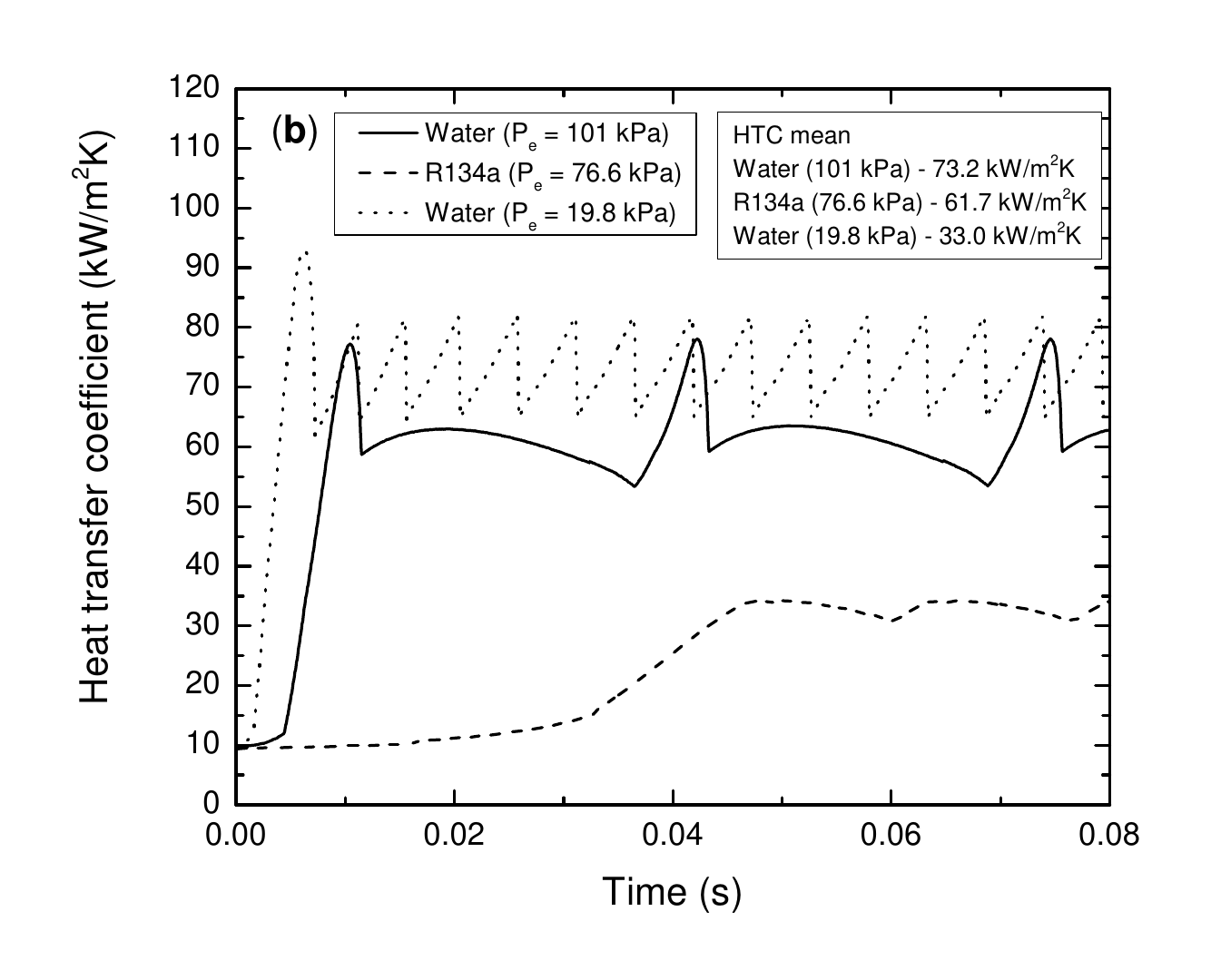}}
	
	\caption{Effect of fluid properties on pressure  drop and heat transfer coefficient, 0.24 \(\times\) 0.50 \(\times\) 40 mm with \(G = 500 \,kg/m^{2}s\), \(q = 100 \,kW/m^{2}\).} \label{fig:13}
\end{figure}

Fig. \ref{fig:13} shows the effect of fluid properties on the channel transient pressure drop and heat transfer coefficient for flow boiling of water (\(P_{e}\) = 101 and 19.8 kPa) and R134a (\(P_{e}\) = 76.6 kPa). For R134a, the expressions for nucleation frequency and initial film thickness are taken from \citet{Thome2004}.  The pressure drop and its amplitude for R134a is almost negligible as compared to that for water, due to the higher value of the product of vapour density and latent heat of vaporization for R134a. Hence, the assumption of constant property for R134a can be justified, but the same assumption cannot be used for water. The amplitude of pressure fluctuation for water  with \(P_{e}\) = 101 kPa is lower than that for water with \(P_{e}\) = 19.8 kPa , as the product of vapour density and latent heat of vaporization is higher for water with \(P_{e}\) = 101 kPa. However, the heat transfer coefficient for R134a is comparable, though lower than that for water.  The heat transfer coefficient for water with \(P_{e}\) = 76.6 kPa is higher than that for water with \(P_{e}\) = 101 kPa due to the higher nucleation frequency and acceleration for the former.  

\section{Conclusions}
In the present study, a 1-dimensional semi-mechanistic model combining pressure and heat transfer coefficient is proposed for flow boiling in a rectangular mini/micro-channel. The major contributions of the present study, based on the available literature, are the evaluation of transient local heat transfer coefficient in conjunction with the local transient pressure, inclusion of the effect of shear stress at liquid-vapour interface on heat transfer coefficient, addition of partial-confined bubble zone to heat transfer model and applying the developed model to evaluate the heat transfer characteristics under flow reversal condition, for flow boiling in a rectangular microchannel. The proposed model is validated with the transient and time-averaged data available in the literature.  The major conclusions that can be drawn from the present work are as follows.
\begin{enumerate}
	\item For high aspect ratio (\(w>>h\)) and shorter channels, the duration over which the partial confinement occurs can be significant and therefore the present model incorporates the same into the heat transfer model. The local heat transfer coefficient is evaluated considering the heat transfer through the thin film and the surrounding bulk liquid. 
	\item The proposed model takes in account the evaporation as well as the shear stress at liquid-vapour interface for the evaluation of rate of thin film depletion during flow boiling in a rectangular mini/micro-channel.  Results indicate that the inclusion of shear stress increases the heat transfer coefficient and also leads to early local dryout. 
	\item The present model has been applied to determine the heat transfer characteristics under flow reversal condition caused by inlet compressibility. Results show that the heat transfer coefficient slightly increases due to the larger contribution from the thin film evaporation resulting from the longer residence period of the elongated bubble. The model also demonstrates that the flow reversal can lead to partial / full dryout based on the operating conditions.   
	\item	The proposed model incorporates the variation of fluid properties with local pressure. The estimated transient and time-averaged pressure drop and heat transfer coefficients are lower than those evaluated with constant pressure equal to the channel exit pressure.  The effect of variable vapour properties is significant for water and is found to be negligible for refrigerants. 
	\item	The present model for a rectangular channel neglects the transient conduction in the thin film. This can be a future study. The present model does not impose thermodynamic equilibrium, hence the estimated exit qualities are lower than the thermodynamic qualities, which seem to match with the CFD prediction in the literature. The presence of superheated liquid slugs, responsible for lowering the quality during flow boiling in mini/microchannels, needs further investigation.
\end{enumerate}

\clearpage
\section*{Nomenclature}
\begin{tabbing}
	xxxxxxxxxxx \= xxxxxxxxxxxxxxxxxxxxxxxxxxxxxxxxxxxxxxxxxxxxxxxx \kill
	xxxxxxxxxxx \= xxxxxxxxxxxxxxxxxxxxxxxxxxxxxxxxxxxxxxxxxxxxxxxx \kill
		\(a\) \> major dimension in hyperellipse \\
	\(A\) \> area,  \(m^{2}\) \\
	\(b\) \> minor dimension in hyperellipse \\
	\(Bo\) \> Boiling number \((Bo = q / G \, i_{lv})\) \\
	\(Bd\) \> Bond number \((Bd = \frac{g \, \Delta \rho \, D_{h}^{2}}{\sigma})\) \\
	\(C1\) \> growth factor \\
	\(C_{p}\) \> specific heat, \(J/kg \, K\) \\
	\(C_{\delta}\) \> factor for initial film thickness \\
	\(Co\) \> confinement number \(\Big(Co = \frac{(\sigma / g \Delta \rho)^{0.5}}{D_{h}})\)\\
	\(cp\) \> channel perimeter, \(m\) \\
	\(D_{h}\) \> hydraulic diameter, \(m\) \\
	\(\Delta T_{sup}\) \> wall superheat \((T_{wall} - T_{sat})\), K \\
	\(dt\) \> time step, \(s\) \\
	\(dz\) \> grid size, \(m\) \\
	
	\(f\) \> Fanning friction factor \\
	\(F_{new}\) \> factor used in correlation \\
	\(g\) \> acceleration due to gravity, \(m/s^{2}\) \\
	\(G\) \> mass flux, \(kg/m^{2}s\) \\
	\(h\) \> height of the channel, \(m\) \\
	
	\(i_{lv}\) \> latent heat of vaporization, \(J/kg\) \\

	\(Ja\) \> Jacob number \((Ja =  \frac{C_{p,l} \, \Delta T_{sup} \, \rho_{l}}{\rho_{v} \, i_{lv}})\) \\
	\(k\) \> conductivity, \(W/m \, K\) \\
	\(K\) \> flow reversal factor \\
	
	\(L\) \> length, \(m\) \\
\(l_{a}\,\,  l_{b}, \, l_{c}\) \> liquid length intercepts, \(m\) \\
\(L1\) \> nucleation site, \(m\) \\
\(L_{ch}\) \> length of channel, \(m\) \\
\(m\) \> last bubble near the exit \\
\(M_{w}\) \> molecular weight, \(kg/mol\) \\
\(n\) \> newest bubble in channel \\
\(n1\) \> factor used in shape determination \\
\(N_{co}\) \> convection number \((N_{co} = (\frac{1-x}{x})^{0.8} \, (\frac{\rho_{v}}{\rho_{l}})^{0.5})\) \\
\(Nu\) \> Nusselt number \\
\(p\)	\> perimeter \\
\(P\) \> pressure, \(Pa\) \\
\(\Delta P\) \> pressure drop, \(Pa\) \\
\(Pr\) \> Prandtl number \\

	\(P_{crit}\) \> critical pressure , \(Pa\) \\
	\(P_{r}\) \> reduced pressure \((P_{r} = P_{sat} / P_{crit})\) \\ 
	\(P_{e}\) \> exit pressure,\(Pa\) \\
	\(q\) \> heat flux, \(W/m^{2}\) \\
	\(r\) \> radial distance, \(m\) \\
	\(R\) \> radius of bubble, \(m\) \\
	\(R_{a}\) \> average roughness, \(\mu \, m\) \\

	\(Re\) \> Reynolds number \((Re = \frac{\rho \, U \, D_{h}}{\mu})\) \\
\(R_{p}\) \>  maximum valley depth, \(\mu \, m\) \\
\(S_{new}\) 	\> factor \\
\(T\) \> 	temperature, \(K\) \\
\(t\) \> 	time, \(s\) \\
\(\overline{T}\) \> mean temperature, \(K\) \\ 
\(u^{*}\)	\> non dimensional velocity, Eq.\ref{eq:38} \\

	\(U\) \> velocity, \(m/s\) \\
	\(V\) \> volume, \(m^{3}\) \\
	\(\dot{V}_{in}\) \> inlet volume flow rate, \(m^{3}/s\) \\
	\(w\) \> width of the channel, \(m\) \\
	\(We\) \> Weber number \((We = \frac{G^{2}\, D_{h}}{\sigma \rho})\) \\
	\(x\) \> quality \\
	\(X\) \> Matinelli parameter \\
	\(z\) \> position, \(m\) \\
\end{tabbing}

\subsection*{Greek letters}
\begin{tabbing}
	xxxxxxxxxxx \= xxxxxxxxxxxxxxxxxxxxxxxxxxxxxxxxxxxxxxxxxxxxxxxx \kill
	xxxxxxxxxxx \= xxxxxxxxxxxxxxxxxxxxxxxxxxxxxxxxxxxxxxxxxxxxxxxx \kill
	$\alpha$	\>aspect ratio \((h/w)\)\\
	$\delta$	\> thin film thickness, \(m\)\\ 
	$ \Gamma$ 	\> gamma function \\
	$\rho$	\> density, $kg/m^{3}$\\
	$\mu$		\>dynamic viscosity, $Pa-s$\\
	$\theta$	\> angle, $rad$\\
	$\sigma$	\> surface tension, $N/m$ \\
	$\tau$		\> time constant or shear stress\\
	$\nu$	\> kinematic viscosity, $m^{2}/s$
\end{tabbing}
\subsection*{Subscripts}
\begin{tabbing}
	xxxxxxxxxxx \= xxxxxxxxxxxxxxxxxxxxxxxxxxxxxxxxxxxxxxxxxxxxxxxx \kill
	\(0\)	\>stagnation point\\
	\(1\)	\> inlet point\\
	\(3\)	\> three sided heating\\
	\(4\)	\> four sided heating\\
	\(4,\infty\) \> developed four sided heating\\
	\(3,\infty\)	\>developed three sided heating\\
	\(acc\)			\>acceleration pressure drop\\
	\(Base\) 		\> base\\
	\(bub\)			\> bubble\\
	\(c\)	\> compressible\\
	\(corner\)	\>corner\\
	\(ch\)			\>channel\\
	\(cs\)	\> cross-section\\
	\(d\)			\>downstream end\\
	\(e\)        \>exit\\
	\(eq\)	\> equivalent \\
	\(elb\) \> elongated bubble \\
	\(est\) \> estimated \\
	\(film\)	\> film\\
	\(h\)	\> heated length\\
	\(grow\) \> growth \\
	\(heat\) \> heated \\
	\(i\)	\> channel inlet\\
	\(ini\) \> initial \\
	
	\(l\)        \>liquid\\
	\(lv\) \> liquid-vapor \\
	\(lo\)        \>liquid only\\
	\(max\)        \>maximum\\
	\(mean\)        \>mean\\
	\(pc\) 	\> partially confined \\
	\(pdr\) \> partial dryout region \\
	\(slug\)        \>liquid slug between two bubbles\\
	\(sp\)        \>single phase\\
	\(sat\)        \>saturated\\
	\(sub\)        \>sub-cooled\\
	\(sup\)        \>superheat\\
	\(tp\)        \>two-phase\\
	\(th\) \> thermodynamic \\
	\(u\)        \>upstream end\\
	\(v\)        \>vapor\\
	\(vis\)        \>viscous pressure drop\\
	\(wall\)        \>wall\\
	\(wait\)	\> waiting 
\end{tabbing}

\subsection*{Abbreviations}
\begin{tabbing}
	xxxxxxxxxxx \= xxxxxxxxxxxxxxxxxxxxxxxxxxxxxxxxxxxxxxxxxxxxxxxx \kill
	\(CFD\)	\> Computational fluid dynamics \\
	\(HTC\) \> Heat transfer coefficient
\end{tabbing}

\newpage

\begin{table}[H]
	\caption{Various correlations for heat transfer coefficient}
	\begin{center}
		\label{table}
		\renewcommand\arraystretch{0.9}
		\resizebox{1.1\textwidth}{!}{
			
			\begin{tabular}{|C{2.5cm}|c|c|}
				\hline
				Reference & Correlation & Comments \\
				\hline
				\citet{lazarek1982evaporative} &  \(HTC_{tp} = \frac{k_{l}}{D_{h}} \, (30 \, Re_{lo}^{0.857} \, Bo^{0.714})\) & Mini channel correlations based on  728 points,\\ & & \(D_{h} = 3.1 \, mm\) \\ & & \(G = 125 - 750 \, kg/m^{2}s, Bo = 2.3 - 76  \times 10^{-4}\)	\\ & & \(P = 1.3 - 4.1 \, bar\)	\\ 
				\hline
				\citet{Kew1997} & \(HTC_{tp} = \frac{k_{l}}{D_{h}} \, (30 \, Re_{lo}^{0.857} \, Bo^{0.714}) \,  (1-x)^{-0.143} \) & Mini/micro channel correlation  \\ & & Based on circular data \(D_{h} = 1.39 - 3.69 \, mm\) , \\ & & working fluid R141b \\
				\hline
				\citet{Sun2009} & \(HTC_{tp} = \frac{6 \, Re_{lo}^{1.05} \, Bo^{0.54}}{We_{l}^{0.191}  (\rho_{l}/\rho_{v})^{0.142}} \, \frac{k_{l}}{D_{h}}\) &  Mini/micro channel correlation \\ & & Based on  11 different fluid and 2505 data points \\ & & \(D_{h} = 0.21 - 6.05 \, mm\) \\ & & \(G = 44 - 1500\, kg/m^{2}s\, , q = 5 - 109 \, kW/m^{2}\) \\
				\hline 
				\citet{li2010general} & \(HTC_{tp} = 334 \, Bo^{0.3} \, \left(Bd \, Re_{l}^{0.36} \right)^{0.4} \, \frac{k_l}{D_{h}}\)& Mini-micorchannel correlation \\ & & Based on 3700 data points, 13 different fluids \\ & & \(D_{h} = 0.2 - 3 \, mm\) \\
				\hline
				\citet{mohamed2012statistical} & For \(D_{h} = 0.52 \, mm\) , \(x \leq 0.3\) & Mini/micro channel correlation \\ & \(HTC_{tp} = 3320 \frac{Bo^{0.63} \, We_{l}^{0.2} \, Re_{l}^{0.11}}{Co^{0.6}} \, \frac{k_{l}}{D_{h}}\) & Based on 8561 data points and R134a - working fluid 
				\\ &  For \(D_{h} = 0.52 \, mm\) , \(x > 0.3\) &  \(D_{h} =  0.52 - 4.26 \, mm \), \(G = 100-500 \, kg/m^{2}s\)
				\\ & \(HTC_{tp} = 5324 \, \left[\frac{Bo^{0.3} \, We_{l}^{0.25}}{N_{co}^{0.25}}\right]^{1.79} \, \frac{k_{l}}{D_{h}}\) & \(q = 2.4 - 175.4 \, kW/m^{2}\), \(P = 6- 14 \, bar\) \\
				\hline
				\citet{mahmoud2013heat} & \(HTC_{tp} = S_{new} \, HTC_{Cooper} + F_{new} \, HTC_{l}\) &  Mini/micro channel correlation 
				\\ & & Based on 5152 data points and R134a-working fluid
				\\ & \(HTC_{Cooper} =  55 \, P_{r}^{0.12- 0.434\, ln (R_{p})} (-log\, P_{r})^{-0.55} \, M_{w}^{-0.5} \, q^{0.67} \) &  \(D_{h} =  0.52 - 4.26 \, mm \), \(G = 100-700 \, kg/m^{2}s\) 
				\\ & \(HTC_{l} = 4.36 \, k_{l}/D_{h} \) for \(Re_{l} < \, 2000\) & \(q = 1.7 - 158 \, kW/m^{2}\), \(P = 6- 14 \, bar\)
				\\ & \(HTC_{l} = 0.023 \, Re_{l}^{0.8} \, Pr_{l}^{0.4} \, \frac{k_{l}}{D_{h}} \) for \(Re_{l} > \, 3000\) &
				\\ & \(Re_{l} = \frac{(1-x) \, G \, D_{h}}{\mu_{l}}\) &
				\\ & \(F_{new} = (1 + \frac{2.812 \, Co^{-0.408}}{X})^{0.64}\) &
				\\ & \(S = \frac{1}{1 + 2.56 \times 10^{-6}  (Re_{l} F_{new}^{1.25})^{1.17}}\) &
				\\ & \(X = (\frac{f_{l}}{f_{v}})^{0.5} \, (\frac{\rho_{v}}{\rho_{l}})^{0.5}  \, \frac{(1-x)}{x}\) & \\
				\hline
				\citet{Lee2005} & For \(0 < \,x \leq \,0.05\), \(HTC_{tp} = 3.856 X^{0.267} \, HTC_{l}\) & Microchannel correlation \\
				& \(HTC_{l} = Nu_{3,\infty} \, k / D_{h}\) & Based on water - 207 data points, \\
				& \(X_{vv} = (\frac{\mu_{l}}{\mu_{v}})^{0.5} \, (\frac{1-x}{x})^{0.5} \, (\frac{\rho_{v}}{\rho_{l}})^{0.5}\) &  R134a - 111 data points \\ 
				& \(X_{vt} = (\frac{f_{l} \, Re^{0.2}_{v} }{0.079})^{0.5} \, (\frac{1-x}{x}) \, (\frac{\rho_{v}}{\rho_{l}})^{0.5}\)& \(D_{h} = 349 \, \mu m\) \\
				& \(Re_{v} = G \, x \, D_{h}/ \mu_v\) & \\
				& For \(0.05 < x \leq 0.55\) , \(HTC_{l} = 436.48 \, Bo^{0.522} \, We^{0.351}_{l} \, X^{0.665} \, HTC_{l}  \) & \\
				& For \(0.55 < x \leq 1, \, HTC_{l}  = MAX(108.6 \, X^{1.665} \, HTC_{v} , \, HTC_{v}) \) &\\
				&\(HTC_{v} = Nu_{v} k_{v}/ D_{h}\) & \\
				& \(Nu_{v} = 0.023 \, Re^{0.8}_{v} \, Pr^{0.4}_{v}\) & \\
				\hline
		\end{tabular}}
	\end{center}
\end{table}
\clearpage
\bibliography{mybibfile}

\end{document}